\documentclass[paper, 12pt, letterpaper, epsf]{JHEP}
\input{epsf.tex}
\usepackage{graphics}
\usepackage{epsfig}
\usepackage{amsmath}
\usepackage{amssymb}
\usepackage{multirow}
\usepackage{anysize}

\usepackage{graphics}
\usepackage{latexsym,amsmath,amssymb,amscd,cite}
\usepackage{pdfsync}
\usepackage{epsf}
\usepackage[all]{xy}
\usepackage{amsthm}
\DeclareSymbolFont{extraup}{U}{zavm}{m}{n}
\DeclareMathSymbol{\varheart}{\mathalpha}{extraup}{86}
\DeclareMathSymbol{\vardiamond}{\mathalpha}{extraup}{87}
\newcommand{\bdiamond}{{\scriptstyle \vardiamond}}
\newcommand{\cinf}{{{\cal \cC}^\infty(M,\K)}}
\newcommand{\cpinf}{{{\cal \cC}^\infty_\perp({\hat M},\K)}}

\newcommand{\bcE}{\boldsymbol{\check{E}}}

\newtheorem{Theorem}{Theorem}[section]
\newtheorem{Lemma}[Theorem]{Lemma}
\newtheorem{Proposition}[Theorem]{Proposition}
\theoremstyle{definition}
\newtheorem{Definition}[Theorem]{Definition}
\theoremstyle{remark}

\newcommand{\vectornorm}[1]{||#1||}
\newcommand{\Hom}{{\rm Hom}}

\def\ad{\mathrm{ad}}
\newcommand{\D}{{\check{D}^\ad}}

\newcommand{\im}{\mathrm{im}}
\newcommand{\cyl}{\mathrm{cyl}}
\newcommand{\cone}{\mathrm{cone}}
\newcommand{\vt}{\mathrm{vert}}

\newcommand{\bp}{\begin{Proposition}}
\newcommand{\ep}{\end{Proposition}}
\newcommand{\bl}{\begin{Lemma}}
\newcommand{\el}{\end{Lemma}}
\newcommand{\bt}{\begin{Theorem}}
\newcommand{\et}{\end{Theorem}}
\newcommand{\bd}{\begin{Definition}}
\newcommand{\ed}{\end{Definition}}
\newcommand{\End}{\mathrm{End}}

\newcommand{\eqdef}{\stackrel{{\rm def.}}{=}}

\DeclareFontFamily{U}{rsf}{}
\DeclareFontShape{U}{rsf}{m}{n}{<5> <6> rsfs5 <7> <8> <9> rsfs7 <10-> rsfs10}{}
\DeclareMathAlphabet\Scr{U}{rsf}{m}{n}

\newcommand{\KA}{K\"{a}hler-Atiyah~}

\def\Z{{\Bbb Z}}
\def\C{{\Bbb C}}
\def\R{{\Bbb R}}

\def\K{{\Bbb K}}

\def\dd{\mathrm{d}}

\def\vol{\mathrm{vol}}
\def\AdS{\mathrm{AdS}}

\def\w{\Omega}

\def\ccyl{{\tiny \begin{array}{c}\cyl\\\cone\end{array}}}

\newcommand{\be}{\begin{equation*}}
\newcommand{\ee}{\end{equation*}}
\newcommand{\ben}{\begin{equation}}
\newcommand{\een}{\end{equation}}
\newcommand{\beqa}{\begin{eqnarray*}}
\newcommand{\eeqa}{\end{eqnarray*}}
\newcommand{\beqan}{\begin{eqnarray}}
\newcommand{\eeqan}{\end{eqnarray}}

\newcommand{\nn}{\nonumber}

\newcommand{\id}{\mathrm{id}}

\def\cC{{\mathcal C}}

\def\cB{\Scr B}

\def\Cl{\mathrm{Cl}}

\def\cK{\mathrm{\cal K}}

\def\Spin{\mathrm{Spin}}

\def\SO{\mathrm{SO}}

\def\cE{\mathcal{E}}
\def\cI{\mathcal{I}}

\def\cC{\mathcal{C}}

\def\G_2{\mathrm{G_2}}
\def\cO{\mathcal{O}}
\def\cL{\mathcal{L}}

\title{Geometric algebra techniques in flux compactifications (II)}
\author{Calin-Iuliu Lazaroiu$^{1,2}$, Elena-Mirela Babalic$^1$\\
$^1$Horia Hulubei National Institute of Physics and Nuclear Engineering (IFIN-HH),
Bucharest-Magurele, Romania
\\$^2$Center for Geometry and Physics, Institute for Basic Science and 
Department of Mathematics, POSTECH, Pohang, Gyeongbuk 790-784, Korea\\
\vspace{0.2cm}calin@ibs.re.kr, mbabalic@theory.nipne.ro, icoman@theory.nipne.ro\\}

\abstract{We study constrained generalized Killing
spinors over the metric cone and cylinder of a (pseudo-)Riemannian manifold, 
developing a toolkit which can be used to investigate certain problems
arising in supersymmetric flux compactifications of
supergravity theories. Using geometric algebra techniques, we give
conceptually clear and computationally effective methods for
translating supersymmetry conditions for the metric and fluxes of the unit section of such cylinders
and cones into differential and algebraic
constraints on collections of differential forms defined on the cylinder or cone. 
In particular, we
give a synthetic description of Fierz identities, which are an
important ingredient of such problems. As a non-trivial application,
we consider the most general ${\cal N}=2$ compactification of
eleven-dimensional supergravity on eight-manifolds.}

\preprint{}

\begin{document}

\tableofcontents

\pagebreak

\vskip .6in

\section{Introduction}
\label{sec:intro}

A central problem in the study of flux compactifications of supergravity and string
theories is that of finding  geometric descriptions of supersymmetry conditions
for various backgrounds in the presence of fluxes. This leads to beautiful and
highly non-trivial connections with various subjects in differential geometry \cite{GauntlettWaldram, Agricola,
  MartelliSparks, Tsimpis, WittGenG2, GabellaSparks}. As pointed out in
\cite{ga1}, the general problem of re-formulating supersymmetry conditions for flux backgrounds admits a powerful resolution
based on geometric algebra techniques \cite{Chevalley, Riesz, Graf, GA0,GA1,GA2}, an approach which is highly
advantageous from a conceptual and computational standpoint.

The purpose of this paper is to combine the methods and ideas of \cite{ga1}
with an extension of the cone formalism of \cite{Bar}, providing a re-formulation of the
latter within the theory of \KA algebras and bundles and developing a toolkit
which can be used to solve a series of problems arising in the study of
certain classes of flux compactifications. In particular, we show how the geometric re-formulation of generalized Killing spinor equations which
was given in \cite{ga1} can be lifted from a 
(pseudo-)Riemannian manifold to its metric cylinder and metric cone --- a 
correspondence which is of particular interest in certain situations when one cannot
encode the supersymmetry conditions through a reduction of the structure group
of the compactification manifold itself. 
As a non-trivial example, Section \ref{sec:application} applies such
techniques and results to the study of the most general flux
compactifications of M-theory on eight-manifolds preserving ${\cal N}=2$
supersymmetry in 3 dimensions --- a class of solutions which was not
analyzed in full generality before (our generalization compared to the
celebrated work of \cite{Becker} being that we do {\em not} impose any
chirality conditions on the internal part of the supersymmetry
generators).  In that example, we have a single algebraic condition
$Q\xi=0$, with $Q=\frac{1}{2}\gamma^m\partial_m\Delta
-\frac{1}{288}F_{m p q r}\gamma^{m p q r}- \frac{1}{6}f_p \gamma^p
\gamma^{(9)} -\kappa\gamma^{(9)}$ and $A_m=\frac{1}{4}f_p\gamma_{m}{}^
{p}\gamma^{(9)}+ \frac{1}{24}F_{m p q r}\gamma^{ p q r}+\kappa
\gamma_m\gamma^{(9)}$. Using our methods, we extract a highly
non-trivial system of differential and algebraic relations for the
associated spinor bilinears, which encodes the geometric constraints
imposed on such backgrounds by the requirement that they preserve the
stated amount of supersymmetry. For reasons of conceptual and
computational convenience, we express such equations in terms of
certain combinations of iterated contractions and wedge products which
are known as `generalized products', whose role and
origin was explained in \cite{ga1}. The reader is encouraged at this point 
to take a look at Section \ref{sec:application}, which
should provide an illustration of the results and techniques developed in the present
paper. A full analysis of those equations and of their physical
consequences, as well as certain other applications of this formalism,
are taken up in subsequent work.

The paper is organized as follows. In Section \ref{sec:cone}, we show how 
a variant of the cone formalism of \cite{Bar} and of
its cylinder version can be constructed within the geometric algebra approach.  
In Section \ref{sec:application}, we apply this formalism to the study of the
most general ${\cal N}=2$ compactifications of M-theory on eight-manifolds. We
conclude in Section \ref{sec:conclusions} with a few remarks on further
directions. The physics-oriented reader can start with Section
\ref{sec:application}, before delving into the technical and
theoretical details of the other sections. This paper is written as a
companion to \cite{ga1}, to which we shall refer repeatedly. Therefore, the
reader should have a copy of \cite{ga1} at hand when approaching the formal
developments of Section \ref{sec:cone}. 

\paragraph{Notations.} As in \cite{ga1}, we let $\K$ denote one of the fields $\R$ and $\C$ of
real and complex numbers, respectively. We work in the smooth
differential category, so all manifolds, vector bundles, maps,
morphisms of bundles, differential forms etc. are taken to be
smooth. We further assume that our connected and smooth manifolds $M$
are paracompact, so that we have partitions of unity subordinate to
any open cover. If $V$ is a $\K$-vector bundle over $M$, we let
$\Gamma(M,V)$ denote the space of smooth ($\cC^\infty$) sections of
$V$. We also let $\End(V)=\Hom(V,V)\approx V\otimes V^\ast $ denote the
$\K$-vector bundle of endomorphisms of $V$, where
$V^\ast=\Hom(V,\cO_\K)$ is the dual vector bundle to $V$ while
$\cO_\K$ denotes the trivial $\K$-line bundle on $M$. The unital ring
of smooth $\K$-valued functions defined on $M$ is denoted by
$\cinf=\Gamma(M,\cO_\K)$. The tensor product of $\K$-vector spaces
and $\K$-vector bundles is denoted by $\otimes$, while the tensor
product of modules over $\cC^\infty(M,\K)$ is denoted by
$\otimes_{\cinf}$; hence $\Gamma(M,V_1\otimes
V_2)\approx \Gamma(M,V_1)\otimes_{\cinf} \Gamma(M,V_2)$.  Setting $T_\K
M\eqdef T M\otimes \cO_\K$ and $T^\ast_\K M\eqdef T^\ast M\otimes
\cO_\K$, the space of $\K$-valued smooth inhomogeneous
globally-defined differential forms on $M$ is denoted by
$\Omega_\K(M)\eqdef \Gamma(M,\wedge T^\ast_\K M)$ and is a $\Z$-graded
module over the commutative ring $\cinf$.  The fixed rank
components of this graded module are denoted by
$\Omega^k_\K(M)=\Gamma(M,\wedge^k T^\ast_\K M)$ ($k=0\ldots d$, where
$d$ is the dimension of $M$).

The kernel and image of any $\K$-linear map
$T:\Gamma(M,V_1)\rightarrow \Gamma(M,V_2)$ will be
denoted by $\cK(T)$ and $\cI(T)$; these are $\K$-linear subspaces of
$\Gamma(M,V_1)$ and $\Gamma(M,V_2)$, respectively.  In the particular
case when $T$ is a $\cinf$-linear map (i.e. when it is a morphism of
$\cinf$-modules), the subspaces $\cK(T)$ and $\cI(T)$ are 
$\cinf$-submodules of $\Gamma(M,V_1)$ and $\Gamma(M,V_2)$,
respectively --- even in those cases when $T$ is not induced by any
bundle morphism from $V_1$ to $V_2$. We always denote a morphism
$f:V_1\rightarrow V_2$ of $\K$-vector bundles and the 
$\cinf$-linear map $\Gamma(M,V_1)\rightarrow \Gamma(M,V_2)$ induced by
it between the modules of sections by the same
symbol. Because of this convention, we clarify that the notations
$\cK(f)\subset \Gamma(M,V_1)$ and $\cI(f)\subset \Gamma(M,V_2)$ denote
the kernel and the image of the corresponding map on sections
$\Gamma(M,V_1)\stackrel{f}{\rightarrow} \Gamma(M,V_2)$, which in this
case are $\cinf$-submodules of $\Gamma(M,V_1)$ and $\Gamma(M,V_2)$,
respectively.  In general, there does {\em not} exist any sub-bundle
$\ker f$ of $V_1$ such that $\cK(f)=\Gamma(M,\ker f)$ nor any
sub-bundle $\im f$ of $V_2$ such that $\cI(f)=\Gamma(M,\im f)$ ---
though there exist sheaves $\ker f$ and $\im f$ with the corresponding
properties.

Given a pseudo-Riemannian metric $g$ on $M$ of signature $(p,q)$,
we let $(e_a)_{a=1\ldots d}$ (where $d=\dim M$) denote a local frame
of $T M$, defined on some open subset $U$ of $M$. We let $e^a$ be the
dual local coframe ($=$ local frame of $T^\ast M$), which satisfies
$e^a(e_b)=\delta^a_b$ and ${\hat g}(e^a,e^b)=g^{ab}$, where $(g^{ab})$
is the inverse of the matrix $(g_{ab})$. The contragradient frame
$(e^a)^\hash$ and contragradient coframe $(e_a)_\hash$ are given by:
\be
(e^a)^\hash=g^{a b}e_b~~,~~(e_a)_\hash=g_{ab}e^b~~,
\ee
where the $\hash$ subscript and superscript denote the (mutually
inverse) musical isomorphisms between $T_\K M$ and $T^\ast_\K M$ given
respectively by lowering and raising indices with the metric $g$.  We
set $e^{a_1\ldots a_k}\eqdef e^{a_1}\wedge \ldots \wedge e^{a_k}$ and
$e_{a_1\ldots a_k}\eqdef e_{a_1}\wedge \ldots \wedge e_{a_k}$ for any
$k=0\ldots d$. A general $\K$-valued inhomogeneous form $\omega\in
\Omega_\K(M)$ expands as:
\ben
\label{FormExpansion}
\omega=\sum_{k=0}^d\omega^{(k)}=_{U}\sum_{k=0}^{d}\frac{1}{k!}\omega^{(k)}_{a_1\ldots
a_k} e^{a_1\ldots a_k}~~,
\een
where the symbol $=_{U}$ means that the equality holds only after
restriction of $\omega$ to $U$ and we have used the expression:
\ben
\label{HomFormExpansion}
\omega^{(k)}=_U\frac{1}{k!}\omega^{(k)}_{a_1\ldots a_k} e^{a_1\ldots a_k}~~.
\een
The locally-defined smooth functions $\omega^{(k)}_{a_1\ldots a_k}\in
\cC^\infty(U,\K)$ (the `strict coefficient functions' of $\omega$) are
completely antisymmetric in $a_1\ldots a_k$. Given a pinor bundle on
$M$ with underlying fiberwise representation $\gamma$ of the Clifford
bundle of $T^\ast_\K M$, the corresponding gamma `matrices' in the
coframe $e^a$ are denoted by $\gamma^a\eqdef \gamma(e^a)$, while the
gamma matrices in the contragradient coframe $(e_a)_\hash$ are denoted
by $\gamma_a\eqdef \gamma((e_a)_\hash)=g_{ab}\gamma^b$. We will
occasionally assume that the frame $(e_a)$ is {\em pseudo-orthonormal}
in the sense that $e_a$ satisfy:
\be
g(e_a,e_b)~(=g_{ab})~=\eta_{ab}~~,
\ee
where $(\eta_{ab})$ is a diagonal matrix with $p$ diagonal
entries equal to $+1$ and $q$ diagonal entries equal to $-1$.

\section{The geometric algebra of metric cylinders and cones}
\label{sec:cone}

In this section, we study the \KA algebra (see \cite{ga1} for background) 
of metric cylinders and cones $({\hat M}, g_\cyl)$ and $({\hat M}, g_\cone)$ over pseudo-Riemannian manifolds $(M,g)$ as well as pin bundles over such spaces, paying special
attention to the manner in which constrained generalized Killing pinor
equations \cite{ga1} behave in such cases. Our treatment is motivated by the application considered
in Section \ref{sec:application}, where it is convenient to consider the metric cone or cylinder over a compactification space for reasons related to giving
an interpretation through reductions of structure group and intrinsic torsion. Though those aspects of the model considered in Section \ref{sec:application}
fall largely outside of the scope of the present paper --- being, instead, discussed in detail in subsequent work --- we encourage the reader to refer to Section \ref{sec:application}
for one of our motivations for developing the formalism discussed below. We start in Subsection \ref{sec:conebasics} by recalling some basic facts about the
geometry of metric cones and cylinders over a pseudo-Riemannian manifold of even dimension --- the case which will form the focus of our considerations, given the application
considered in Section \ref{sec:application}. In Subsection \ref{sec:coneKA}, we discuss the \KA algebra of metric cones and cylinders, paying special attention to certain
subalgebras which play a crucial role in the study of pinors over such
spaces. In particular, we consider the subalgebras of twisted (anti-)selfdual
forms \cite{ga1}, the subalgebra
of so-called {\em special forms} and the subalgebra of forms which are orthogonal to the (dual of the) generating vector field of the cylinder and cone, respectively. We show that the intersection of
the subalgebra of special forms with the subalgebra of forms orthogonal to the generator (an intersection which we call the {\em subalgebra of vertical forms})
is isomorphic with the \KA algebra of the unit section $(M,g)$ of the cone or cylinder
via a natural geometric  isomorphism --- a result which allows one to easily
lift problems and results from the \KA algebra of the unit section to the
metric cone or cylinder. We also discuss various isomorphic models of this subalgebra. Subsection \ref{sec:coneconnections} considers ---
within the geometric algebra framework --- the
Levi-Civita connections of the cylinder and cone as well as the connections
induced by them on the exterior bundle. Subsection \ref{sec:conepin}  discusses pin bundles over metric cones and cylinders (in the case when the Clifford
algebra associated with the dimension of the cone and with the signature of the cone metric is non-simple --- which, once again, is the case relevant for the application considered
in Section \ref{sec:application}). We give an explicit construction of the module structure on such pin bundles, a result
which will be useful later.  In Subsection \ref{sec:conefierz}, we discuss
some basic properties of the Fierz isomorphism \cite{ga1} of cylinders and cones --- in particular, we explain its relation
with the Fierz isomorphism of the unit section. In Subsections \ref{sec:conespinconnections} and \ref{sec:conedeq}, we discuss the lift of connections from the pin bundle of $(M, g)$ to the pin
bundle of the cylinder and cone as well as the `dequantization' of such lifted 
connections --- which (as in \cite{ga1}) results in certain `geometric
connections' on the \KA algebra of the cylinder and cone. Subsection
\ref{sec:coneliftingalg} considers the lift of algebraic constraints on pinors
and forms to the metric cylinder and cone while Subsection
\ref{sec:coneliftingGK} discusses the similar lift of generalized Killing
conditions on pinors and forms.  Subsection \ref{sec:coneCGK} combines these
results to treat the lift of
constrained generalized Killing  conditions from $(M,g)$ to its metric cylinder and cone --- a process which will be used in the example of Section \ref{sec:application} in order to simplify
the analysis of CGK pinor equations on $(M,g)$. Finally, Subsection
\ref{sec:conetruncated} considers certain truncated models which --- as in
\cite{ga1} --- turn out to be especially amenable to implementation in various
symbolic computation systems such as  {\tt Ricci} \cite{Ricci} or Cadabra
\cite{Cadabra}. 

{\bf Simplifying assumptions.} Throughout this section, we let $(M,g)$ be a pseudo-Riemannian manifold of {\em even}
dimension $d$ and ${\hat M}$ be a manifold diffeomorphic with $\R
\times M$ (on which we shall consider the cylinder and cone metrics, respectively). We let $(p,q)$ be the signature type of the metric $g$
on $M$; then $\dim {\hat M}=d+1$ and both the cone and cylinder metric which we shall consider on ${\hat M}$ will have
signature type $(p+1,q)$. We also assume that the Clifford algebra
$\Cl_\K(p+1,q)$ is {\em non-simple} and that the Schur algebra \cite{ga1} of $\Cl_\K(p+1,q)$
equals the base field $\K$, which amounts to assuming that one of the following assumptions hold:

\

\noindent (A) ~$\K=\C$~,

\noindent or

\noindent (B) ~$\K=\R$ and $p-q\equiv_8 0$.

\

\noindent With these assumptions, it follows that the Clifford algebra
$\Cl_\K(p,q)$ which is relevant for $(M,g)$ is simple and that its Schur algebra also equals the base field. We further
assume that $M$ is oriented and that on ${\hat M}$ we have chosen the orientation compatible with that of $M$.

\subsection{Preparations}
\label{sec:conebasics}

On ${\hat M}$, consider the cylinder metric $g_\cyl$ whose
squared line element takes the form:
\be
\dd s_\cyl^2=\dd u^2 + \dd s^2~~~~~(u\in \R)~~,
\ee
where $\dd s^2$ is the squared line element of $g$. This is related by
a conformal transformation to the cone metric $g_\cone$ on ${\hat M}$, whose squared
line element is given by:
\be
\dd s_\cone^2=\dd r^2 +r^2 \dd s^2=r^2\dd s_\cyl^2~~~~(r\eqdef e^u\in (0,+\infty))~~.
\ee
We have $g_\cone=r^2g_\cyl$ and\footnote{As usual, ${\hat g}_\cone$ and ${\hat g}_\cyl$ denote the metrics induced by $g_\cone$ and $g_\cyl$ on $T^\ast_\K {\hat M}$.}
${\hat g}_\cone=\frac{1}{r^2}{\hat g}_\cyl$, where we view $u$ and $r=e^u$ as smooth functions defined on ${\hat M}$,
namely $u\in \cC^\infty({\hat M},\R)$ and $ r\in \cC^\infty({\hat M},(0,+\infty))\subset \cC^\infty({\hat M},\R)$.
The transformation $u\rightarrow r$ maps the limit $u\rightarrow
-\infty$ to the limit $r\rightarrow 0$.  Unless $M$ is a sphere, the cone metric is
not complete due to the conical singularity which arises when one attempts to add the point at
$r=0$. For any vector field $V\in \Gamma({\hat M},T_\K {\hat M})$ and any one-form
$\eta\in \Gamma({\hat M},T^\ast_\K {\hat M})=\Omega^1_\K({\hat M})$, we have
$V_{\hash_\cone}=r^2 V_{\hash_\cyl}$ and $\eta^{\hash_\cone}=\frac{1}{r^2}\eta^{\hash_\cyl}$, where
$\hash_\cyl$ and $\hash_\cone$ are the musical isomorphisms of the cylinder and cone, respectively.

\begin{figure}[htbp]
\begin{center}
\scalebox{0.35}{\input{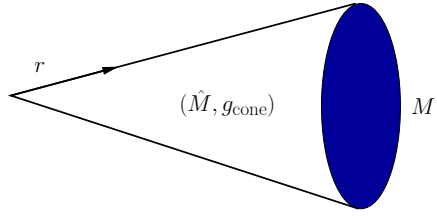}}
\caption{Metric cone over $M$}
\label{fig:cone}
\end{center}
\end{figure}

\begin{figure}[htbp]
\begin{center}
\scalebox{0.35}{\input{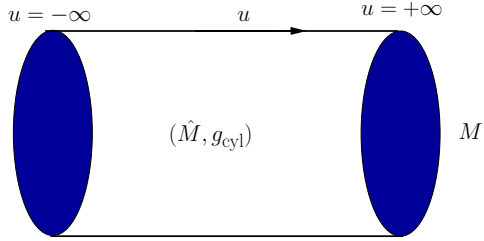}}
\caption{Metric cylinder over $M$}
\label{fig:cyl}
\end{center}
\end{figure}

\paragraph{The ring $\cpinf$.} We let $\Pi:{\hat M}\rightarrow M$ be the projection on the second factor of the Cartesian product ${\hat M}=\R\times M$. 
For later reference, consider the following 
unital subring of the commutative ring $\cC^\infty({\hat M},\K)$: 
\be
\cC^\infty_\perp({\hat M},\K)\eqdef \{f\circ \Pi|f\in \cC^\infty(M,\K)\}\subset \cC^\infty({\hat M},\K)~~.
\ee
It coincides with the image $\Pi^\ast(\cC^\infty(M,\K))$ through the pullback map $\Pi^\ast$, which acts as follows 
on smooth functions defined on $M$:
\be
\Pi^\ast(f)=f\circ \Pi\in \cC^\infty_\perp({\hat M},\K)~~,~~\forall f\in \cinf~~.
\ee
In fact, $\Pi^\ast$ corestricts to a unital isomorphism of rings: 
\be
\cinf\stackrel{\Pi^\ast|^{~{\tiny \cC^\infty_\perp({\hat M},\K)}}~}{\longrightarrow } \cpinf~~,
\ee
which allows us to identify $\cC^\infty_\perp({\hat M},\K)$ with $\cinf$. The pullback $\Pi^\ast:\Omega_\K(M)\rightarrow \Omega_\K({\hat M})$ 
of $\K$-valued differential forms satisfies: 
\be
\Pi^\ast(f\omega)=\Pi^\ast(f)\Pi^\ast(\omega)~~,~~\forall f\in \cC^\infty(M,\K)~~,~~\forall \omega\in \Omega_\K(M,\K)~~
\ee
and maps wedge products into wedge products. It can therefore be viewed as a morphism of $\cC^\infty(M,\K)$-algebras from 
the exterior algebra of $M$ to that of ${\hat M}$, provided that we identify $\cC^\infty_\perp({\hat M},\K)$ with $\cinf$ as explained 
above.   

\paragraph{The lift of vector fields.} Notice that the cone
can be viewed as the warped product \cite{warped} $({\hat M}, g_\cone)\approx
((0,\infty),\dd r^2) \times_r (M, \dd s^2)$ (of warp factor equal to
$r$) of the positive axis (endowed with the flat metric of squared
length element $\dd r^2$) with $(M,g)$, while the cylinder is, of
course, the direct metric product $({\hat M}, g_\cyl)\approx (\R, \dd
u^2) \times (M, \dd s^2)$ of the real axis (endowed with the flat metric of
squared length element $\dd u^2$) with $(M,g)$. The latter is the same
as the warped product $(\R, \dd u^2) \times_1 (M, g)$ with constant
warp factor equal to one. The pulled-back bundle $\Pi^\ast(T_\K M)$
can be identified with the sub-bundle $T^\perp_\K {\hat M}$ whose
fiber at a point ${\hat x} \in {\hat M}$ is the orthogonal complement
in $T_{\K,{\hat x}} {\hat M}$ of the tangent vector
$(\partial_r)_{{\hat x}}\in T_{\K,{\hat x}} {\hat M}$ with respect to
$g_\cone$; of course, this coincides with the orthogonal complement of
the vector $(\partial_u)_{{\hat x}}$ with respect to $g_\cyl$. A
vector field $X\in \Gamma(M,T_\K M)$ pulls back to the section
$\Pi^\ast(X)$ of the bundle $\Pi^\ast( T_\K M)$, which in turn can be
viewed as a section $X_\ast$ of the sub-bundle $T^\perp_\K {\hat
M}\subset T_\K {\hat M}$, i.e. as a vector field on ${\hat M}$ which
is everywhere orthogonal to $\partial_r$ (and thus to
$\partial_u$). Of course, $X_\ast$ coincides with the well-known
 lift of $X$ along a warped product. Using the identification
$\cC^\infty_\perp({\hat M},\K)\approx \cinf$ discussed above, the
pullback of sections can be viewed as a (non-surjective)
$\cinf$-linear map $\Gamma(M,T_\K M)\ni
X\stackrel{\Pi^\ast}{\longrightarrow} X_\ast\in \Gamma(M,T^\perp_\K
{\hat M})$.

\paragraph{The canonical normalized one-forms.}
The one-form:
\be
\psi=\dd u=  \frac{1}{r}\dd r~~
\ee
has unit norm with respect to the cylinder metric, being dual  to the unit norm vector field
$\psi^{\hash_\cyl}=\partial_u =r\partial_r$ with respect to the metric $g_\cyl$:
\be
\psi=\partial_u \lrcorner g_\cyl~~.
\ee
Similarly, the one-form:
\be
\theta=\dd r= r \psi~~
\ee
has unit norm with respect to the cone metric, being dual  to the unit norm vector field
$\theta^{\hash_\cone}=\partial_r$ with respect to the metric $g_\cone$:
\be
\theta=\partial_r \lrcorner g_\cone~~.
\ee
The pairings (inner products) induced by $g_\cone$ and $g_\cyl$ on $\Omega_\K({\hat M})$ are related
through:
\be
\langle \omega,\eta\rangle_\cone=\frac{1}{r^{2k}}\langle \omega,
\eta\rangle_\cyl~~,~~\forall \omega,\eta\in\Omega^k_\K({\hat M})~~.
\ee
Together with the definition of the (left) interior product, the last relation implies:
\ben
\label{iotacylcone}
\iota^\cone_{\omega}=\frac{1}{r^{2k}}\iota^\cyl_\omega~~,~~\forall \omega\in \Omega^k_\K({\hat M})~~.
\een
In turn, this gives: 
\be
\label{iotacc}
\iota^\cone_\theta=\frac{1}{r}\iota_\psi^\cyl~~,
\ee
a relation which will be used below. For any vector field $V$ on ${\hat M}$, we let $\cL_V$ denote the Lie
derivative with respect to $V$.  We have $\cL_{f V}\omega =\dd f\wedge (V
\lrcorner\omega)+f\cL_V\omega$ for any $f\in \cC^\infty({\hat M},\K)$ and any $\omega\in \Omega_\K({\hat M})$, a relation
which gives:
\ben
\label{Lpsitheta}
\cL_{\partial_u}\omega=r \cL_{\partial_r}\omega+ \theta \wedge (\partial_r \lrcorner \omega)~~,~~\forall \omega\in \Omega_\K({\hat M})~~.
\een
In turn, this implies:
\ben
\label{LieRelation}
\cL_{\partial_r}\omega=\frac{1}{r}(\cL_{\partial_u}\omega -\psi \wedge (\partial_u \lrcorner\omega)) ~~,~~\forall \omega\in \Omega_\K({\hat M})~~,
\een
where we noticed that $ \theta \wedge (\partial_r \lrcorner \omega)= \psi \wedge (\partial_u \lrcorner\omega)$.

\paragraph{The Euler operator.}
Consider the Euler operator $\cE=\oplus_{k=0}^{d+1}{k~\id_{\Omega^k_\K({\hat M})}}$ on
$\Omega_\K({\hat M})$ associated with the rank decomposition
$\Omega_\K({\hat M})=\oplus_{k=0}^{d+1} \Omega^k_\K({\hat M})$.  This
acts as follows on a general inhomogeneous form:
\be
\cE(\omega)=\sum_{k=0}^{d+1}{k \omega^{(k)}}~~,~~\forall \omega=
\sum_{k=0}^{d+1}\omega^{(k)}\in \Omega_\K({\hat M})~~{\rm with}~~\omega^{(k)}\in
\Omega^k_\K({\hat M})~~.
\ee
Notice that $\cE$ has eigenvalues $0,\ldots, d+1$, with eigenspaces given by: 
\be
\cK(\cE-k~\id_{\Omega_\K({\hat M})})=\Omega^k_\K({\hat M})~~.
\ee
Since $\cE$ is induced by an endomorphism of the exterior bundle, the
exponential $e^\cE=\sum_{n=0}^\infty{\frac{1}{n!}\cE^{\circ n}}$
is well-defined, being itself induced by a bundle endomorphism of $\wedge T^\ast_\K {\hat M}$.
We obtain well-defined $\cC^\infty({\hat M},\K)$-linear automorphisms $\lambda^\cE\eqdef e^{(\ln \lambda) \cE}$
of $\Omega_\K({\hat M})$ for any positive real number $\lambda>0$ (and
similar operators for any non-vanishing complex number $\lambda$, provided that we choose a
branch of the logarithm). Of course, we have
$(\lambda_1\lambda_2)^\cE=\lambda_1^\cE\circ \lambda_2^\cE$ for any
$\lambda_1,\lambda_2>0$ and $1^\cE=\id_{\Omega_\K({\hat
M})}$, so the map $(0,+\infty)\ni \lambda \rightarrow \lambda^\cE\in \End_{\cC^\infty({\hat M},\K)}(\Omega_\K({\hat M}))$ gives
a representation of the multiplicative group of positive reals through $\cC^\infty({\hat M},\K)$-linear
endomorphisms of $\Omega_\K({\hat M})$. These operators act as:
\be
\lambda^\cE(\omega)=\sum_{k=0}^{d+1} \lambda^k\omega^{(k)}~~,~~\forall
\omega=\sum_{k=0}^{d+1}\omega^{(k)}\in \Omega_\K({\hat M})~~{\rm with}~~\omega^{(k)}\in
\Omega^k_\K({\hat M})~~.
\ee
Notice the following relations which hold on $\Omega_\K({\hat M})$:
\beqan
\label{rexpcom}
r^\cE\circ \partial_r\lrcorner~=\frac{1}{r} (\partial_r\lrcorner~) \circ r^{\cE}~~&,&~~
r^\cE\circ \partial_u\lrcorner~=\frac{1}{r} (\partial_u\lrcorner~) \circ r^{\cE}~~,\nn\\
r^\cE\circ \wedge_\theta=r \wedge_\theta \circ ~r^{\cE}~~&,&~~ r^\cE\circ \wedge_\psi~=r \wedge_\psi \circ ~r^{\cE}~~.
\eeqan
They follow from the fact that contraction with a vector field lowers the rank of a homogeneous form by one while wedge product with a one form increases the rank by one.
Also notice that the obvious relations:
\ben
\label{Lrk}
\cL_{\partial_r}(r^k)=\frac{k}{r}r^k\Longleftrightarrow \cL_{\partial_u}r^k=k r^k~~,~~\forall k\in \Z~~
\een
imply:
\ben
\label{Lrexp}
(\cL_{\partial_u}-\cE)\circ r^\cE=r^\cE\circ \cL_{\partial_u}~~({\rm on}~~\Omega_\K({\hat M}))~~,
\een
where we used identity \eqref{LieRelation}. 

\paragraph{Forms parallel and orthogonal to the canonical one-forms.}
As in \cite{ga1}, let
$P_\parallel^\cyl=\wedge_\psi \circ \iota^\cyl_\psi$,
$P_\perp^\cyl=\iota_\psi^\cyl \circ \wedge_\psi$ and
$P_\parallel^\cone=\wedge_\theta \circ \iota^\cone_\theta$,
$P_\perp^\cone=\iota^\cone_\theta \circ \wedge_\theta$ be the
projectors on forms parallel and orthogonal to $\psi$ and $\theta$ on
the cylinder and cone, respectively. Relation \eqref{iotacc} implies
$P_\parallel^\cone =P_\parallel^\cyl$ and $P_\perp^\cone
=P_\perp^\cyl$, so we define:
\be
P_\parallel\eqdef P_\parallel^\cone=P_\parallel^\cyl=\wedge_\psi\circ (\partial_u\lrcorner~)=\wedge_\theta\circ (\partial_r \lrcorner~)
~~,~~P_\perp\eqdef P_\perp^\cone =P_\perp^\cyl=(\partial_u\lrcorner~)\circ \wedge_\psi=(\partial_r\lrcorner~)\circ \wedge_{\theta}~~.
\ee
In particular, the decomposition of a form $\omega\in \Omega_\K({\hat M})$ into its part $\omega_\parallel=P_\parallel(\omega)$ parallel to $\theta$ (and thus also to $\psi$)
and its part $\omega_\perp=P_\perp(\omega)$ orthogonal to $\theta$ (and thus also to $\psi$) is the same on the cylinder and cone. We let:
\beqa
\Omega^\parallel_\K({\hat M})&\eqdef & P_\parallel(\Omega_\K({\hat M}))=\{\omega\in \Omega_\K({\hat M})|\omega=\omega_\parallel\}~~,\\
\Omega^\perp_\K({\hat M}) &\eqdef &P_\perp(\Omega_\K({\hat M}))=\{\omega\in \Omega_\K({\hat M})|\omega=\omega_\perp\}~~,
\eeqa
obtaining the same decomposition $\Omega_\K({\hat M})=\Omega^\parallel_\K({\hat M})\oplus \Omega^\perp_\K({\hat M})$ for the cylinder and cone.
Notice the natural isomorphism $\Omega^\perp_\K({\hat M})\approx \Gamma({\hat M}, \wedge (T_\K^\perp {\hat M})^\ast)$, which we shall often use tacitly later on. 
Relations \eqref{rexpcom} imply:
\be
[r^\cE, P_\parallel]_{-,\circ}=[r^\cE, P_\perp]_{-,\circ}=0\Longleftrightarrow r^\cE(\omega)_{\parallel}=r^\cE(\omega_\parallel)~~{\rm and}~~r^\cE(\omega)_{\perp}=r^\cE(\omega_\perp)~~.
\ee
In particular, we have:
\ben
\label{rexpinv}
r^\cE(\Omega^\parallel_\K({\hat M}))=\Omega^\parallel_\K({\hat M})~~,~~r^\cE(\Omega^\perp_\K({\hat M}))=\Omega^\perp_\K({\hat M})~~.
\een
On the other hand, the relation $\partial_r\lrcorner \theta=1$ and the
fact that $\theta$ is closed imply $\cL_{\partial_r} \theta=0$ upon
using the identity $\cL_V\omega =\dd (V\lrcorner\omega )+ V\lrcorner
(\dd\omega)$, which holds for any vector field $V$ and any
inhomogeneous form $\omega$ defined on ${\hat M}$. In turn, this implies that
$\cL_{\partial_r}$ commutes with the operator $\wedge_\theta$ and
(since it commutes with the operator $\partial_r\lrcorner $) also with
the projectors $P_\parallel$ and $P_\perp$:
\be
[\cL_{\partial_r}, \partial_r \lrcorner~]_{-,\circ} =[\cL_{\partial_r}, \wedge_\theta]_{-,\circ}=0~~
\Longrightarrow [\cL_{\partial_r}, P_\parallel]_{-,\circ}= [\cL_{\partial_r},
P_\perp]_{-,\circ}=0~~.
\ee
Similarly, we have the commutation relations:
\be
[\cL_{\partial_u}, \partial_u \lrcorner~]_{-,\circ}=[\cL_{\partial_u}, \wedge_\psi]_{-,\circ}=0~~
\Longrightarrow [\cL_{\partial_u}, P_\parallel]_{-,\circ}= [\cL_{\partial_u},
P_\perp]_{-,\circ}=0~~,
\ee
as a consequence of the identity $\partial_u \lrcorner \psi=1$, which implies $\cL_{\partial u}\psi=0$. 
Also notice that relation \eqref{Lpsitheta} reads: 
\be
\cL_{\partial_u}=r\cL_{\partial_r}+P_\parallel~~.
\ee
The commutation relations given above imply:
\beqan
\label{Lparperp}
&&(\cL_{\partial_r}\omega)_\parallel=
\cL_{\partial_r}(\omega_\parallel)~~,~~(\cL_{\partial_r}\omega)_\perp=
\cL_{\partial_r}(\omega_\perp)~~,\nn\\
&&(\cL_{\partial_u}\omega)_\parallel=\cL_{\partial_u}(\omega_\parallel)~~,~~(\cL_{\partial_u}\omega)_\perp=
\cL_{\partial_u}(\omega_\perp)~~,
\eeqan
for all $\omega\in \Omega_\K({\hat M})$. In particular, we have:
\beqa
&&\cL_{\partial_r}(\Omega^\perp_\K({\hat M}))\subset \Omega^\perp_\K({\hat M})~~,~~\cL_{\partial_r}(\Omega^\parallel_\K({\hat M}))\subset \Omega^\parallel_\K({\hat M})~~,\\
&&\cL_{\partial_u}(\Omega^\perp_\K({\hat M}))\subset \Omega^\perp_\K({\hat M})~~,~~\cL_{\partial_u}(\Omega^\parallel_\K({\hat M}))\subset \Omega^\parallel_\K({\hat M})~~.\nn
\eeqa

\paragraph{(Conformal) Killing properties.}
Finally, note that $\partial_r $ is a normalized conformal Killing vector field for $g_\cone$ and a Killing vector field for
$g_\cyl$, while $\partial_u=r\partial_r$ is a homothety for $g_\cone$ and a normalized Killing vector field for
$g_\cyl$:
\beqan
\label{Lg}
\cL_{\partial_r} g_\cone=\frac{2}{r}g_\cone~~&,&~~\cL_{\partial_r} g_\cyl=0~~,\\
\cL_{\partial_u} g_\cone=2g_\cone~~&,&~~\cL_{\partial_u} g_\cyl=0~~.\nn
\eeqan
Thus $\theta$ and $\psi$ are Killing-Yano one-forms (of various types)
with respect to both metrics.

\subsection{The \KA algebra of metric cones and cylinders over pseudo-Riemannian manifolds}
\label{sec:coneKA}

\paragraph{Relating the \KA algebras of the cylinder and cone.}
We let $\diamond$ and $\bigtriangleup_p$~($p=0\ldots d$) denote the geometric and
generalized products constructed on $\Omega_\K(M)$ using the metric
$g$.  Similarly, we let $\diamond^\cyl$, $\bigtriangleup^\cyl_p$ and
$\diamond^\cone$, $\bigtriangleup^\cone_p$~($p=0\ldots d+1$) denote the
geometric and generalized products on $\Omega_\K({\hat M})$ induced by the metrics
$g_\cyl$ and $g_\cone$ respectively. Using the definition of
generalized products, we find:
\ben
\label{GenProdScaling}
\bigtriangleup_p^\cone=\frac{1}{r^{2p}}\bigtriangleup^\cyl_p~~,~~\forall p=0\ldots d+1~~.
\een
Since the generalized product $\bigtriangleup_p$ is homogeneous of degree $-2p$ when viewed as a map
$\bigtriangleup_p:\Omega_\K({\hat M})\otimes_{\cC^\infty({\hat M},\K)} \Omega_\K({\hat M})\rightarrow \Omega_\K({\hat M})$ from
the tensor product $\Omega_\K({\hat M})\otimes_{\cC^\infty({\hat M},\K)} \Omega_\K({\hat M})$ (endowed with the grading induced by the rank
grading of the exterior algebra) to $\Omega_\K({\hat M})$, the following identities hold for all $p=0\ldots d+1$:
\beqan
\label{Econecyl}
(\cE+2p~\id_{\Omega_\K({\hat M})})\circ \bigtriangleup_p^\cyl~&=&\bigtriangleup_p^\cyl\circ (\cE\otimes \id_{\Omega_\K({\hat M})} +\id_{\Omega_\K({\hat M})}\otimes \cE) ~~,\nn\\
(\cE+2p~\id_{\Omega_\K({\hat M})})\circ \bigtriangleup_p^\cone&=&\bigtriangleup_p^\cone\circ (\cE\otimes \id_{\Omega_\K({\hat M})} +\id_{\Omega_\K({\hat M})}\otimes \cE)~~,
\eeqan
i.e.:
\beqa
(\cE+2p~\id_{\Omega_\K({\hat M})})
(\omega \bigtriangleup^\cyl_p \eta) ~&=& \cE(\omega) \bigtriangleup^\cyl_p \eta+\omega \bigtriangleup^\cyl_p \cE(\eta)~~,~~\forall \omega,\eta\in \Omega_\K({\hat M})~~,\nn\\
(\cE+2p~\id_{\Omega_\K({\hat M})})
(\omega \bigtriangleup^\cone_p \eta) &=& \cE(\omega) \bigtriangleup^\cone_p \eta+\omega \bigtriangleup^\cone_p \cE(\eta)~~,~~\forall \omega,\eta\in \Omega_\K({\hat M})~~.\nn
\eeqa
These identities imply:
\beqa
&&r^\cE\circ \bigtriangleup^\cyl_p=\frac{1}{r^{2p}}~\bigtriangleup^\cyl_p\circ (r^\cE\otimes r^\cE)\Longleftrightarrow
r^\cE(\omega \bigtriangleup^\cyl_p \eta)=\frac{1}{r^{2p}} [r^\cE(\omega) \bigtriangleup^\cyl_p r^\cE(\eta)]~,~\forall \omega,\eta\in \Omega_\K({\hat M})~,\nn\\
\!\!\!\!\!\!&&r^\cE\circ \bigtriangleup^\cone_p=\frac{1}{r^{2p}}~\bigtriangleup^\cone_p\circ (r^\cE\otimes r^\cE)\Longleftrightarrow
r^\cE(\omega \bigtriangleup^\cone_p \eta)=\frac{1}{r^{2p}} [r^\cE(\omega) \bigtriangleup^\cone_p r^\cE(\eta)]~,~\forall \omega,\eta\in \Omega_\K({\hat M})~.~~~~~~~~\nn
\eeqa
Combining the first of these relations with \eqref{GenProdScaling} gives:
\be
r^\cE\circ \bigtriangleup^\cyl_p=\bigtriangleup^\cone_p\circ (r^\cE\otimes r^\cE)
\Longleftrightarrow r^\cE(\omega \bigtriangleup^\cyl_p \eta)=r^\cE(\omega) \bigtriangleup^\cone_p r^\cE(\eta)~~,~~\forall \omega,\eta\in \Omega_\K({\hat M})~~,
\ee
which in turn implies:
\ben
\label{rexpmorphism}
r^\cE\circ \diamond^\cyl=\diamond^\cone\circ (r^\cE\otimes r^\cE)
\Longleftrightarrow ~r^\cE(\omega \diamond^\cyl \eta)=r^\cE(\omega) \diamond^\cone r^\cE(\eta)~~,~~\forall \omega,\eta\in \Omega_\K({\hat M})~~.
\een
Together with the obvious relation $r^\cE(1_{\hat M})=1_{\hat M}$ (which follows from ${\cE}(1_{\hat M})=0$), this
shows that the \KA algebras of the cylinder and cone can be identified through appropriate rescalings of their fixed rank subspaces:

\paragraph{Proposition.} The maps $r^\cE$ and $r^{-\cE}$ are mutually inverse $\cC^\infty({\hat M},\K)$-linear unital isomorphisms of algebras between the \KA algebras of
the cylinder and cone:
\ben
\label{diagram:risom}
\scalebox{1.2}{
\xymatrix@1{(\Omega_\K(\hat M),\diamond^\cyl){~}  \ar@<0.7ex>[r]^{r^{\cE}} {~~} & {~~}
(\Omega_\K ({\hat M}),\diamond^\cone) {~~} \ar@<0.7ex>[l]^{r^{-\cE}~}
}}~.
\een
For later reference, we note the following identities which are an easy consequence of the previous proposition: 
\ben
\label{RLcylcone}
L^\cone_{r^\cE(\omega)}=r^\cE\circ L^\cyl_\omega\circ r^{-\cE}~~,~~R^\cone_{r^\cE(\omega)}=r^\cE\circ R^\cyl_\omega\circ r^{-\cE}~~,~~\forall \omega\in \Omega_\K({\hat M})~~,
\een
where $L^\cyl_\omega,R^\cyl_\omega$ and $L^\cone_\omega,R^\cone_\omega$ are the operators of left and right multiplication with $\omega$ in the \KA algebras of the cylinder and cone. 

\paragraph{Relating $(\Omega_\K^\perp({\hat M}),\diamond^\cyl)$ and $(\Omega_\K^\perp({\hat M}), \diamond^\cone)$. }
Recalling properties \eqref{rexpinv} and the fact that $\Omega^\perp_\K({\hat M})$ is a unital subalgebra of the \KA algebras of the cylinder and cone (see \cite{ga1}),
the previous proposition implies:

\paragraph{Corollary.} The maps $r^\cE$ and $r^{-\cE}$ restrict to mutually inverse unital isomorphisms between the algebras $(\Omega^\perp_\K({\hat M}),\diamond^\cyl)$ and
$(\Omega^\perp_\K({\hat M}),\diamond^\cone)$:
\ben
\label{diagram:risom_perp}
\scalebox{1.2}{
\xymatrix@1{(\Omega_\K^\perp(\hat M),\diamond^\cyl){~~}  \ar@<0.7ex>[r]^{r^{\cE}|_{\Omega_\K^\perp(\hat M)}} {~~} & {~~~}
(\Omega_\K^\perp({\hat M}),\diamond^\cone) {~~} \ar@<0.7ex>[l]^{r^{-\cE}|_{\Omega_\K^\perp({\hat M})}~}
}}~.
\een

\paragraph{The special and vertical subalgebras.}
Using the fact that $\cL_{\partial_u}$ is a degree zero $\K$-linear derivation of the exterior algebra,
the last of relations \eqref{Lg} and the identity $[\cL_{\partial_u},\partial_u\lrcorner ]_{\circ,-}=0$ imply that $\cL_{\partial_u}$ is a $\cpinf$-linear derivation of degree zero of all generalized products of the cylinder:
\be
\cL_{\partial_u}\circ \bigtriangleup_p^\cyl= \bigtriangleup_p^\cyl\circ(\cL_{\partial_u}\otimes \id_{\Omega_\K({\hat M})}+\id_{\Omega_\K({\hat M})}\otimes \cL_{\partial_u})~~,
\ee
a relation which encodes the fact that $\bigtriangleup_p$ are invariant under the translations $u\rightarrow u+\epsilon$ ($\epsilon\in \R$) along the generator of the cylinder. 
This implies that $\cL_{\partial_u}$ is an even $\cpinf$-linear derivation of the \KA algebra $(\Omega_\K({\hat M}), \diamond^\cyl)$:
\ben
\label{new1}
\cL_{\partial_u}\circ \diamond^\cyl= \diamond^\cyl\circ(\cL_{\partial_u}\otimes \id_{\Omega_\K({\hat M})}+\id_{\Omega_\K({\hat M})}\otimes \cL_{\partial_u})~~.
\een
Using \eqref{GenProdScaling} and \eqref{Lrk}, the relations given above imply:
\be
(\cL_{\partial_u}+2p~\id_{\Omega_\K({\hat M})})\circ 
\bigtriangleup_p^\cone= \bigtriangleup_p^\cone\circ(\cL_{\partial_u}\otimes \id_{\Omega_\K({\hat M})}+\id_{\Omega_\K({\hat M})}\otimes \cL_{\partial_u})~~.
\ee
Combining this with \eqref{Econecyl} shows that the operator
$\cL_{\partial_u}-\cE$ is a degree zero $\cpinf$-linear 
derivation of all generalized products of the cone:
\be
(\cL_{\partial_u}-\cE)\circ \bigtriangleup_p^\cone= \bigtriangleup_p^\cone\circ \left[(\cL_{\partial_u}-\cE)\otimes \id_{\Omega_\K({\hat M})}+\id_{\Omega_\K({\hat M})}\otimes
(\cL_{\partial_u}-\cE) \right]~~
\ee
and hence this operator is an even $\cpinf$-linear derivation of the \KA algebra  $(\Omega_\K({\hat M}), \diamond^\cone)$:
\ben
\label{new2}
(\cL_{\partial_u}-\cE)\circ \diamond^\cone= \diamond^\cone\circ \left[(\cL_{\partial_u}-\cE)\otimes \id_{\Omega_\K({\hat M})}+\id_{\Omega_\K({\hat M})}\otimes
(\cL_{\partial_u}-\cE) \right]~~.
\een
In particular, we have:
\be
(\cL_{\partial_u}-\cE)(\omega \diamond^\cone \eta)=\left[(\cL_{\partial_u}-\cE) \omega\right] \diamond^\cone \eta+
\omega \diamond^\cone \left[(\cL_{\partial_u} -\cE)\eta\right]~~,~~\forall \omega,\eta\in \Omega_\K({\hat M})~~.
\ee
Notice that $\cL_{\partial_u}$ is {\em not} a derivation of the \KA algebra of
the cone. The observations above imply that the following subspaces of
$\Omega_\K({\hat M})$:
\be
\boxed{\Omega^{\cyl}_\K({\hat M})\eqdef \cK(\cL_{\partial_u})~~,~~\Omega^{\cone}_\K({\hat M})\eqdef \cK\left(\cL_{\partial_u}-\cE\right)}
\ee
are unital $\cpinf$-subalgebras of the \KA algebras $(\Omega_\K({\hat M}),
\diamond^\cyl)$ and $(\Omega_\K({\hat M}), \diamond^\cone)$, which we shall call the {\em special subalgebras}
of the cylinder and cone, respectively.  The previous proposition and relation \eqref{Lrexp} give: 

\paragraph{Proposition.} The appropriate restrictions of the maps
$r^{\pm \cE}$ give mutually inverse $\cpinf$-linear unital isomorphisms of algebras between
the special subalgebras of the cylinder and cone:
\ben
\label{diagram:risom_special}
\scalebox{1.1}{
\xymatrix@1{(\Omega_\K^{\cyl}({\hat M}),\diamond^\cyl){~~} \ar@<0.7ex>[r]^{r^{\cE}|_{\Omega_\K^{\cyl}({\hat M})} } {~~~} &
{~~~~}(\Omega_\K^{\cone} ({\hat M}),\diamond^\cone) {~~}\ar@<0.7ex>[l]^{r^{-\cE}|_{\Omega_\K^{\cone} ({\hat M})}~}
}}~.
\een

We also know from \cite{ga1} that
the subspace:
\be
\Omega^\perp_\K({\hat M})\eqdef \{\omega\in \Omega_\K({\hat M})|\partial_u\lrcorner \omega=0\}=
\{\omega\in \Omega_\K({\hat M})|\partial_r\lrcorner \omega=0\}
\ee
is a unital $\cC^\infty({\hat M},\K)$-subalgebra of both $(\Omega_\K({\hat M}), \diamond^\cyl)$
and $(\Omega_\K({\hat M}), \diamond^\cone)$. Therefore, the
intersections:
\be
\boxed{\Omega^{\perp,\cyl}_\K({\hat M})\eqdef \Omega^\perp_\K({\hat M})\cap \Omega^{\cyl}_\K({\hat M})~~,~~
\Omega^{\perp,\cone}_\K({\hat M}) \eqdef \Omega^\perp_\K({\hat M})\cap \Omega^{\cone}_\K({\hat M})}~~
\ee
are unital $\cpinf$-subalgebras $(\Omega_\K({\hat M}),
\diamond^\cyl)$ and $(\Omega_\K({\hat M}), \diamond^\cone)$ respectively, which we
shall call the {\em vertical subalgebras} of the cylinder and cone.
Since $\partial_r\lrcorner ~|_{\Omega^\perp_\K({\hat M})}=0$ , equations 
\eqref{Lpsitheta} and \eqref{Lrexp} give:
\ben
(\cL_{\partial_r}-\frac{\cE}{r})\circ r^\cE|_{\Omega^\perp_\K({\hat M})}=
\frac{1}{r} r^\cE\circ \cL_{\partial_u}|_{\Omega^\perp_\K({\hat M})}~~.
\een
The observations made above imply that the operator $r^\cE$ maps the vertical subalgebra of the cylinder into that of the cone:
\be
r^\cE(\Omega^{\perp,\cyl}_\K({\hat M}))=\Omega^{\perp,\cone}_\K({\hat M})~~.
\ee
Combining this with the previous proposition, we find:

\paragraph{Proposition.} The appropriate restrictions of the maps
$r^{\pm \cE}$ give mutually inverse $\cpinf$-linear unital isomorphisms of algebras between
the vertical subalgebras of the cylinder and cone:
\ben
\label{diagram:risom_perpspecial}
\scalebox{1.1}{
\xymatrix@1{(\Omega_\K^{\perp,\cyl}({\hat M}),\diamond^\cyl){~~~} \ar@<0.7ex>[r]^{r^{\cE}|_{\Omega_\K^{\perp,\cyl}({\hat M})} } {~~} &
{~~~~~}(\Omega_\K^{\perp,\cone} ({\hat M}),\diamond^\cone) {~~}\ar@<0.7ex>[l]^{r^{-\cE}|_{\Omega_\K^{\perp,\cone} ({\hat M})}~}
}}~.
\een

\paragraph{The modified volume forms of the cylinder and of the cone.}
Since $\Cl_\K(p+1,q)$ is assumed to be non-simple, the volume forms $\nu^\cone$ and $\nu^\cyl$ defined by $g_\cone$ and 
$g_\cyl$ on ${\hat M}$ square to $1_{\hat M}$ and are central in $(\Omega_\K({\hat M}),\diamond^\cone)$ and in $(\Omega_\K({\hat
M}),\diamond^\cyl)$, respectively; they are given explicitly by:
\ben
\label{nu0cylcone}
\nu^\cyl=\psi \wedge \nu_\top^\cyl=\psi \diamond^\cyl \nu_\top^\cyl~~,
~~\nu^\cone=\theta\wedge \nu_\top^\cone=\theta\diamond^\cone \nu_\top^\cone=r^{d+1}\nu^\cyl=r^{\cE}(\nu^\cyl)~~.
\een
Here:
\ben
\label{pullbacks}
\nu_\top^\cyl=\iota^\cyl_\psi \nu_\cyl=\partial_u\lrcorner \nu^\cyl=\Pi^\ast(\nu)~~,
~~\nu_\top^\cone=\iota^\cone_\theta\nu^\cone=\partial_r\lrcorner\nu^\cone= r^d \Pi^\ast(\nu)=r^d\nu_\top^\cyl=r^\cE(\nu_\top^\cyl)~~,
\een
where $\nu$ is the volume form of $(M,g)$. 

\paragraph{Special twisted (anti-)selfdual forms on cylinders and cones.}
The last relation in \eqref{nu0cylcone} and the second relation in \eqref{RLcylcone}
imply that the twisted Hodge operators ${\tilde \ast}_\cyl=R^\cyl_{\nu^\cyl}$ of the cylinder and 
${\tilde \ast}_\cone=R^\cone_{\nu^\cone}$ of the cone are related through:
\ben
\label{Hodger}
{\tilde \ast}_\cone\circ r^\cE=r^\cE\circ {\tilde \ast}_\cyl
\Longleftrightarrow P_\pm^\cone\circ  r^\cE=r^\cE\circ P_\pm^\cyl~~.
\een
This implies that $r^\cE$ identifies the subalgebras:
\be
\boxed{
\begin{split}
& \Omega^\pm_{\K,\cyl}({\hat M})\eqdef P_\pm^\cyl(\Omega_\K({\hat M}))=\{\omega\in \Omega_\K({\hat M})|{\tilde \ast}_\cyl \omega=\pm \omega\}~~,\nn\\
&\Omega^\pm_{\K,\cone}({\hat M})\eqdef  P_\pm^\cone(\Omega_\K({\hat M}))=\{\omega\in \Omega_\K({\hat M})|{\tilde \ast}_\cone \omega=\pm \omega\
\end{split}}
\ee
of twisted (anti-)selfdual forms on the cylinder and cone:
\be
r^\cE(\Omega^{\pm}_{\K,\cyl}({\hat M}))=\Omega^{\pm}_{\K,\cone}({\hat M})~~.
\ee
In fact, we have mutually inverse $\cC^\infty({\hat M},\K)$-linear unital isomorphisms of algebras:
\ben
\label{diagram:risom_asd}
\scalebox{1.2}{
\xymatrix@1{(\Omega^\pm_{\K,\cyl}({\hat M}),\diamond^\cyl){~~~} \ar@<0.7ex>[r]^{r^{\cE}|_{\Omega^\pm_{\K,\cyl}({\hat M})} } {~~} &
{~~~~~}(\Omega_{\K,\cone}^\pm ({\hat M}),\diamond^\cone) {~~}\ar@<0.7ex>[l]^{r^{-\cE}|_{\Omega_{\K,\cone}^\pm ({\hat M})}~}
}}~,
\een
where unitality follows from \eqref{nu0cylcone}, which implies
$r^\cE(p_\pm^\cyl)=p_\pm^\cone$ (recall from \cite{ga1} that
$p_\pm^\cyl=\frac{1}{2}(1\pm \nu^\cyl)$ and
$p_\pm^\cone=\frac{1}{2}(1\pm \nu^\cone)$ are the unit elements of the
algebras $(\Omega^{\pm}_{\K,\cyl}({\hat M}),\diamond^\cyl)$ and
$(\Omega^{\pm}_{\K,\cone}({\hat M}), \diamond^\cone)$, since $d+1$ is
odd and thus $\nu^\cyl,\nu^\cone$ are central in the corresponding \KA
algebras).

On the other hand, relations \eqref{Lg} imply:
\be
\cL_{\partial_u}\nu^\cyl=0~~,
\ee
and (since $\nu^\cone=r^{d+1}\nu^\cyl=r^\cE(\nu^\cyl)$):
\be
(\cL_{\partial_u}-\cE)\nu^\cone =(\cL_{\partial_u}-(d+1))\nu^\cone=0~~.
\ee
In particular, we have $\nu^\cyl\in \Omega_\K^\cyl({\hat M})$ and $\nu^\cone\in \Omega_\K^\cone({\hat M})$. 
Since ${\tilde \ast}_\cyl(\omega)=\omega\diamond^\cyl\nu^\cyl$ and  ${\tilde \ast}_\cone(\omega)=\omega\diamond^\cone\nu^\cone$ for all $\omega\in \Omega_\K({\hat M})$,
the properties listed above imply:
\ben
\label{new3}
[\cL_{\partial_u}, {\tilde \ast}_\cyl]_{-,\circ}=0\Longleftrightarrow  [\cL_{\partial_u}, P^\cyl_\pm]_{-,\circ}=0~~,
~~[\cL_{\partial_u}-\cE, {\tilde \ast}_\cone]_{-,\circ}=0~~\Longleftrightarrow  [\cL_{\partial_u}-\cE, P^\cone_\pm]_{-,\circ}=0~~,
\een
where we used the fact that $\cL_{\partial_u}$ and
$\cL_{\partial_u}-\cE$ are even derivations of the \KA
algebras of the cylinder and cone, respectively.  In particular, the
operators $\cL_{\partial_u}$ and $\cL_{\partial_u}-\cE$ preserve
the subspaces of twisted (anti-)selfdual forms on the cylinder and cone,
respectively: 
\be
\cL_{\partial_u}(\Omega_{\K,\cyl}^\pm({\hat M}))\subset \Omega_{\K,\cyl}^\pm({\hat M})~~,
~~(\cL_{\partial_u}-\cE)(\Omega_{\K,\cone}^\pm({\hat M}))\subset \Omega_{\K,\cone}^\pm({\hat M})
\ee

\paragraph{Definition.} The subalgebras of {\em special twisted (anti-)selfdual
forms} are the following $\cpinf$-subalgebras of the \KA algebras of the
cylinder and of the cone:
\be
\boxed{\Omega^{\pm,\cyl}_\K({\hat M})\eqdef \Omega^{\pm}_{\K,\cyl}({\hat M})\cap \Omega^{\cyl}_\K({\hat M})~~,
~~\Omega^{\pm,\cone}_\K({\hat M})\eqdef \Omega^{\pm}_{\K,\cone}({\hat M})\cap \Omega^{\cone}_\K({\hat M})}~~.
\ee
These algebras have units $p_\pm^\cyl=\frac{1}{2}(1\pm \nu^\cyl)$ and $p_\pm^\cone=\frac{1}{2}(1\pm \nu^\cone)$, respectively. 
Combining the observations above gives:

\paragraph{Proposition.} The appropriate restrictions of the maps
$r^{\pm \cE}$ give mutually inverse $\cpinf$-linear unital
isomorphisms of algebras between the subalgebras of special twisted
selfdual/anti-selfdual forms of the cylinder and cone:
\ben
\label{diagram:risom_specialasd}
\scalebox{1.2}{
\xymatrix@1{(\Omega_\K^{\pm,\cyl}({\hat M}),\diamond^\cyl) ~~~\ar@<0.7ex>[r]^{r^{\cE}|_{\Omega_\K^{\pm,\cyl}({\hat M})} ~} {~~~} &
{~~~~~}(\Omega_\K^{\pm,\cone} ({\hat M}),\diamond^\cone)  \ar@<0.7ex>[l]^{r^{-\cE}|_{\Omega_\K^{\pm,\cone} ({\hat M})}~~}
}}~~.
\een

\paragraph{Recovering the \KA algebra of $(M,g)$.}  The $\cpinf$-algebras
$(\Omega_\K^{\perp,\cyl} ({\hat M}),\diamond^\cyl)$ and $(\Omega_\K^{\perp,\cone} ({\hat M}),\diamond^\cone)$ can be identified with the
\KA algebra $(\Omega_\K(M),\diamond)$ as follows.  Let $\Pi:{\hat
M}\rightarrow M$ be the projection on the second factor. Then one has
the following quite obvious statement:

\paragraph{Proposition.} The pullback map
$\Pi^\ast:\Omega_\K(M)\rightarrow \Omega_\K({\hat M})$ has image equal
to $\Omega^{\perp,\cyl}_\K ({\hat M})$. Furthermore, its corestriction
to this image (which we again denote by $\Pi^\ast$) is a unital
$\cinf$-linear isomorphism of algebras from $(\Omega_\K(M),\diamond)$
to the vertical subalgebra $(\Omega^{\perp,\cyl}_\K({\hat
M}),\diamond^\cyl)$ of the cylinder, provided that we identify
$\cpinf\approx \cinf$.  The inverse of this isomorphism is the
pullback map $j^\ast$, where $j:M\hookrightarrow {\hat M}$ is the
embedding of $M$ as the section $r=1$ of ${\hat M}$. Thus, we have
mutually inverse unital isomorphisms of $\cinf\approx\cpinf$-algebras:
\ben
\label{diagram:jPi}
\scalebox{1.2}{
\xymatrix@1{(\Omega_\K(M),\diamond) ~~~\ar@<0.7ex>[r]^{\Pi^{\ast}|^{\Omega_\K^{\perp,\cyl}({\hat M})}~} {~~~} &
{~~~~}(\Omega_\K^{\perp,\cyl}({\hat M}),\diamond^\cyl)  \ar@<0.7ex>[l]^{j^{\ast}|_{\Omega_\K^{\perp,\cyl}({\hat M})}~~~}
}}~.
\een
\noindent The proof is easy and left to the reader. Combining with the
previous proposition gives the relation between the \KA algebra of $M$
and the vertical subalgebra of the cone:

\paragraph{Proposition.} We have mutually-inverse unital isomorphisms of
$\K$-algebras:
\ben
\label{diagram:rjPi}
\scalebox{1.2}{
\xymatrix@1{(\Omega_\K(M),\diamond) ~~~~~ \ar@<0.7ex>[r]^{~~r^{\cE}\circ ~\Pi^{\ast}|^{\Omega_\K^{\perp,\cone} ({\hat M})}~~~~~} {~~~} &
{~~~~~~~}(\Omega_\K^{\perp,\cone} ({\hat M}),\diamond^\cone)  \ar@<0.7ex>[l]^{~~j^\ast \circ ~r^{-\cE}|_{\Omega_\K^{\perp,\cone} ({\hat M})}~~~~~}
}}~.
\een
Thus $\Omega^{\perp,\cyl}_\K({\hat M})$ consists of those inhomogeneous forms
on ${\hat M}$ which are $\Pi$-pullbacks of inhomogeneous forms $\omega$ on $M$; this pullback will be called the
{\em cylinder lift} $\omega_\cyl$ of $\omega$:
\ben
\label{cyl_lift}
\boxed{\omega_\cyl\eqdef \Pi^\ast(\omega)\in \Omega^{\perp,\cyl}_\K({\hat M})~~,~~\forall \omega\in \Omega_\K(M)}~~.
\een
On the other hand, $\Omega^{\perp,\cone}_\K({\hat M})$ consists of forms obtained by  rescaling the various rank components of such pullbacks
with appropriate non-negative powers of $r$. Explicitly, $\Omega^{\perp,\cone}_\K(M)$ consists of {\em cone lifts}:
\ben
\label{cone_lift}
\boxed{\omega_\cone\eqdef r^\cE(\Pi^\ast(\omega))\in \Omega^{\perp,\cone}_\K({\hat M})~~,~~\forall \omega\in \Omega_\K(M)}~~,
\een
which are inhomogeneous forms of the type:
\be
\omega_\cone=r^{\cE}(\Pi^\ast(\omega))=\sum_{k=0}^d r^k \Pi^\ast(\omega^{(k)})~~,
~~\forall \omega=\sum_{k=0}^d\omega^{(k)}\in \Omega_\K(M)~~{\rm with}~~\omega^{(k)}\in \Omega^k_\K(M)~~.
\ee
For example, relations \eqref{pullbacks} show that $\nu_\top^\cyl, \nu_\top^\cone$ are the cylinder and cone lifts of $\nu$:
\be
\nu_\top^\cyl=\nu_\cyl~~,~~\nu_\top^\cone=\nu_\cone~~.
\ee 

\paragraph{Truncated models.}

As in \cite{ga1}, we consider the complementary idempotent operators
$P_<:\Omega_\K({\hat M})\rightarrow \Omega^<_\K({\hat M})$ and $P_>:\Omega_\K({\hat M})\rightarrow \Omega^>_\K({\hat M})$
which associate to an inhomogeneous form the sum of its components of rank smaller, respectively bigger than the half-integer number $\frac{1}{2}\dim {\hat M}=\frac{d+1}{2}$.
Since both $r^\cE$ and the Lie derivative with respect to a vector field preserve the rank of differential forms,
we have the commutation relations:
\ben
\label{new4}
[r^\cE,P_<]_{-,\circ}=[r^\cE,P_>]_{-,\circ}=[\cL_{\partial_u},P_<]_{-,\circ}=[\cL_{\partial_u},P_>]_{-,\circ}=
[\cL_{\partial_u}-\cE,P_<]_{-,\circ}=[\cL_{\partial_u}-\cE,P_>]_{-,\circ}=0
\een
In particular, these give:
\be
r^\cE(\Omega^<_\K({\hat M}))=\Omega^<_\K({\hat M})~~,~~\cL_{\partial_u}(\Omega^<_\K({\hat M}))\subset \Omega^<_\K({\hat M})~~,~~
(\cL_{\partial_r}-\cE)(\Omega^<_\K({\hat M}))\subset \Omega^<_\K({\hat M})~~.
\ee
Recall from \cite{ga1} that the subspace $\Omega^<_\K({\hat M})$ carries associative binary products defined through:
\be
\bdiamond_\pm^\cyl\eqdef 2P_<\circ P_\pm^\cyl \circ \diamond^\cyl|_{\Omega_\K^<({\hat M})\otimes_{\cC^\infty({\hat M},\K)} \Omega_\K^<({\hat M})}~~,~~
\bdiamond_\pm^\cone\eqdef 2P_<\circ P_\pm^\cone \circ \diamond^\cone|_{\Omega_\K^<({\hat M})\otimes_{\cC^\infty({\hat M},\K)} \Omega_\K^<({\hat M})}~~.
\ee
Combining these relations with \eqref{rexpmorphism}, \eqref{Hodger} and with the commutation relations given above, one easily checks the identity:
\be
r^\cE\circ \bdiamond_\pm^\cyl=\bdiamond_\pm^\cone\circ (r^\cE\otimes r^\cE)~~,
\ee
which shows that we have mutually-inverse $\cC^\infty({\hat M},\K)$-linear unital isomorphisms of algebras:
\ben
\label{diagram:rbdiamond}
\scalebox{1.2}{
\xymatrix@1{(\Omega_\K^<({\hat M}),\bdiamond_\pm^\cyl) ~\ar@<0.7ex>[r]^{r^{\cE}|_{\Omega_\K^<({\hat M})} } {~~~} &
{~~~}(\Omega_\K^< ({\hat M}),\bdiamond_\pm^\cone)  \ar@<0.7ex>[l]^{r^{-\cE}|_{\Omega_\K^< ({\hat M})}~}
}}~~.
\een
Relations \eqref{new3} and \eqref{new4} together with \eqref{new1} and \eqref{new2} show that $\cL_{\partial_u}$ and  $\cL_{\partial_u}-\cE$ are derivations of the the algebras $(\Omega^<_\K({\hat M}), \bdiamond_\pm^\cyl)$ and  $(\Omega^<_\K({\hat M}), \bdiamond_\pm^\cone)$, respectively. The observations made above show that the subspaces of {\em truncated special inhomogeneous forms}:
\be
\boxed{\Omega^{<,\cyl}_\K({\hat M})\eqdef\Omega^<_\K({\hat M})\cap \Omega_\K^{\cyl}({\hat M})~~,~~
\Omega^{<,\cone}_\K({\hat M})\eqdef\Omega^<_\K({\hat M})\cap \Omega_\K^{\cone}({\hat M})}~~
\ee
are unital $\cpinf$-subalgebras of the algebras $(\Omega^<_\K({\hat M}),\bdiamond^\cyl_\pm)$ and $(\Omega^<_\K({\hat M}),\bdiamond^\cone_\pm)$.  
Furthermore:

\paragraph{Proposition.} We have mutually-inverse unital isomorphisms of
$\K$-algebras:
\ben
\label{diagram:truncspecial}
\scalebox{1.2}{
\xymatrix@1{(\Omega_\K^{<,\cyl}({\hat M}),\bdiamond^\cyl_\pm) ~~~~\ar@<0.7ex>[r]^{r^{\cE}|_{\Omega_\K^{<,\cyl}({\hat M})}~~ } {~~} & 
{~~~~~}(\Omega_\K^{<,\cone} ({\hat M}),\bdiamond^\cone_\pm) {~~} \ar@<0.7ex>[l]^{r^{-\cE}|_{\Omega_\K^{<,\cone} ({\hat M})}~~ } {~~}
}}~.
\een
The situation for the cylinder is summarized in the following commutative diagram:
\ben
\label{diagram:cylforms}
\scalebox{1.2}{\xymatrix{
(\Omega^{<,\cyl}_\K({\hat M}),\bdiamond_\pm^\cyl) \ar@<0.5ex>[d]^{\Xi^\cyl_\pm} ~ \ar@<0.5ex>[r]^{P_\pm^\cyl} ~ & ~
(\Omega^{\pm, \cyl}_\K (\hat{M}),\diamond^\cyl) ~ \ar@<0.5ex>[l]^{2 P_<} ~ \ar@<0.5ex>[d]^{2P_\perp} ~\\
~ (\Omega_\K(M),\diamond) ~ \ar@<0.5ex>[u]^{(\Xi^\cyl_\pm)^{-1}}  \ar@<0.5ex>[r]^{\Pi^{\ast}~~~}  & ~
(\Omega^{\perp,\cyl}_\K (\hat{M}),\diamond^\cyl) ~ \ar@<0.5ex>[l]^{ j^{\ast}~~~~} ~ \ar@<0.5ex>[u]^{P_\pm^\cyl} ~
}}
\een
where $\Xi^\cyl_\pm\eqdef 2 j^{\ast}\circ P_\perp\circ P_\pm^\cyl $
and $(\Xi^\cyl_\pm)^{-1}=2 P_<\circ P_\pm^\cyl \circ \Pi^{\ast}$,
while the situation for the cone is summarized in the commutative diagram:
\ben
\label{diagram:coneforms}
\scalebox{1.2}{\xymatrix{
(\Omega^{<,\cone}_\K({\hat M}),\bdiamond_\pm^\cone) \ar@<0.5ex>[d]^{\Xi^\cone_\pm} ~ \ar@<0.5ex>[r]^{P_\pm^\cone} ~ & ~
(\Omega^{\pm, \cone}_\K (\hat{M}),\diamond^\cone) ~ \ar@<0.5ex>[l]^{2 P_<} ~ \ar@<0.5ex>[d]^{2P_\perp} ~\\
~ (\Omega_\K(M),\diamond) ~ \ar@<0.5ex>[u]^{(\Xi^\cone_\pm)^{-1}}  \ar@<0.5ex>[r]^{r^{\cE}\circ ~\Pi^{\ast}~~~~}  & ~
(\Omega^{\perp,\cone}_\K (\hat{M}),\diamond^\cone) ~ \ar@<0.5ex>[l]^{ j^{\ast}\circ ~r^{-\cE}~~~~} ~ \ar@<0.5ex>[u]^{P_\pm^\cone} ~
}}
\een
where $\Xi^\cone_\pm\eqdef 2 j^{\ast}\circ ~r^{-\cE} \circ P_\perp\circ P_\pm^\cone $ and
$(\Xi^\cone_\pm)^{-1}=2 P_<\circ P_\pm^\cone \circ r^{\cE}\circ \Pi^{\ast}$. The
full collection of isomorphic models of the \KA algebra of $(M,g)$ which arise from the cone and cylinder constructions is summarized in the commutative diagram below:
\be
\scalebox{1.02}{
\!\!\!\!\!\!\!\!\!\!\! \xymatrix{
& (\Omega_\K^{<,\cyl}(\hat{M}),\bdiamond_\pm^{\cyl})  \ar@<0.5ex>^{P_\pm^{\cyl}}[rr]  \ar@{.}[d]<0.5ex> \ar@<-0.5ex>@{<.}_{(\Xi^{\cyl}_\pm)^{-1}}[d] \ar@<0.5ex>[dl]^{~~r^\cE}
& & (\Omega_\K^{\pm,\cyl}(\hat{M}),\diamond^{\cyl})  \ar@<0.5ex>^{2P_\perp}[dd] \ar@<0.5ex>^{r^\cE}[dl] \ar@<0.5ex>^{2P_<}[ll] \\
~~~~~~~(\Omega_\K^{<,\cone}(\hat{M}),\bdiamond_\pm^{\cone})  \ar@<0.5ex>[ur]^{~r^{-\cE}}\ar@<0.5ex>^{~~~~~~~~~~~~~~~~~~~P_\pm^{\cone}}[rr] \ar@<0.5ex>^{\Xi^{\cone}_\pm}[dd]
& &  (\Omega_\K^{\pm,\cone}(\hat{M}),\diamond^{\cone})   \ar@<0.5ex>^{r^{-\cE}}[ur]{~~}\ar@<0.5ex>^>>>>>>{2P_\perp}[dd] \ar@<0.5ex>^{2P_<~~~~~~~~~~~~~~~}[ll] \\
& (\Omega_\K(M),\diamond)  \ar@<0.5ex>@{.}[r]\ar@<-0.5ex>@{<.}[r]_{j^\ast}\ar@<0.5ex>@{.}[u]\ar@<-0.5ex>@{<.}_{\Xi^{\cyl}_\pm}[u]
& &  (\Omega_\K^{\perp,\cyl}(\hat{M}),\diamond^{\cyl})  \ar@<0.5ex>@{.}[l]\ar@<-0.5ex>@{<.}_{\Pi^\ast}[l] \ar@<0.5ex>^{P_\pm^{\cyl}}[uu] \ar@<0.5ex>^{~~r^{\cE}}[dl]\\
(\Omega_\K(M),\diamond) \ar@<0.5ex>^{r^\cE \circ~\Pi^\ast}[rr] \ar@<0.5ex>^{(\Xi^{\cone}_\pm)^{-1}}[uu] \ar@{.>}^{\id}[ur]<0.5ex>\ar@<-0.5ex>@{<.}_{\id}[ur]
& & (\Omega_\K^{\perp,\cone}(\hat{M}),\diamond^{\cone})  \ar@<0.5ex>^{r^{-\cE}}[ur] \ar@<0.5ex>^>>>>>>{P_\pm^{\cone}}[uu] \ar@<0.5ex>^{j^\ast\circ~r^{-\cE}}[ll]
}} ~~~ \nn
\ee

\subsection{The Levi-Civita connections of the cylinder and cone } 
\label{sec:coneconnections}

\paragraph{The Levi-Civita connections.}

Using the formulas for the Levi-Civita connection of a warped product, one finds that the Levi-Civita connection of the cylinder is given by: 
\be
\nabla^\cyl_V W=V(g_\cyl(\partial_u,W))\partial_u + \nabla_V^\ast (W^\perp)~~,~~\forall V,W\in \Gamma({\hat M},T_\K {\hat M})~~,
\ee
where $W^\perp\in \Gamma({\hat M}, T_\K^\perp {\hat M})$ is the part of $W$ orthogonal to $\partial_u$ (and thus to $\partial_r$): 
\be
W^\perp=W-g_\cyl(\partial_u, W)\partial_u=W-g_\cone(\partial_r, W)\partial_r~~
\ee
while $\nabla^\ast$ (a connection on the bundle $T_\K^\perp {\hat M}\approx \Pi^\ast(T_\K M)$) 
is the pullback along $\Pi$ of the Levi-Civita connection $\nabla$ of $(M,g)$. Notice the properties: 
\ben
\label{nablacyl_props}
\nabla^\cyl_V(W)^\perp=\nabla^\cyl_V(W^\perp)=\nabla_V^\ast(W^\perp)~~,~~\nabla_V^\cyl(\partial_u)=0~~,~~\forall V, W\in \Gamma({\hat M},T_\K {\hat M})~~.
\een
On the other hand, the Levi-Civita connection of the cone can be expressed as:
\ben
\label{nablacone}
\nabla^\cone_V W=\frac{1}{r} \nabla^\cyl_V (r W)+\lambda (V,W)~~,~~\forall~V, W\in \Gamma({\hat M}, T_\K {\hat M})~~,
\een
where $\lambda \in \Gamma({\hat M}, T^\ast_\K {\hat M}\otimes T^\ast_\K {\hat M}\otimes T_\K {\hat M})$ is a tensor of type
$(2,1)$ (i.e. with two covariant indices and one contravariant index) defined through the formula:
\ben
\label{lambda}
\lambda(V,W)=g_\cyl(\partial_u, W)V -g_\cyl(V,W)\partial_u
=\frac{1}{r} (g_\cone(\partial_r, W)V -g_\cone(V,W)\partial_r)~~,
\een
for all $V, W\in \Gamma({\hat M}, T_\K {\hat M})$. In particular, the Levi-Civita connections of the cylinder and cone satisfy the following relations for all 
vector fields $X,Y$ on $M$:
\beqan
\label{pbc}
&&\nabla^\cyl_{X_\ast} Y_\ast =(\nabla_{X} Y)_\ast~~,~~
\nabla^\cyl_{\partial_u}\partial_u =0~~,~~\nabla^\cyl_{\partial_u}X_\ast =\nabla^\cyl_{X_\ast}\partial_u= 0~~,\\
&&\nabla^\cone_{X_\ast} Y_\ast=(\nabla_X Y)_\ast- (g(X,Y)\circ \Pi)  r\partial_r~~,~~\nabla^\cone_{\partial_r}\partial_r =0~~,
~~\nabla^\cone_{\partial_r}X_\ast=\nabla_{X_\ast}\partial_r=\frac{1}{r} X_\ast~~.\nn
\eeqan

\paragraph{Remark.} We note that $\lambda$ can also be written in the form:
\be
\lambda(V,W)=g_\cyl(\partial_u, W)V^\perp -g_\cyl(V^\perp,W)\partial_u
=\frac{1}{r} (g_\cone(\partial_r, W)V^\perp
-g_\cone(V^\perp,W)\partial_r)~~,
\ee
where $V, W\in \Gamma({\hat M}, T_\K {\hat M})$ and $V^\perp$ is the part of $V$ which is orthogonal to $\partial_r$ (and hence to $\partial_u$).

\paragraph{The connections induced on differential forms.}
Direct computation using \eqref{nablacone} shows that the connections
induced on $\Omega_\K({\hat M})$ by the Levi-Civita connections of the
cone and cylinder are related through:
\!\!\!\!\!\! \ben
\label{formconnections1}
\nabla^\cone_V \omega=(r^\cE\circ \nabla_V^\cyl\circ
r^{-\cE})(\omega)+\frac{1}{r}\left[V_{\hash_\cone}\wedge (\partial_r
\lrcorner \omega)- \theta\wedge (V\lrcorner
\omega)\right]~,~\forall~V\in \Gamma({\hat M}, T_\K{\hat
M})~,~\forall \omega\in \Omega_\K({\hat M})~~~~~~~~~~
\een
where $V_{\hash_\cone}$ is the one-form dual to $V$ with respect to the cone metric:
\be
V_{\hash_\cone}=V\lrcorner g_\cone~~.
\ee

\paragraph{Remark.} One has the obvious identity:
\be
V^\parallel_{\hash_\cone}\wedge (\partial_r \lrcorner \omega)=\theta\wedge (V^\parallel\lrcorner \omega)~~,~~\forall \omega\in \Omega_\K({\hat M})~~,
\ee
where $V\in \Gamma({\hat M},T_\K {\hat M})$ and $V^\parallel=g_\cyl(\partial_u, V)\partial_u=g_\cone(\partial_r, V)\partial_r$ is the part of $V$ which is parallel to $\partial_u$ (and thus 
to $\partial_r$). This implies that \eqref{formconnections1} can also be written as:
\ben
\label{formconnections2}
\!\!\!  \nabla^\cone_V \omega=(r^\cE\circ \nabla_V^\cyl\circ
r^{-\cE})(\omega)+\frac{1}{r}\left[V_{\hash_\cone}^\perp \wedge
(\partial_r \lrcorner \omega)- \theta\wedge (V^\perp \lrcorner
\omega)\right]~,~\forall~V\in \Gamma({\hat M}, T_\K {\hat
M})~,~\forall \omega\in \Omega_\K({\hat M})~~~~~~~~~~~
\een
where $V^\perp_{\hash_\cone}=V^\perp\lrcorner g_\cone$. Using the expansion of
the graded $\diamond$-commutator (see Section 3 of \cite{ga1}), it is not very
hard to check that the following identities hold for all $\omega\in
\Omega_\K({\hat M})$ and any vector field $V^\perp \in \Gamma({\hat M},
T^\perp_\K {\hat M})$ which is everywhere orthogonal to $\partial_r$ (and
thus to $\partial_u$ as well):
\be
\!\!\!   V^\perp_{\hash_\cone}\wedge (\partial_r \lrcorner \omega)-\theta\wedge (V^\perp\lrcorner \omega)
=-(V^\perp_{\hash_\cone}\wedge \theta) \bigtriangleup_1^\cone \omega
=\frac{1}{2} [[V^\perp_{\hash_\cone}\wedge \theta, \omega]]_{-,\diamond^\cone}=
\frac{1}{2}[V^\perp_{\hash_\cone}\wedge \theta, \omega]_{-,\diamond^\cone}~~
\ee
and:
\be
\label{laterid}
V^\perp_{\hash_\cone}\wedge (\partial_r \lrcorner \omega)-\theta\wedge (V^\perp \lrcorner \omega)=
-r(V^\perp_{\hash_\cyl}\wedge \psi)\bigtriangleup_1^\cyl \omega=\frac{r}{2} [[V^\perp_{\hash_\cyl}\wedge \psi, \omega]]_{-,\diamond^\cyl}=
\frac{r}{2}[V_{\hash_\cyl}\wedge \psi, \omega]_{-,\diamond^\cyl}~~.
\ee
Combining the last identity with \eqref{formconnections2} gives the following relation which will be used below:
\ben
\label{mainrelation}
\nabla^\cone_V \omega=(r^\cE\circ \nabla_V^\cyl\circ r^{-\cE})(\omega)+\frac{1}{2}[V^\perp_{\hash_\cyl}\wedge \psi, \omega]_{-,\diamond^\cyl}=
(r^\cE\circ \nabla_V^\cyl\circ r^{-\cE})(\omega)+\frac{1}{2r}[V^\perp_{\hash_\cone}\wedge \theta, \omega]_{-,\diamond^\cone}
\een
Equation \eqref{mainrelation} expresses $\nabla^\cone$ in terms of
$\nabla^\cyl$ using operations from the \KA algebra of the cylinder or
of the cone. 

\paragraph{Some useful identities.}
The identity: 
\be
\nabla_V^\cyl(W\lrcorner \omega)=(\nabla_V^\cyl W)\lrcorner\omega+W\lrcorner \nabla_V^\cyl \omega
\ee
(which is also satisfied by any linear connection on $T_\K{\hat M}$) and the second relation in \eqref{nablacyl_props} imply: 
\be
[\nabla_V^\cyl, \partial_u\lrcorner ~~]_{-,\circ}=0\Longleftrightarrow [\nabla_V^\cyl, \iota_\psi^\cyl]_{-,\circ}=0~~,~~\forall V\in \Gamma({\hat M},T_\K {\hat M})~~,
\ee
while the fact that $\nabla_V^\cyl$ is an even derivation of the exterior algebra of ${\hat M}$ and the obvious relation $\nabla_V^\cyl\psi=0$ imply: 
\be
[\nabla_V^\cyl, \wedge_\psi]_{-,\circ}=0~~,~~\forall V\in \Gamma({\hat M},T_\K {\hat M})~~.
\ee 
These two properties of $\nabla^\cyl$ imply that the following identities hold for all $V, W\in \Gamma({\hat M},T_\K{\hat M})$: 
\be
[\nabla_V^\cyl, P_\parallel]_{-,\circ}=[\nabla_V^\cyl, P_\perp]_{-,\circ}=0\Longrightarrow \nabla_V^\cyl(\omega_\parallel)=\nabla_V^\cyl(\omega)_\parallel~~
,~~\nabla_V^\cyl(\omega_\perp)=\nabla_V^\cyl(\omega)_\perp~~,~~\forall \omega\in \Omega_\K({\hat M})
\ee
as well as: 
\be
\nabla_V^\cyl(\omega_\top^\cyl)=\nabla_V^\cyl(\omega)^\cyl_\top~~,~~\forall \omega\in \Omega_\K({\hat M})~~.
\ee
Using these observations and the fact that $\nabla^\cyl_V\nu^\cyl=0$, we find
that the morphisms $\varphi_\pm^\cyl$ constructed using $\diamond^\cyl$ and
$\nu^\cyl$ as in Section 3.10 of \cite{ga1} satisfy the following relation
which will be used later on:
\ben
\label{nablacyl_varphi}
[\nabla^\cyl_V,\varphi_\pm^\cyl]_{-,\circ}=0\Longleftrightarrow \nabla_V^\ast\circ \varphi_\pm^\cyl=\varphi_\pm^\cyl\circ \nabla_V^\cyl~~,~~\forall V\in \Gamma({\hat M},T_\K {\hat M})~~,
\een
where we used the fact that the restriction $\nabla^\cyl|_{\wedge
(T_\K^\perp {\hat M})^\ast }$ equals the pullback $\nabla^\ast$
through $\Pi$ of the connection induced by $\nabla$ on $\wedge
T_\K^\ast M$ (as usual, we identify $\wedge (T_\K^\perp {\hat M})^\ast
\approx \Pi^\ast(\wedge T_\K^\ast M)$.

\subsection{Pinors on metric cylinders and cones}
\label{sec:conepin}

\paragraph{The pin bundle of ${\hat M}$.}
Let $S$ be a pin bundle of $(M,g)$ and $\gamma:(\wedge T^\ast_\K
M,\diamond)\rightarrow (\End(M),\circ)$ be its fiberwise
representation.  Let ${\hat S}\eqdef \Pi^\ast(S)$ be the pullback bundle and $\gamma_\ast\eqdef \Pi^\ast(\gamma):\wedge (T^\perp_\K {\hat M})^\ast
\rightarrow \End({\hat S})$ be the pullback of $\gamma$ to the bundle $\Pi^\ast(\wedge T^\ast_\K M)\approx \wedge (T^\perp_\K {\hat M})^\ast$.  
Recall that our assumptions imply that $\gamma$ induces a bijection from
$\Omega_\K(M)=\Gamma(M,\wedge T^\ast_\K M)$ to $\Gamma(M,\End(S))$. In
turn, this implies that the map induced by $\gamma_\ast$ on sections is an isomorphism of $\cC^\infty({\hat M},\K)$-algebras:
\be
\gamma_\ast:\Omega^\perp_\K({\hat M})\stackrel{\sim}{\longrightarrow} \Gamma({\hat M},{\End(\hat S}))~~.
\ee
We have the basic property: 
\ben
\label{pbprop}
\gamma_\ast\circ\Pi^\ast=\Pi^\ast\circ \gamma~~,
\een
where, in the left hand side, $\Pi^\ast$ denotes the pullback of
differential forms while in the right hand side it denotes the
pullback of sections of $\End(S)$ to sections of $\End({\hat S})$. This gives a commutative square of 
unital morphisms of $\cinf$-algebras which constitutes the rightmost part of the diagram \eqref{diagram:gammaPi} 
(as usual, we identify $\cpinf\approx \cinf$). 
\ben
\label{diagram:gammaPi}
\scalebox{1.2}{
\xymatrix{
\Omega_\K (\hat{M}) \ar[d]_{r^\cE} ~  \ar[dr]^{\gamma_\cyl} \ar[r]^{\varphi^\cyl_\epsilon}
& ~~\Omega^\perp_\K(\hat{M}) ~ \ar[d]^{\gamma_\ast}~ &~ \Omega_\K(M) ~\ar[l]_{\Pi^\ast} \ar[d]^\gamma \\
 \Omega_\K(\hat{M}) \ar[r]^{\gamma_\cone~~~~~}~ & ~\Gamma(\hat{M},\End(\hat{S})) ~& ~\Gamma(M,\End(S))  ~\ar[l]_{~~\Pi^\ast}
  }}
\een 

\paragraph{The morphisms $\gamma_\cyl$ and $\gamma_\cone$.}
Let us fix a sign factor $\epsilon\in \{-1,1\}$. Considering the
unital morphisms of $\cC^\infty({\hat M},\K)$-algebras: 
\beqa
\varphi_\epsilon^\cyl~~&=&~~2P_\perp \circ
P_\epsilon^\cyl:(\Omega_\K({\hat M}),\diamond^\cyl)\rightarrow~
(\Omega_\K^\perp({\hat M}),\diamond^\cyl)~~,\nn\\
\varphi_\epsilon^\cone&=&2P_\perp \circ
P_\epsilon^\cone:(\Omega_\K({\hat M}),\diamond^\cone)\rightarrow
(\Omega_\K^\perp({\hat M}),\diamond^\cone)
\eeqa 
as in \cite{ga1}, we define unital morphisms of $\cC^\infty({\hat M},\K)$-algebras:
$\gamma_{\ccyl}:(\Omega_\K({\hat M}), \diamond^\ccyl ) \rightarrow
(\Gamma({\hat M},\End({\hat S})),\circ)$ through (see diagram \eqref{diagram:gammaPi}):
\ben
\label{gammacylconedef}
\gamma_\cyl \eqdef \gamma_\ast\circ \varphi^\cyl_\epsilon~~,
~~\gamma_\cone\eqdef \gamma_\cyl \circ r^{-\cE}~~.
\een
It is clear that $\gamma_\ccyl $ are $\cC^\infty({\hat M},\K)$-linear,
so they are induced by corresponding morphisms of bundles of algebras,
which are easily seen to be irreducible on the fibers. 
Since $r^\cE$ commutes with $P_\perp$ and satisfies \eqref{Hodger}, we find:
\be
r^\cE\circ \varphi_\epsilon^\cyl=\varphi_\epsilon^\cone\circ r^\cE\Longleftrightarrow \varphi_\epsilon^\cyl\circ r^{-\cE}=r^{-\cE}\circ \varphi_\epsilon^\cone~~,
\ee
which means that $\gamma_\cone$ can also be written as (see diagram \eqref{diagram:gammacylconeast}):
\be
\gamma_\cone=\gamma_\ast \circ r^{-\cE}\circ \varphi^\cone_\epsilon~~.
\ee
\ben
\label{diagram:gammacylconeast}
\scalebox{1.2}{
\xymatrix{
\Omega_\K (\hat{M}) \ar[dd]_{\varphi^\cyl_\epsilon} ~  \ar[dr]^{\gamma_\cyl} \ar[rr]^{r^\cE}
&& ~\Omega_\K(\hat{M}) ~ \ar[dd]^{\varphi^\cone_\epsilon}~ \ar[dl]_{ \gamma_\cone}  \\
 & ~\Gamma(\hat{M},\End(\hat{S})) \\
 \Omega^\perp_\K(\hat{M}) \ar[ur]^{\gamma_\ast} \ar[rr]^{r^\cE} 
&&  \Omega^\perp_\K(\hat{M}) \ar[ul]_{~~\gamma_\ast\circ~r^{-\cE}}
  }}
\een
The last relation of Section 3.10 in \cite{ga1}
gives $\varphi^\cyl_\epsilon(\nu^\cyl)=\epsilon 1_{\hat M}$, which implies
$\gamma_\cyl(\nu^\cyl)=\epsilon \gamma_\ast(1_{\hat M})=\epsilon \Pi^\ast(\gamma(1_M))=\epsilon \Pi^\ast(\id_{S})=
\epsilon ~\id_{{\hat S}}$, where we noticed that $1_{\hat M}=\Pi^\ast(1_M)$ and $\id_{{\hat S}}=\Pi^\ast(\id_S)$.
We find: 
\ben
\label{gammanu_cylcone}
\gamma_\cyl(\nu^\cyl)=\gamma_\cone(\nu^\cone)=\epsilon~\id_{{\hat S}}~~,
\een
where we used the fact that $r^{-\cE}(\nu^\cone)=\nu^\cyl$. It follows
that $\gamma_\cyl$ makes ${\hat S}$ into a pin bundle of $({\hat M},
g_\cyl)$ having signature $\epsilon$, while $\gamma_\cone$ makes
${\hat S}$ into a pin bundle of $({\hat M}, g_\cone)$ of the same
signature. Since $\varphi_\epsilon^\cyl(\psi)=\epsilon \nu_\top^\cyl=\epsilon \Pi^\ast(\nu)$, we also have  
$\gamma_\cone(\theta)=\gamma_\cyl(\psi)=\epsilon \gamma_\ast(\Pi^\ast(\nu))=
\epsilon~\Pi^\ast(\gamma(\nu))$, where we noticed that $r^{-\cE}(\theta)=\psi$. Thus:
\ben
\label{gammagen_cylcone}
\gamma_\cone(\theta)=\gamma_\cyl(\psi)=\epsilon~\Pi^\ast(\gamma(\nu))~~.
\een
The form of $\varphi_\epsilon$ given in  \cite{ga1} implies:
\be
\gamma_\cyl(\omega)=\gamma_\ast(\epsilon {\tilde
\ast}^\cyl_{0}(\omega^\cyl_\top)+\omega_\perp)~~,~~
\gamma_\cone(\omega)=(\gamma_\ast\circ r^{-\cE})(\epsilon {\tilde
\ast}^\cone_{0}(\omega^\cone_\top)+\omega_\perp)~~,
\ee
where:
\be
\omega^\cyl_\top\eqdef \iota^\cyl_\psi\omega=\partial_u\lrcorner \omega~~,
~~\omega^\cone_\top\eqdef \iota^\cone_\theta\omega=\partial_r\lrcorner \omega=\frac{1}{r} \omega_\top^\cyl~~,
~~\forall \omega\in \Omega_\K({\hat M})~~.
\ee
Furthermore, we have the following formulas for the cylinder and cone lifts \eqref{cyl_lift} and \eqref{cone_lift}
of a form $\omega\in \Omega_\K(M)$:
\ben
\label{gamma_lifts}
\gamma_\cyl(\omega_\cyl)=\gamma_\cone(\omega_\cone)=\Pi^\ast(\gamma(\omega))~~,~~\forall \omega\in \Omega_\K(M)~~,
\een
where we used the fact that $\Pi^\ast(\omega)\in \Omega^\perp_\K({\hat M})$ while $\varphi_\epsilon^\cyl$ restricts to the identity on
$\Omega^\perp_\K({\hat M})$.

\paragraph{The dequantization maps of the cylinder and cone.}
Since $P_\perp$ restricts to a bijection from $\Omega^\epsilon_{\K,\ccyl}({\hat M})$ to $\Omega^\perp_\K({\hat M})$, the restriction of $\gamma_\ccyl$ 
to $\Omega^\epsilon_{\K,\ccyl}({\hat M})$ is a composition of bijections: 
\be
\gamma_\cyl|_{\Omega^\epsilon_{\K,\cyl }({\hat M})}=2\gamma_\ast|_{\Omega^\perp_\K({\hat M})}\circ
P_\perp|^{\Omega^\perp_\K({\hat M})}_{\Omega^\epsilon_{\K,\cyl}({\hat M})}~~,~~
\gamma_\cone|_{\Omega^\epsilon_{\K,\cone}({\hat M})}=2\gamma_\ast|_{\Omega^\perp_\K({\hat M})}\circ
r^{-\cE}\circ P_\perp|^{\Omega^\perp_\K({\hat M})}_{\Omega^\epsilon_{\K,\cone}({\hat M})}~~
\ee
and hence the partial inverses of $\gamma_\cyl$ and $\gamma_\cone$ are given by: 
\ben
\label{gammainv_cylcone}
\gamma_\cyl^{-1}=P_\epsilon^\cyl\circ \gamma_\ast^{-1}~~,
~~\gamma_\cone^{-1}=P_\epsilon^\cone\circ r^\cE\circ \gamma_\ast^{-1}=r^\cE\circ \gamma_\cyl^{-1}~~,
\een
where we used the fact (see Subsection 3.10 of \cite{ga1}) that the inverse of $2P_\perp|^{\Omega^\perp_\K({\hat M})}_{\Omega^\epsilon_{\K,\ccyl}({\hat M})}$ is given by 
$P_\epsilon^\ccyl|^{\Omega^\epsilon_{\K,\ccyl}({\hat M})}_{\Omega^\perp_\K({\hat M})}$. The situation is summarized in diagram \eqref{diagram:gammacc}. 
\ben
\label{diagram:gammacc}
\scalebox{1.2}{
\xymatrix{
\Gamma(\hat{M},\End(\hat{S}))~ \ar[d]_{\gamma_\ast^{-1}}  \ar[dr]^{\gamma_\cyl^{-1}} \ar[r]^{\gamma_\cone^{-1}}
& ~\Omega^\epsilon_{\K,\cone}(\hat{M})~ \\
\Omega^\perp_\K(\hat{M}) \ar[r]_{P^\cyl_\epsilon}~ & ~\Omega^\epsilon_{\K,\cyl}(\hat{M}) ~ \ar[u]_{r^\cE}~ 
  }}
\een

\subsection{The Fierz isomorphism of cylinders and cones}
\label{sec:conefierz}

\paragraph{The morphisms ${\hat E}$ and $\check{E}_\ast$.}
For the pin bundle $S$ over $M$, let us consider, as in \cite{ga1}, the natural isomorphism $q:S\otimes
S^\ast\stackrel{\sim}{\rightarrow} \End(S)$ as well as the isomorphism
$\rho:S\stackrel{\sim}{\rightarrow} S^\ast$ induced by a
non-degenerate admissible pairing $\cB$ on $S$.  The fiberwise
bilinear pairing $\cB$ pulls-back to a non-degenerate bilinear pairing
${\hat \cB}\eqdef\Pi^\ast(\cB)$ on ${\hat S}$, which is easily seen to be
admissible for both $\gamma_\cyl$ and $\gamma_\cone$. It induces a
bundle isomorphism between ${\hat S}$ and its dual which coincides
with the pullback ${\hat \rho}\eqdef\Pi^\ast(\rho):{\hat
S}\stackrel{\sim}{\rightarrow}{\hat S}^\ast$ of $\rho$. Furthermore,
the pullback ${\hat q}\eqdef\Pi^\ast(q):{\hat S}\otimes {\hat
S}^\ast\stackrel{\sim}{\longrightarrow} \End({\hat S})$ of $q$ coincides with the
natural isomorphism between ${\hat S}\otimes {\hat S}^\ast$ and
$\End({\hat S})$. Combining these, we find that the pullback:
\ben
\label{East}
{\hat E}\eqdef \Pi^\ast(E) ={\hat q}\circ (\id_{\hat S}\otimes {\hat \rho}):{\hat
S}\otimes {\hat S}\stackrel{\sim}{\rightarrow}
\End({\hat S})
\een
of the isomorphism $E=q\otimes(\id_S\otimes \rho):S\otimes S\stackrel{\sim}{\rightarrow} \End(S)$ coincides with
the isomorphism built as in \cite{ga1} from the
admissible bilinear pairing ${\hat \cB}$ on ${\hat S}$. This implies that
the bipinor bundle of algebras $({\hat S}\otimes {\hat S}, {\hat
\bullet})$ built from ${\hat S}$ using the pairing ${\hat \cB}$ is the
pullback of the bipinor bundle of algebras $(S\otimes S,\bullet)$
built from $S$ using the pairing $\cB$. Defining:
\ben
\label{checkEast}
\check{E}_\ast\eqdef\gamma_\ast^{-1}\circ {\hat E}:{\hat S}\otimes {\hat
S}\stackrel{\sim}{\rightarrow} \wedge (T^\perp_\K {\hat M})^\ast \approx
\Pi^\ast (T^\ast _\K M)~~,
\een
we have $\check{E}_\ast=\Pi^\ast(\check{E})$ where $\check{E}:S\otimes
S\rightarrow \wedge T^\ast_\K M$ is the Fierz isomorphism of $(M,g)$.
The situation is summarized in the commutative diagram \eqref{diagram:checkEast}, which 
also encodes the action of $\check{E}_\ast$ on pulled-back forms. 
\ben
\label{diagram:checkEast}
\scalebox{1.1}{
\xymatrix{
& \Gamma(M,S\otimes S)  \ar[rr]^{\id_S\otimes \rho}  \ar@{.>}[dd]_>>>>>>>{\check{E}} \ar[dl]_{\Pi^\ast} \ar@{.>}[ddrr]^>>>>>>>>>>{E}
& & \Gamma(M,S\otimes S^\ast)   \ar[dd]^{q} \ar[dl]_{\Pi^\ast} \\
\Gamma(\hat{M},\hat{S}\otimes \hat{S})   \ar[rr]^>>>>>>>>>>{\id_{\hat{S}}\otimes {\hat \rho}}  \ar[dd]_{\check{E}_\ast}  \ar[ddrr]_>>>>>>>>>>>>>>{{\hat E}}
& &  \Gamma(\hat{M},\hat{S}\otimes \hat{S}^\ast)  \ar@<0.5ex>[dd]^>>>>>>>{{\hat q}}  \\
& \Omega_\K(M)   \ar@{.>}[dl]_{\Pi^\ast}   
& &  \Gamma(M,\End(S)) \ar@{.>}[ll]_>>>>>>>>>>>>>>>>>>{\gamma^{-1}}   \ar[dl]^{\Pi^\ast}\\
\Omega^\perp_\K (\hat{M})
& & \Gamma(\hat{M},\End(\hat{S}))  \ar[ll]^{\gamma_\ast^{-1}}
}}
\een
The relation between $\check{E}_\ast$ and $\check{E}$ is summarized in the smaller commutative 
diagram \eqref{diagram:East}, where the pullback morphisms are non-surjective. 
\ben
\label{diagram:East}
\scalebox{1.1}{
\xymatrix{
\Gamma(\hat{M},\hat{S}\otimes\hat{S}) \ar[rr]^{~~~~~~~\check{E}_\ast}  \ar[dr]^{{\hat E}} & & \Omega^\perp_\K(\hat{M})  \ar[dl]_{\gamma_\ast}  \\
& \Gamma(\hat{M},\End(\hat{S})) \\
\Gamma(M,S\otimes S)  \ar[uu]^{\Pi^\ast} \ar@<1ex>^>>>>>>>>>>>>>>>>>{\check{E}}@{.>}[rr] \ar[dr]^{E} & & \Omega_\K(M)  \ar[uu]^{\Pi^\ast} \ar[dl]_{\gamma} \\
& \Gamma(M,\End(S)) \ar@<1ex>^>>>>>>{\Pi^\ast}[uu]
}}
\een
In particular, we note the relations: 
\ben
\label{hatEPiast}
{\hat E}\circ \Pi^\ast=\Pi^\ast \circ {E}~~
\een
and:
\ben
\label{checkEastPiast}
\check{E}_\ast\circ \Pi^\ast=\Pi^\ast \circ \check{E}~~,
\een
which will be used later on.  

\paragraph{The Fierz isomorphisms $\check{E}^\cyl$ and $\check{E}^\cone$.}
By definition, postcomposing ${\hat E}:{\hat S}\otimes {\hat
S}\rightarrow \End({\hat S})$ with the partial inverses
$\gamma_\ccyl^{-1}:\End({\hat S})\rightarrow \Omega_\K^\epsilon
({\hat M})$ defines the Fierz isomorphisms of the cylinder and cone:
\ben
\label{checkE_cylcone}
\check{E}^\cyl\eqdef \gamma_\cyl^{-1}\circ {\hat E}:{\hat S}\otimes {\hat
S}\stackrel{\sim}{\longrightarrow} \wedge^{\epsilon,\cyl} T_\K^{\ast} {\hat M}~~,~~\check{E}^\cone
\eqdef \gamma_\cone^{-1}\circ {\hat E}: {\hat
S}\otimes {\hat
S}\stackrel{\sim}{\longrightarrow}\wedge^{\epsilon,\cone} T_\K^{\ast} {\hat
M}~~.
\een
Relations \eqref{gammainv_cylcone} imply:
\ben
\label{checkEcheckEast}
\check{E}^\cyl=P^\cyl_\epsilon\circ
\check{E}_\ast~~,~~\check{E}^\cone=P^\cone_\epsilon\circ r^\cE\circ
\check{E}_\ast=r^\cE\circ P^\cyl_\epsilon\circ \check{E}_\ast ~~.
\een
as well as:
\be
\check{E}^\cone=r^\cE\circ \check{E}^\cyl~~.
\ee
Recall that our assumptions imply that $\Cl_\K(p,q)$ is simple and that its Schur algebra equals the base field.
Therefore, the bundle morphism $\gamma$ is a bundle isomorphism and the map which it induces on sections is bijective. As a consequence, 
the $\cC^\infty({\hat M},\K)$-linear map $\Omega^\perp_\K({\hat M}) \approx \Gamma({\hat M},\wedge (T^\perp_\K {\hat M})^\ast)\rightarrow \Gamma({\hat M},\End({\hat S}))$ induced
on sections (which, as usual, we have again denoted by $\gamma_\ast$) is bijective. 

\paragraph{The pullback of pinors.}

The pullback of sections induces an injective but non-surjective $\cinf$-linear map: 
\ben
\label{pullback}
\Gamma(M,S)\stackrel{\Pi^\ast}{\longrightarrow} \Gamma({\hat M},{\hat S})~~,
\een
where, as usual, we identify $\cpinf\approx \cinf$. To characterize the image of this map, consider the following $\cpinf$-submodule of the $\cC^\infty({\hat M},\K)$-module 
$\Gamma({\hat M},{\hat S})$, which we shall call the {\em $\cpinf$-module of vertical sections of ${\hat S}$}:
\be
\Gamma^\vt({\hat M},{\hat S})\eqdef \{{\hat \xi}\in \Gamma({\hat M},{\hat S})|\cL^{\hat S}_{\partial_u}\xi=0\}=
\{{\hat \xi}\in \Gamma({\hat M},{\hat S})|\cL^{\hat S}_{\partial_r}\xi=0\}~~.
\ee
Here and below, the symbol $\cL^{\hat S}_V$ denotes the spinorial Lie derivative (a.k.a. the Kosmann-Schwarzbach derivative) \cite{KS} 
of sections of ${\hat S}$ along a vector field $V\in \Gamma({\hat M}, T_\K {\hat M})$. It is then easy to see that the image of \eqref{pullback} coincides 
with the $\cpinf$-module of vertical pinors on ${\hat M}$:
\be
\Pi^\ast(\Gamma(M,S))=\Gamma^\vt({\hat M},{\hat S})~~.
\ee
Using the identification $\cpinf\approx\cinf$, the pullback of sections corestricts to an isomorphism of $\cinf$-modules from $\Gamma(M,S)$ to $\Gamma^\vt({\hat M},{\hat S})$, whose 
image is the appropriate restriction of the map $j^\ast:\Gamma({\hat M},{\hat S})\rightarrow \Gamma(M,S)$ which restricts sections 
to the closed submanifold $\{r=1\Leftrightarrow u=0\}\approx M$ of ${\hat M}$:
\ben
\label{diagram:Gamma}
\scalebox{1.2}{
\xymatrix@1{
\Gamma(M, S) \ar@<0.7ex>[r]^{{\tiny {\Pi^\ast|^{{\tiny \Gamma^\vt({\hat M},{\hat S})}}}}}{~~~} {~~~} &
{~~~~~}\Gamma^\vt({\hat M},{\hat S}) \ar@<0.7ex>[l]^{j^\ast|_{\Gamma^\vt({\hat M},{\hat S})}~~~} ~~.
}}
\een
Similarly, the pullback of sections gives an injective but non-surjective morphism of $\cinf$-algebras (where we identify $\Pi^\ast(\End(S))\approx \End({\hat S})$ since ${\hat S}=\Pi^\ast(S)$):
\be
(\Gamma(M,\End(S)),\circ)\stackrel{\Pi^\ast}{\longrightarrow} (\Gamma({\hat M},\End({\hat S})),\circ)~~,
\ee
whose image coincides with the following $\cpinf$-subalgebra of $(\Gamma({\hat M},\End({\hat S})),\circ)$: 
\be
\Gamma^\vt({\hat M}, \End({\hat S}))\eqdef \{{\hat T}\in \Gamma({\hat M},\End({\hat S}))|[\cL^{\hat S}_{\partial_u},{\hat T}]_{-,\circ}=0\}=
\{{\hat T}\in \Gamma({\hat M},\End({\hat S}))|[\cL^{\hat S}_{\partial_r}, {\hat T}]_{-,\circ}=0\}~~.~~~~
\ee

\paragraph{Pullback properties.}
Let $\xi,\xi'\in \Gamma(M,S)$ and $\xi_\ast=\Pi^\ast(\xi),~
\xi'_\ast=\Pi^\ast(\xi')\in \Gamma({\hat M},{\hat S})$ be their
pullbacks. Relation \eqref{checkEastPiast} implies: 
\ben
\label{cE_pb}
\check{E}^\cyl\circ \Pi^\ast=P^\cyl_\epsilon \circ \Pi^*\circ \check{E}~~,~~\check{E}^\cone\circ \Pi^\ast=P^\cone_\epsilon \circ r^\cE\circ \Pi^*\circ \check{E}~~. 
\een
This gives:
\be
\!\!\!\! (\check{E}_\ast)_{\xi_\ast,\xi'_\ast}=\Pi^\ast(\check{E})(\Pi^\ast(\xi\otimes \xi'))=\Pi^\ast(\check{E}(\xi\otimes \xi'))=
\Pi^*(\check{E}_{\xi,\xi'})\in \Omega^{\perp, \cyl}_\K({\hat M})\Longrightarrow r^\cE ((\check{E}_\ast)_{\xi_\ast,\xi'_\ast})\in \Omega^{\perp, \cone}_\K({\hat M})~,~~~~
\ee
where we noticed that $\xi_\ast\otimes \xi'_\ast=\Pi^\ast(\xi\otimes\xi')$. It follows that: 
\be
\check{E}^\cyl_{\xi_\ast,\xi'_\ast}= P^\cyl_\epsilon (\Pi^*(\check{E}_{\xi,\xi'}))\in \Omega^{\epsilon,\cyl}_\K({\hat M})~~,~~
\check{E}^\cone_{\xi_\ast,\xi'_\ast}=(P^\cone_\epsilon \circ r^{\cE}) (\Pi^*(\check{E}_{\xi,\xi'}))\in \Omega^{\epsilon,\cone}_\K({\hat M})~~.
\ee
The situation is summarized in the diagram below. 
\ben
\label{diagram:cEcylccone_pullbacks}
\scalebox{1.2}{
\xymatrix{
\Omega_\K (M) \ar[dd]_{\Pi^\ast}~  \ar[rr]^{\id_{\Omega_\K(M)}}
&& ~\Omega_\K(M) ~ \ar[dd]^{\Pi^\ast} \\
 & ~\Gamma(M,S\otimes S) \ar[ul]^{\check{E}} \ar[ur]_{\check{E}} \ar@<0ex>_>>>>>>{\Pi^\ast}[dd] \\
 \Omega^\perp_\K(\hat{M}) \ar@{.>}[rr]^{~~~~~~~~~~~~~r^\cE} \ar[dd]_{P^\cyl_\epsilon} &&  \Omega^\perp_\K(\hat{M}) \ar[dd]^{P^\cone_\epsilon} \\
 & ~\Gamma(\hat{M},\hat{S}\otimes \hat{S}) \ar[ul]^{\check{E}_\ast} \ar[ur]_{r^\cE\circ\check{E}_\ast} \ar[dl]_{\check{E}_\cyl} \ar[dr]^{\check{E}_\cone} \\
 \Omega^\epsilon_{\K,\cyl}(\hat{M}) \ar[rr]^{r^\cE}  &&  \Omega^\epsilon_{\K,\cone}(\hat{M})
}}
\een
Notice that $\check{E}^\ccyl_{\xi_\ast,\xi'_\ast}$ lie in the corresponding $\cpinf$-subalgebras of {\em special} twisted selfdual/anti-selfdual forms. As explained in the previous
subsections, a computationally useful model for the later is provided by the $\cpinf$-algebras $(\Omega^<_\K({\hat M}),\bdiamond_\epsilon^\ccyl)$, which can therefore be used to
implement the formalism of \cite{ga1} in a symbolic computation system. 

\subsection{The connections on the pin bundle induced by the Levi-Civita connections of the cylinder and cone}
\label{sec:conespinconnections}

The connection $\nabla^{{\hat S},\cyl}$ induced by $\nabla^\cyl$ on ${\hat S}$ coincides with the pullback through $\Pi$ of the connection $\nabla^S$ induced by 
$\nabla$ on $S$: 
\beqan
\label{cylspin}
\nabla^{{\hat S},\cyl}=(\nabla^S)^\ast~~,
\eeqan
while the connection $\nabla^{{\hat S},\cone}$ induced by  $\nabla^\cone$ on ${\hat S}$ is given by: 
\ben
\label{conesconn}
  \nabla^{{\hat S},\cone}_V=\nabla^{{\hat S},\cyl}_V+\frac{1}{2}\gamma_\cyl(V^\perp_{\hash_\cyl}\wedge \psi)=
\nabla^{{\hat S},\cyl}_V+\frac{1}{2r}\gamma_\cone(V^\perp_{\hash_\cone}\wedge \theta)~,~\forall V\in
\Gamma({\hat M}, T^\ast_\K M)~,~~~~~
\een
where we used the relation $\gamma_\cone=\gamma_\cyl\circ r^{-\cE}$. 

\paragraph{Remark.} The Clifford connection
property of $\nabla^{{\hat S},\cyl}$: 
\be
[\nabla^{{\hat S},\cyl}_V, \gamma_\cyl(\omega)]_{-,\circ}=\gamma_\cyl(\nabla^\cyl_V\omega)~~,~~\forall \omega\in \Omega_\K({\hat M})~~,~~\forall V\in \Gamma({\hat M},T_\K {\hat M})~~
\ee 
follows from the definition \eqref{gammacylconedef} of $\gamma_\cyl$ upon using relation \eqref{nablacyl_varphi} as well as the identity:
\be
[\nabla^{{\hat S},\cyl}_V, \gamma_\ast(\eta)]=[(\nabla^S)^\ast_V, \gamma_\ast(\eta)]=\gamma_\ast(\nabla^\ast_V \eta)~~,
~~\forall \eta\in \Omega_\K^\perp({\hat M})~~,~~\forall V\in \Gamma({\hat M},T_\K {\hat M})~~, 
\ee
which is a direct consequence of the Clifford connection property of $\nabla^S$: 
\be
[\nabla^S_X, \gamma(\varrho)]_{-,\circ}=\gamma(\nabla_X \varrho)~~,~~\forall \varrho\in \Omega_\K(M)~~,~~\forall X\in \Gamma(M,T_\K M)~~. 
\ee
On the other hand, the Clifford connection property of $\nabla^{{\hat S},\cone}$
follows from that of $\nabla^{{\hat S},\cyl}$ upon using equation
\eqref{mainrelation}. Indeed, for any $\omega\in \Omega_\K({\hat M})$ and any $V\in \Gamma({\hat M}, T_\K {\hat M})$, we compute:
\ben
\label{intermediate}
[\nabla^{{\hat S},\cone}_V,\gamma_\cone(\omega)]_{-,\circ}=[\nabla^{{\hat S},\cyl}_V,\gamma_\cyl(r^{-\cE}\omega)]_{-,\circ}+
\frac{1}{2}\gamma_\cyl( [V^\perp_{\hash_\cyl}\wedge \psi, r^{-\cE} \omega]_{-,\diamond^\cyl})~~,
\een
where we used \eqref{conesconn} and the fact that $\gamma_\cyl$ is a morphism of algebras. The first term in \eqref{intermediate} equals
$\gamma_\cyl(\nabla^\cyl(r^{-\cE}\omega))=\gamma_\cone((r^{\cE}\circ \nabla^\cyl \circ r^{-\cE})(\omega))$ by the Clifford connection property of
$\nabla^{{\hat S},\cyl}$. Using the fact that $r^\cE$ is a morphism of algebras from $(\Omega_\K({\hat M}),\diamond^\cyl)$ to $(\Omega_\K({\hat M}),\diamond^\cone)$ and the 
relation $\gamma_\cone=\gamma_\cyl\circ r^{-\cE}$, the second term of \eqref{intermediate} can be expressed as:
\be
\frac{r^2}{2}\gamma_\cone([V^\perp_{\hash_\cyl}\wedge \psi, \omega]_{-,\diamond^\cone})=
\frac{1}{2r}\gamma_\cone([V^\perp_{\hash_\cone}\wedge \theta, \omega]_{-,\diamond^\cone})~~.
\ee
Using these observations as well as identity
\eqref{mainrelation}, we see that the two terms in the right hand side
of \eqref{intermediate} combine to give:
\be
[\nabla^{{\hat S},\cone}_V,\gamma_\cone(\omega)]_{-,\circ}=\gamma_\cone(\nabla^\cone_V \omega)~~,~~\forall \omega\in \Omega_\K({\hat M})~~,~~\forall V\in \Gamma({\hat M},T_\K {\hat M})~~,
\ee
which is the Clifford property of $\nabla^{{\hat S},\cone}$. Hence the Clifford property of the canonical pin connection of
the cone is a consequence of the rather subtle expression
\eqref{mainrelation} for the connection induced on differential forms
by the Levi-Civita connection of the cone.

\paragraph{Local expressions.}
Let $(e_m)_{m=1\ldots d}$ be an oriented local pseudo-orthonormal frame of
$(M,g)$. Then a convenient choice of oriented local pseudo-orthonormal
frames and hence of their dual coframes for the cylinder and cone is
given by:
\beqan
\label{cylconeframe}
e^\cyl_m\eqdef (e_m)_\ast~~,~~e^\cyl_{d+1}\eqdef \partial_u=r\partial_r~~&\Longleftrightarrow &~~e^m_\cyl=\Pi^\ast(e^m)~~,~~e^{d+1}_\cyl=\psi~~,\\ 
{e}^\cone_m\eqdef \frac{1}{ r}(e_m)_\ast ~~,~~e^\cone_{d+1}\eqdef \partial_r~~&\Longleftrightarrow&~~e^m_\cone= r \Pi^\ast(e^m)~~,~~e^{d+1}_\cone=\theta=r\psi~~,\nn
\eeqan
where $(e^m)$ is the coframe on $M$ dual to $(e_m)$ (thus $e^m(e_n)=\delta^m_n$). 
We have:
\be
e_a^\cone=\frac{1}{r}e_a^\cyl
\Longleftrightarrow e^a_\cone=r e^a_\cyl~~,~~\forall a=1, \ldots, d+1~~.
\ee
Notice that $e^m_\cyl, e^m_\cone$ are the cylinder and cone lifts of the one-forms $e^m\in \Omega^1_\K(M)$ (see equations \eqref{cyl_lift} 
and \eqref{cone_lift}). Let us define: 
\be
\gamma^m\eqdef \gamma(e^m)\in \Gamma(M,\End(S))~~,~~\gamma^{(d+1)}\eqdef \gamma(\nu)\in \Gamma(M,\End(S))~~.
\ee
Since the local pseudo-orthonormal frame $(e_m)$ of $(M,g)$ is oriented, we have: 
\be
\gamma^{(d+1)}=\gamma^1\circ \ldots \circ  \gamma^d~~.
\ee
We also define:
\be
{\hat \gamma}^a\eqdef \gamma_\cone(e^a_\cone)=\gamma_\cyl(e^a_\cyl)\in \Gamma({\hat M},\End({\hat S}))~~,
\ee
where we used relation \eqref{gammacylconedef} and the fact that $r^{-\cE}(e^a_\cone)=e^a_\cyl$. For simplicity of notation, we denote 
$\Pi^\ast(\gamma^m)=\gamma_\ast(e_m^\cyl)$ by $\gamma_\ast^m$ and $\Pi^\ast(\gamma^{(d+1)})=\gamma_\ast(\nu_\top^\cyl)$
by $\gamma_\ast^{(d+1)}$. Identities \eqref{gamma_lifts} and \eqref{gammagen_cylcone} give: 
\be
{\hat \gamma}^m=\Pi^\ast(\gamma^m)=\gamma_\ast^m~~,~~{\hat \gamma}^{d+1}=\epsilon~\Pi^\ast(\gamma^{(d+1)})=\epsilon~\gamma_\ast^{(d+1)}~~,
\ee
where we used the fact that $\varphi_\epsilon^\cyl(e^m_\cyl)=e^m_\cyl$ since $\psi\perp e^m_\cyl$. Relation \eqref{conesconn} gives: 
\ben
\label{conespin}
\nabla^{{\hat S},\cone}=\nabla^{{\hat S},\cyl}+
\frac{1}{2}e^m_\cyl \otimes \gamma_\cyl((e^\cyl_m)_{\hash_\cyl}\wedge \psi)=\nabla^{{\hat S},\cyl}+
\frac{1}{2r}e^m_\cone \otimes \gamma_\cone((e^\cone_m)_{\hash_\cone}\wedge \theta)~~.
\een
Combining with \eqref{cylspin}, we find: 
\ben
\label{spinconncylcone}
\nabla^{{\hat S},\cone}_{\partial_u} = \nabla^{{\hat S},\cyl}_{\partial_u}=\cL^{{\hat S}}_{\partial_u}~~,
~~\nabla^{{\hat S},\cone}_{e^\cyl_m} = \nabla^{{\hat S},\cyl}_{e_m^\cyl}+\frac{1}{2}\epsilon \gamma_{\ast,m} \gamma^{(d+1)}_\ast~~,
\een
where $\cL^{{\hat S}}_{\partial_u}$ is the Kosmann-Schwarzbach derivative with respect to $\partial_u$ on ${\hat S}$ and: 
\be
\gamma_{\ast, m}\eqdef \eta_{mn}\gamma^n_\ast~~.
\ee 
Direct computation shows that the connection one-forms $\w_{m n}\eqdef g (e_m, \nabla e_n)$ of $\nabla$ in the
frame $(e_m)$ of $M$ are related as follows to the connection
one-forms $\w^{\cyl}_{a b}\eqdef g_{\cyl}(e^{\cyl}_a,\nabla^\cyl e^\cyl_b)$ of $\nabla^{\cyl}$ and
$\w^{\cone}_{a b}\eqdef g_{\cone}(e^{\cone}_a,\nabla^\cone
e^{\cone}_b)$ of $\nabla^\cone$ in the frames $(e^{\cyl}_a)$ and $(e^\cone_a)$ of ${\hat M}$, respectively:
\beqan
\label{Omegacylcone}
&&\w^\cyl_{m n}(e^\cyl_{d+1})=0~~,~~\w^\cyl_{m n}(e_p)=\w_{m
n}(e_p)~~,~~\w^\cyl_{d+1 m}(e_n)=0~~,~~\w^\cyl_{d+1 m}(e^\cyl_{d+1})=0~~\\ &&\w^\cone_{m n}(e^\cone_{d+1})=0~~,~~\w^\cone_{m n}(e_p)=\w_{m
n}(e_p)~~,~~\w^\cone_{d+1 m}(e_n)=-\eta_{m n}~~, ~~\w^\cone_{d+1
m}(e^\cone_{d+1})=0~~.\nn
\eeqan
Using \eqref{Omegacylcone}, it is easy to check that equations \eqref{spinconncylcone} agree with the formula $\nabla^{{\hat S},\ccyl}=\dd^{\hat S}+\frac{1}{4}\w^{\ccyl}_{a
b}{\hat \gamma}^{a b}$, where $\dd^{\hat S}\eqdef e^a_\cyl\otimes \cL^{\hat S}_{e_a^\cyl}:\Gamma({\hat M},{\hat S})\rightarrow \Omega^1_\K({\hat M},{\hat S})$ is the 
Kosmann-Schwarzbach differential of ${\hat S}$.  

\subsection{The lift of a general pin connection. Cone and cylinder dequantizations of the lift}
\label{sec:conedeq}

Consider an arbitrary connection $D=\nabla^S+A$ on $S$, where $A\in \Omega^1_\K(M,\End(S))$. 
We define the {\em lift} ${\hat D}$ of $D$ to be the connection on ${\hat S}$ obtained from $D$ by pullback through $\Pi$:
\ben
\label{Dlift}
{\hat D}\eqdef D^\ast~~.
\een
Then ${\hat D}$ can be expressed as:
\be
\label{Dcyl}
{\hat D}=\nabla^{{\hat S},\cyl}+A^\cyl~~,
\ee
where:
\be
A^\cyl\eqdef \Pi^\ast(A)\in \Omega^1_\K({\hat M},\End({\hat S}))
\ee 
and we used the fact that $(\nabla^S)^\ast=\nabla^{{\hat S},\cyl}$. Relation \eqref{conespin} implies that ${\hat D}$ can also be written in the form:
\ben
\label{Dcone}
{\hat D}=\nabla^{{\hat S},\cone}+A^\cone~~,
\een
where: 
\be
\label{Acone}
A^\cone \eqdef A^\cyl -\frac{1}{2}e^m_\cyl \otimes \gamma_\cyl( (e^\cyl_m)_{\hash_\cyl}\wedge \psi)= 
A^\cyl - \frac{1}{2r} e^m_\cone \otimes \gamma_\cone((e^\cone_m)_{\hash_\cone}\wedge \theta) \in \Omega^1_\K({\hat M},\End({\hat S}))~~.
\ee
The last relation amounts to:
\be
\!\!\!  A^\cone  (V)=A^\cyl (V)-\frac{1}{2} \gamma_\cyl((V^\perp)_{\hash_\cyl}\wedge \psi)=
A^\cyl (V)-\frac{1}{2r} \gamma_\cone((V^\perp)_{\hash_\cone}\wedge \theta)~~,
~~\forall V\in \Gamma({\hat M},T_\K {\hat M})~.~~~~~
\ee
Let us define:
\be
\nabla^S_m\eqdef \nabla^S_{e_m}~~,~~D_m\eqdef D_{e_m}~~
\ee
(which are derivations of the $\cinf$-module $\Gamma(M,S)$) and:
\be
A_m\eqdef A(e_m)\in \Gamma(M,\End(S))~~.
\ee
Then: 
\be
\nabla^S=\sum_{m=1}^d e^m\otimes \nabla^S_m~~,~~D=\sum_{m=1}^d e^m\otimes D_m~~,~~A=\sum_{m=1}^d e^m\otimes A_m~~{\rm and}~~D_m=\nabla_m^S+A_m~~.
\ee
Similarly, we define: 
\be
{\hat D}_a \eqdef {\hat D}_{e_a^\cyl}~~,~~\nabla^{{\hat S},\cyl}_a \eqdef \nabla^{\hat S,\cyl}_{e^\cyl_a}~~
,~~\nabla^{{\hat S},\cone}_a \eqdef \nabla^{\hat S,\cone}_{e^\cyl_a}~~,
\ee
(which are derivations of the $\cC^\infty({\hat M},\K)$-module $\Gamma({\hat M},{\hat S})$) and: 
\be
A_a^\cyl \eqdef A^\cyl(e_a^\cyl)\in \Gamma({\hat M},\End({\hat S}))~~,~~A_a^\cone \eqdef A^\cone(e_a^\cyl)\in \Gamma({\hat M},\End({\hat S}))~~.
\ee
Then:
\be
\nabla^{{\hat S},\cyl}=\sum_{a=1}^{d+1}e^a_\cyl\otimes \nabla^{{\hat S},\cyl}_a~~
,~~\nabla^{{\hat S},\cone}=\sum_{a=1}^{d+1}e^a_\cyl\otimes \nabla^{{\hat S},\cone}_a~~,
~~{\hat D}=\sum_{a=1}^{d+1}e^a_\cyl\otimes {\hat D}_a~~
\ee
and:
\be
A^\cyl=\sum_{m=1}^{d+1} e^a_\cyl\otimes A_a^\cyl~~,~~A^\cone=\sum_{m=1}^{d+1} e^a_\cyl\otimes A_a^\cone~~,
\ee  
where:
\ben
\label{Acyl}
A_{d+1}^\cyl=0~~,~~A_m^\cyl=\Pi^\ast(A_m)~~
\een
and:
\ben
\label{Aacone}
A_{d+1}^\cone=0~~,~~A_m^\cone= A^\cyl_m -\frac{1}{2}\gamma_\cyl( (e^\cyl_m)_{\hash_\cyl}\wedge \psi)= 
A^\cyl_m - \frac{1}{2r} \gamma_\cone((e^\cone_m)_{\hash_\cone}\wedge \theta)=A_m^\cyl-\frac{\epsilon}{2} \gamma_{\ast,m}\gamma^{(d+1)}_\ast
\een
Notice that:
\ben
\label{Kosmann}
D^\cyl_{\partial_u}=\nabla^{{\hat S},\cyl}_{\partial_u}=\cL^{\hat S}_{\partial_u}~~,~~D^\cone_{\partial_r}=\nabla^{{\hat S},\cone}_{\partial_r}=\cL^{\hat S}_{\partial_r}~~.
\een

\paragraph{Cone and cylinder dequantizations of the lift of a general pin connection.}
As before, consider the lift ${\hat D}=D^\ast$ of a general linear connection $D=e^a \otimes D_a$ on $S$. Recall from \cite{ga1} 
that the adjoint dequantization of $D$ is given by: 
\be
\D_m\omega\eqdef\nabla_m\omega+[\check{A}_m,\omega]_{-,\diamond}~~,~~\forall \omega\in \Omega_\K(M)~~,
\ee
where: 
\be
\check{A}_m\eqdef \gamma^{-1}(A_m)\in \Omega_\K(M)~~.
\ee
As in \cite{ga1}, we define the {\em adjoint cylinder and cone dequantizations} of the lift ${\hat D}$ of $D$ through:
\be
(\D)^\cyl\eqdef \sum_{a=1}^{d+1}{e^a_\cyl\otimes (\D_a)^\cyl}~~,~~(\D)^\cone\eqdef \sum_{a=1}^{d+1}{e^a_\cyl\otimes (\D_a)^\cone}~~,
\ee
where: 
\ben
\label{checkDacylcone}
(\D_a)^\cyl\omega\eqdef\nabla^\cyl_a\omega+[\check{A}^\cyl_a, \omega]_{-,\diamond^\cyl}~~,~
(\D_a)^\cone\omega\eqdef\nabla^\cone_a\omega+[\check{A}^\cone_a, \omega]_{-,\diamond^\cone}~~,~~\forall \omega\in \Omega_\K({\hat M})~~,
\een
and:
\ben
\check{A}^\cyl_a\eqdef \gamma_\cyl^{-1}(A_a^\cyl)\in \Omega^{\epsilon,\cyl}_\K({\hat M})~~,~~\check{A}^\cone_a\eqdef \gamma_\cone^{-1}(A_a^\cone)\in \Omega^{\epsilon,\cone}_\K({\hat M})~~.
\een
Relations \eqref{gammainv_cylcone} and \eqref{Acyl}, \eqref{Aacone} give:
\ben
\label{Amdec_cylcone}
\check{A}_m^\cyl=P_\epsilon^\cyl (\check{A}_{m,\cyl})~~,~~\check{A}_m^\cone=P_\epsilon^\cone \left(\check{A}_{m,\cone}-\frac{1}{2}(e^m_\cone)_{\hash_\cone}\wedge \theta\right)~~,
~~\check{A}_{d+1}^\cyl=\check{A}_{d+1}^\cone=0~~,
\een
where: 
\be
\check{A}_{m,\cyl}\eqdef \Pi^\ast(\check{A}_m)~~{\rm and}~~\check{A}_{m,\cone}\eqdef r^\cE\Pi^\ast(\check{A}_m)~~
\ee
are the cylinder and cone lifts of the inhomogeneous differential forms $\check{A}_m$ (see equations \eqref{cyl_lift} and \eqref{cone_lift}) and we used the identities 
$\gamma_\cyl^{-1}\circ \gamma_\cyl=P_\epsilon^\cyl$ and $\gamma_\cone^{-1}\circ \gamma_\cone=P_\epsilon^\cone$. In particular, we have:
\ben
\label{checkAcylcone}
\check{A}^\cone_m=r^\cE(\check{A}^\cyl_m)-\frac{1}{2}P_\epsilon^\cone[(e^m_\cone)_{\hash_\cone}\wedge \theta]~~,
\een
where we used relation \eqref{Hodger}. Equations \eqref{checkAcylcone}, \eqref{checkDacylcone} and \eqref{mainrelation} imply that the two 
dequantized connections on ${\hat M}$ are related through:
\ben
\label{Dacc}
(\D_a)^\cone=r^\cE\circ (\D_a)^\cyl\circ r^{-\cE}~~.
\een
The pullback of $\D$ gives an operator $(\D)^\ast:\Omega_\K^{\perp,\cyl}({\hat M})\rightarrow \Omega^1_\K({\hat M})\otimes_{\cC^\infty({\hat M},\K)} \Omega_\K^{\perp,\cyl}({\hat M})$ which expands 
as $(\D)^\ast=e_a^\cyl\otimes (\D_a)^\ast$, where $(\D_{d+1})^\ast=0$ while $(\D_m)^\ast$ are derivations of the algebra $(\Omega_\K^{\perp,\cyl}({\hat M}),\diamond^\cyl)$ 
which act as:
\be
(\D_m)^\ast\omega=\nabla^\ast_m\omega+[\Pi^\ast(\check{A}_m),\omega]_{-,\diamond^\cyl}=\nabla^\cyl_m\omega+[\check{A}_{m,\cyl},\omega]_{-,\diamond^\cyl}~~,~~\forall \omega\in 
\Omega_\K^{\perp,\cyl}({\hat M})\approx \Pi^\ast(\Omega_\K(M))~~
\ee
and satisfy:
\ben
\label{DastPiast}
(\D_m)^\ast\circ \Pi^\ast=\Pi^\ast \circ \D_m~~.
\een
Since $\nabla^\cyl$ commutes with $P_\pm^\cyl$, relations \eqref{Amdec_cylcone} imply: 
\ben
\label{DdecPiast}
(\D_a)^\cyl\circ P_\epsilon^\cyl=P_\epsilon^\cyl\circ (\D_a)^\ast~~,~~(\D_a)^\cone\circ P_\epsilon^\cone\circ r^\cE=P_\epsilon^\cone\circ r^\cE \circ (\D_a)^\ast~~.
\een 
Together with \eqref{DastPiast}, this gives the identities: 
\ben
\label{DcylconePiast}
(\D_a)^\cyl\circ P_\epsilon^\cyl\circ \Pi^\ast=P_\epsilon^\cyl\circ \Pi^\ast \circ \D_a~~,~~(\D_a)^\cone\circ P_\epsilon^\cone\circ r^\cE\circ \Pi^\ast=
P_\epsilon^\cone\circ r^\cE \circ \Pi^\ast \circ \D_a~~,
\een
which will be used later on. 

\paragraph{Remark.} The Clifford connection properties of $\nabla^{{\hat S},\cyl}$ and $\nabla^{{\hat S},\cone}$:
\be
\gamma_\cyl \circ \nabla_a^\cyl=(\nabla^{{\hat S},\cyl}_a)^\ad\circ \gamma_\cyl~~,
~~\gamma_\cone\circ \nabla_a^\cone=(\nabla^{{\hat S},\cone}_a)^\ad\circ \gamma_\cone
\ee
imply:
\be
\gamma_\cyl\circ (\D_a)^\cyl={\hat D}_a^\ad\circ \gamma_\cyl~~,
~~\gamma_\cone\circ (\D_a)^\cone={\hat D}_a^\ad\circ \gamma_\cone~~,
\ee
i.e.:
\be
P^\cyl_\epsilon \circ (\D_a)^\cyl=\gamma_\cyl^{-1}\circ {\hat D}_a^\ad\circ \gamma_\cyl~~,
~~P^\cone_\epsilon\circ (\D_a)^\cone=\gamma_\cone^{-1}\circ {\hat D}_a^\ad\circ \gamma_\cone~~.
\ee

\subsection{Lifting algebraic constraints on pinors. Dequantizations of lifted algebraic constraints}
\label{sec:coneliftingalg}

\paragraph{The lift of algebraic constraints.}
Given an endomorphism $Q\in \Gamma(M,\End(S))$, we define its lift ${\hat Q}$ to ${\hat M}$ to be the pullback of $Q$ to a globally-defined endomorphism of
the pin bundle ${\hat S}$ of ${\hat M}$:
\ben
\label{hatQ}
{\hat Q}\eqdef \Pi^\ast(Q)\in\Gamma({\hat M}, \End({\hat S}))~~.
\een
Since ${\hat Q}\circ \Pi^\ast=\Pi^\ast \circ Q$, the pullback map \eqref{pullback} induces an injective but non-surjective $\cinf$-linear map from the 
space $\cK(Q)=\{\xi\in \Gamma(M,S)|Q\xi=0\}$ to the space  $\cK({\hat Q})=\{{\hat \xi}\in \Gamma({\hat M},{\hat S})|{\hat Q}{\hat \xi}=0\}$:
\be
\cK(Q)\stackrel{\Pi^\ast|_{\cK(Q)}}{\longrightarrow}\cK({{\hat Q}})~~,
\ee
where, as usual, we identify $\cpinf\approx \cinf$. The image of this map is the following $\cpinf$-submodule of $\Gamma({\hat M},{\hat S})$: 
\be
\Pi^\ast(\cK(Q))=\cK({\hat Q})\cap \Gamma^\vt({\hat M},{\hat S})\eqdef \cK^\vt({\hat Q})~~,
\ee 
which we shall call the {\em $\cpinf$-module of vertical ${\hat Q}$-constrained pinors on ${\hat M}$}. Identifying $\cpinf\approx \cinf$, the appropriate restrictions 
of $\Pi^\ast$ and $j^\ast$ give mutually-inverse isomorphisms of $\cinf$-modules:  
\ben
\label{diagram_KQ}
\scalebox{1.2}{
\xymatrix@1{
\cK(Q) \ar@<0.7ex>[r]^{{\tiny {\Pi^\ast|^{{\tiny \cK^\vt({\hat Q})}}_{{\tiny \cK(Q)}}}}~~~~~} {~~~~~} &
{~~~~~} \cK^\vt({\hat Q}) \ar@<0.7ex>[l]^{j^\ast|_{\cK^\vt({\hat Q})}^{\cK(Q)}~~~~~} ~~.
}}
\een
Hence a pinor ${\hat \xi}\in \Gamma({\hat M},{\hat S})$ on ${\hat M}$
satisfies ${\hat Q}{\hat \xi}=\cL^{\hat S}_{\partial_r}{\hat \xi}=0$
iff. it is the pullback ${\hat \xi}=\Pi^\ast(\xi)$ of a pinor $\xi\in
\Gamma(M,S)$ which satisfies $Q\xi=0$. This allows one to translate
between algebraic constraints on pinors defined on $M$ and on pinors
defined on ${\hat M}$.

\paragraph{Cone and cylinder dequantizations of the lift of an algebraic constraint.}
As in \cite{ga1}, consider the dequantization
\be
\check{Q}\eqdef \gamma^{-1}(Q)\in \Omega_\K(M)
\ee
of $Q\in \End(S)$ as well as the {\em cone and cylinder dequantizations} of the lift ${\hat Q}$ of $Q$:
\beqan
\label{Qdec_cylcone}
\check{Q}^\cyl ~~\eqdef \gamma_\cyl^{-1}({\hat Q})~&=& (P_\epsilon^\cyl \circ \Pi^\ast) (\check{Q})=P_\epsilon^\cyl (\check{Q}_\cyl)\in \Omega_\K^{\epsilon,\cyl}({\hat M})~~,\\
\check{Q}^\cone\eqdef \gamma_\cone^{-1}({\hat Q}) &=& (P_\epsilon^\cone \circ r^\cE \circ \Pi^\ast) (\check{Q})=P_\epsilon^\cone(\check{Q}_\cone)=r^\cE(\check{Q}^\cyl)\in \Omega_\K^{\epsilon,\cone}({\hat M})~~,~~\nn
\eeqan
where we used relations \eqref{gammainv_cylcone} and where
\be
\check{Q}_\cyl\eqdef \Pi^\ast(\check{Q})\in \Omega_\K^\cyl({\hat M})~~,~~\check{Q}_\cone=r^\cE(\Pi^\ast(\check{Q}))=r^\cE(\check{Q}_\cyl)\in \Omega_\K^\cone({\hat M})
\ee 
are the cylinder and cone lifts of the inhomogeneous form $\check{Q}\in \Omega_\K(M)$ (see relations \eqref{cyl_lift} and \eqref{cone_lift}). 

\paragraph{Lifting the algebra of constrained differential forms.}
As in \cite{ga1}, consider the $\cinf$-algebra of constrained inhomogeneous forms on $M$:
\be
\check{\cK}_Q\eqdef \check{\cK}(Q)=\check{E}(\cK({\hat Q})\otimes_{\cinf}\cK(Q))=
\cK(L_{\check{Q}})\cap \cK(R_{\tau_{\cB}(\check{Q})})
\ee
as well as the $\cC^\infty({\hat M},\K)$-algebras of constrained inhomogeneous forms on the cylinder and cone: 
\beqan
\label{KQdec_cylcone}
\check{\cK}_{{\hat Q},\cyl}~&\eqdef& \check{\cK}^\cyl({\hat Q})~=~~\check{E}^\cyl(\cK({\hat Q})\otimes_{\cC^\infty({\hat M},\K)}\cK({\hat Q}))~~=
\cK(L^\cyl_{\check{Q}^\cyl})\cap \cK(R^\cyl_{\tau_{{\hat \cB}}(\check{Q}^\cyl)})\cap \Omega^\epsilon_{\K,\cyl}({\hat M})~~\nn\\
\check{\cK}_{{\hat Q},\cone}&\eqdef& \check{\cK}^\cone({\hat Q})=~~\check{E}^\cone(\cK({\hat Q})\otimes_{\cC^\infty({\hat M},\K)}\cK({\hat Q}))=
\cK(L^\cone_{\check{Q}^\cone})\cap \cK(R^\cone_{\tau_{{\hat \cB}}(\check{Q}^\cone)})\cap \Omega^\epsilon_{\K,\cone}({\hat M})~~.~~~~~~~~~
\eeqan
We also define the $\cpinf$-algebras of {\em special constrained inhomogeneous forms} on the cylinder and cone through:
\beqan
\label{specialKQdec_cylcone}
\check{\cK}^\cyl_{{\hat Q}}~&\eqdef& \check{\cK}_{{\hat Q},\cyl}\cap \Omega_\K^\cyl({\hat M})~~~=
\cK(L^\cyl_{\check{Q}^\cyl})\cap \cK(R^\cyl_{\tau_{{\hat \cB}}(\check{Q}^\cyl)})\cap \Omega^{\epsilon,\cyl}_\K({\hat M})~~\nn\\
\check{\cK}^\cone_{{\hat Q}}&\eqdef& \check{\cK}_{{\hat Q},\cone}\cap \Omega_\K^\cone({\hat M})=
\cK(L^\cone_{\check{Q}^\cone})\cap \cK(R^\cone_{\tau_{{\hat \cB}}(\check{Q}^\cone)})\cap \Omega^{\epsilon,\cone}_\K({\hat M})~~.~~~~~~~~~
\eeqan
The relation $\check{Q}^\cone=r^\cE(\check{Q}^\cyl)$ and identities \eqref{RLcylcone} imply: 
\be
\cK(L^\cone_{\check{Q}^\cone})=r^\cE(\cK(L^\cyl_{\check{Q}^\cyl}))~~,~~\cK(R^\cone_{\tau_{{\check \cB}}(\check{Q}^\cone)})=r^\cE(\cK(R^\cyl_{\tau_{{\check \cB}}(\check{Q}^\cyl)}))~~.
\ee
Together with $r^\cE(\Omega_\K^{\epsilon,\cyl}({\hat M}))=\Omega_\K^{\epsilon,\cone}({\hat M})$, this gives: 
\be
\check{\cK}^\cone_{{\hat Q}}=r^\cE(\check{\cK}^\cyl_{{\hat Q}})~~.
\ee
Relations \eqref{checkEcheckEast} imply: 
\beqa
\check{\cK}_{{\hat Q},\cyl}=P_\epsilon^\cyl(\check{\cK}_{{\hat Q}})~~&,&~~\check{\cK}_{{\hat Q},\cone}=(P_\epsilon^\cone\circ r^\cE)(\check{\cK}_{{\hat Q}})~~,\\
\check{\cK}^\cyl_{{\hat Q}}~~=~P_\epsilon^\cyl(\check{\cK}^\vt_{{\hat Q}})~~&,&~~\check{\cK}^\cone_{{\hat Q}}~~=~(P_\epsilon^\cone\circ r^\cE)(\check{\cK}^\vt_{{\hat Q}})~~,
\eeqa
where we defined: 
\be
\check{\cK}_{{\hat Q}}\eqdef \check{E}_\ast (\cK({\hat Q})\otimes_{\cC^\infty({\hat M},\K)}\cK({\hat Q}))\subset \Omega^\perp_\K({\hat M})~~,
~~\check{\cK}_{{\hat Q}}^\vt\eqdef \check{E}_\ast (\cK^\vt({\hat Q})\otimes_{\cC^\infty({\hat M},\K)}\cK^\vt({\hat Q}))\subset \Omega^{\perp,\cyl}_\K({\hat M})~~.
\ee
Relation \eqref{checkEastPiast} implies: 
\ben
\label{cKvertPiast}
\check{\cK}_{{\hat Q}}^\vt=\Pi^\ast(\check{\cK}(Q))~~,
\een
which in turn gives:
\be
\check{\cK}^\cyl_{{\hat Q}}=(P_\epsilon^\cyl \circ \Pi^\ast)(\check{\cK}_Q)~~,~~ \check{\cK}^\cone_{{\hat Q}}=(P_\epsilon^\cyl \circ r^\cE \circ \Pi^\ast)(\check{\cK}_Q)~~.
\ee
Hence the isomorphisms of $\cpinf\approx \cinf$-algebras: 
\ben
\label{PepsilonPiast}
P^\cyl_\epsilon\circ \Pi^\ast:(\Omega_\K(M),\diamond) \stackrel{\sim}{\rightarrow} (\Omega_\K^{\epsilon,\cyl}({\hat M}),\diamond^\cyl)~~
,~~
P^\cone_\epsilon\circ r^\cE\circ \Pi^\ast:(\Omega_\K(M),\diamond) \stackrel{\sim}{\rightarrow} (\Omega_\K^{\epsilon,\cone}({\hat M}),\diamond^\cone)~~
\een
restrict to isomorphism of $\cinf$-algebras from $\cK_Q$ to $\check{\cK}^\cyl_{{\hat Q}}$ and $\check{\cK}^\cone_{{\hat Q}}$, respectively, whose inverses are given by 
the appropriate restrictions of $j^\ast\circ P_\perp$. It follows that the $\cpinf$-algebras $(\check{\cK}_{{\hat Q}}^\cyl,\diamond^\cyl)$ and $(\check{\cK}_{{\hat Q}}^\cone,\diamond^\cone)$ 
give models for the $\cinf$-algebra $(\check{\cK}_Q,\diamond)$, allowing to translate between constrained inhomogeneous differential forms defined on $M$ and those defined on 
the cylinder and cone over $M$. 
\ben
\label{diagram:ccKQ}
\scalebox{1.2}{\xymatrix{
(\check{\cK}_Q,\diamond) \ar@<0.5ex>[d]^{} ~ \ar@<0.5ex>[r]^{\Pi^\ast} ~ & ~
(\check{\cK}^\vt_{{\hat Q}},\diamond^\cyl) ~ \ar@<0.5ex>[l]^{j^\ast} ~ \ar@<0.5ex>[d]^{P_\epsilon^\cyl} ~\\
~ (\check{\cK}_{{\hat Q}}^\cone,\diamond^\cone) ~ \ar@<0.5ex>[u]^{}  \ar@<0.5ex>[r]^{~~~~r^{-\cE}~~~}  & ~
(\check{\cK}_{{\hat Q}}^\cyl,\diamond^\cyl) ~ \ar@<0.5ex>[l]^{~~~~r^{\cE}~~~~} ~ \ar@<0.5ex>[u]^{P_\perp} ~
}}
\een

\subsection{Lifting generalized Killing pinors and generalized Killing forms}
\label{sec:coneliftingGK}

\paragraph{The lift of generalized Killing pinors.} 
Relations \eqref{Kosmann} imply that the Killing pinor equations
with respect to ${\hat D}$ for a pinor ${\hat \xi}\in \Gamma({\hat
M},{\hat S})$ amount to the condition $\cL^{\hat S}_{\partial_u}{\hat
\xi}=0\Leftrightarrow \cL_{\partial_r}^{\hat S}{\hat \xi}=0$ (which
says that ${\hat \xi}$ coincides with the pullback
$\xi_\ast=\Pi^\ast(\xi)$ of some pinor $\xi\in \Gamma(M,S)$ defined
on $M$) together with the conditions $(\nabla^S_m)^\ast{\hat
\xi}+(A_m)_\ast{\hat \xi}=0\Leftrightarrow \Pi^\ast(D_m\xi)=0$ (which are equivalent with the
generalized Killing pinor equations $D_m\xi=0$ on $M$). In particular, the space: 
\be
\cK({\hat D})\eqdef \cap_{a=1}^{d+1}\cK({\hat D}_a)=\{{\hat \xi}\in \Gamma({\hat M},{\hat S})|{\hat D}_a\xi=0~~,~~\forall a=1,\ldots, d+1\}
\ee
is a subspace of $\Gamma^\vt({\hat M},{\hat S})$ which coincides with the $\Pi$-pullback of the space: 
\be
\cK(D)\eqdef \cap_{m=1}^{d}\cK(D_m)=\{\xi\in \Gamma(M,S)|D_m\xi=0~~,~~\forall m=1,\ldots, d\}~~.
\ee
The relation: 
\be
\cK({\hat D})=\Pi^\ast(\cK(D))
\ee
shows that $\Pi^\ast$ induces a isomorphism of $\K$-vector spaces between $\cK(D)$ and $\cK({\hat D})$, whose inverse is given by the appropriate restriction of 
$j^\ast$:
\ben
\label{diagram:cKD}
\scalebox{1.2}{
\xymatrix@1{
\cK(D) ~~\ar@<0.7ex>[r]^{~~\Pi^\ast|_{\cK(D)}~~} &~~ \cK({\hat D}) \ar@<0.7ex>[l]^{~~~~~j^\ast|_{\cK({\hat D})}~~~~~}}} 
~~.
\een
This allows us to translate between generalized Killing pinor equations on $M$ and ${\hat M}$. 

\paragraph{Lifting the flat Fierz $\K$-algebra of generalized Killing pinors.} 
As in \cite{ga1}, consider the flat Fierz $\K$-algebra of $D$ on $M$: 
\be
\check{\cK}(D)\eqdef \check{E}(\cK(D)\otimes_{\cinf}\cK(D))\subset \Omega_\K(M)~~,
\ee
which is a $\K$-subalgebra of the \KA algebra $(\Omega_\K(M),\diamond$). Let us define:
\be
\check{\cK}({\hat D})\eqdef \check{E}_\ast(\cK({\hat D})\otimes_{\cpinf}\cK({\hat D}))\subset \Omega^{\perp,\cyl}_\K({\hat M})~~,
\ee
which is a $\K$-subalgebra of the $\cpinf$-algebra $(\Omega^{\perp,\cyl}_\K({\hat M}),\diamond^\cyl)$. Relations \eqref{hatEPiast} imply: 
\be
\label{checkcKPiast}
\check{\cK}({\hat D})=\Pi^\ast(\check{\cK}(D))~~,
\ee
so the appropriate co-restriction of $\Pi^\ast$ gives a unital isomorphism of $\K$-algebras from $(\check{\cK}(D),\diamond)$ to $(\check{\cK}({\hat D}),\diamond^\cyl)$. 
As in \cite{ga1}, consider now the flat Fierz $\K$-algebras determined on the cylinder and cone by 
the ${\hat \cB}$-flat subspace $\cK({\hat D})\subset \Gamma({\hat M},{\hat S})$: 
\beqa
\check{\cK}^\cyl({\hat D})~~&\eqdef&~ \check{E}^\cyl(\cK({\hat D})\otimes_{\cC^\infty({\hat M},\K)}\cK({\hat D}))\subset \Omega_\K^{\epsilon,\cyl}({\hat M})~~,\\
\check{\cK}^\cone({\hat D})&\eqdef& \check{E}^\cone(\cK({\hat D})\otimes_{\cC^\infty({\hat M},\K)}\cK({\hat D}))\subset \Omega_\K^{\epsilon,\cyl}({\hat M})~~.
\eeqa
Relations \eqref{checkEcheckEast} imply: 
\ben
\label{FierzK_cylcone}
\check{\cK}^\cyl({\hat D})=P_\epsilon^\cyl(\check{\cK}({\hat D}))~~,~~\check{\cK}^\cone({\hat D})=
(P_\epsilon^\cone\circ r^\cE)(\check{\cK}({\hat D}))=r^\cE(\check{\cK}^\cyl({\hat D}))~~.
\een
When combined with \eqref{checkcKPiast}, the last equations show that the appropriate restrictions of the morphisms of algebras \eqref{PepsilonPiast} give 
isomorphisms of algebras between $\check{\cK}(D)$ and $\check{\cK}^\cyl({\hat D})$, respectively $\check{\cK}^\cone({\hat D})$, whose inverses are given by 
$j^\ast\circ P_\perp$ and $j^\ast\circ r^{-\cE}\circ P_\perp$, respectively. The situation is summarized in the commutative diagram: 
\ben
\label{diagram:ccKD}
\scalebox{1.2}{\xymatrix{
(\check{\cK}(D),\diamond) \ar@<0.5ex>[d]^{} ~ \ar@<0.5ex>[r]^{\Pi^\ast} ~ & ~
(\check{\cK}({\hat D}),\diamond^\cyl) ~ \ar@<0.5ex>[l]^{j^\ast} ~ \ar@<0.5ex>[d]^{P_\epsilon^\cyl} ~\\
~ (\check{\cK}^\cone({\hat D}),\diamond^\cone) ~ \ar@<0.5ex>[u]^{}  \ar@<0.5ex>[r]^{~~~~r^{-\cE}~~~}  & ~
(\check{\cK}^\cyl({\hat D}),\diamond^\cyl) ~ \ar@<0.5ex>[l]^{~~~~r^{\cE}~~~~} ~ \ar@<0.5ex>[u]^{P_\perp} ~
}}
\een

\paragraph{The lift of generalized Killing forms.}
As in \cite{ga1}, consider the generalized Killing $\K$-algebra determined by $D$ on $M$: 
\be
\check{\cK}_D\eqdef \cK(\D)\subset \Omega_\K(M)~~
\ee
as well as the generalized Killing $\K$-algebras determined by ${\hat D}$ on the cylinder and cone:  
\beqa
\check{\cK}_{{\hat D}}^\cyl~~&\eqdef&~ \cK((\D)^\cyl)\cap \Omega^\epsilon_{\K,\cyl}({\hat M})=\cK((\D)^\cyl)\cap \Omega^{\epsilon,\cyl}_\K({\hat M})~~,\\
~~\check{\cK}_{{\hat D}}^\cone &\eqdef& \cK((\D)^\cone)\cap \Omega^\epsilon_{\K,\cone}({\hat M})=\cK((\D)^\cone)\cap \Omega^{\epsilon,\cyl}_\K({\hat M})~~,
\eeqa
where we noticed that  $\cK((\D)^\cyl)\subset \Omega^\cyl_\K({\hat M})$ and
$\cK((\D)^\cone)\subset \Omega^\cone_\K({\hat M})$.  Also consider the following subalgebra of $(\Omega^{\perp,\cyl}_\K({\hat M}),\diamond^\cyl)$: 
\be
\check{\cK}_{{\hat D}}\eqdef \cK((\D)^\ast)=\cap_{m=1}^{d}\cK((\D_m)^\ast)\subset  \Omega^{\perp,\cyl}_\K({\hat M})~~.
\ee
Relation \eqref{DastPiast} implies: 
\ben
\label{GKPiast}
\check{\cK}_{{\hat D}}=\Pi^\ast(\check{\cK}_D)~~. 
\een
Relations \eqref{checkEcheckEast} give:
\ben
\label{GKcylcone}
\check{\cK}_{{\hat D}}^\cyl=P_\epsilon^\cyl(\check{\cK}_{{\hat D}})~~,~~\check{\cK}_{{\hat D}}^\cone=(P_\epsilon^\cone\circ r^\cE)(\check{\cK}_{\hat D})=
r^\cE(\check{\cK}_{{\hat D}}^\cyl)~~.
\een
Combining this with \eqref{GKPiast}, we find that the morphisms of algebras \eqref{PepsilonPiast} restrict to isomorphisms of $\K$-algebras between 
$\check{\cK}_D$ and $\check{\cK}_{\hat D}^\cyl$, respectively $\check{\cK}_{\hat D}^\cone$. The situation is summarized in the commutative diagram: 
\ben
\label{diagram:ccGK}
\scalebox{1.2}{\xymatrix{
(\check{\cK}_D,\diamond) \ar@<0.5ex>[d]^{} ~ \ar@<0.5ex>[r]^{\Pi^\ast} ~ & ~
(\check{\cK}_{{\hat D}},\diamond^\cyl) ~ \ar@<0.5ex>[l]^{j^\ast} ~ \ar@<0.5ex>[d]^{P_\epsilon^\cyl} ~\\
~ (\check{\cK}^\cone_{{\hat D}},\diamond^\cone) ~ \ar@<0.5ex>[u]^{}  \ar@<0.5ex>[r]^{~~~~r^{-\cE}~~~}  & ~
(\check{\cK}^\cyl_{{\hat D}},\diamond^\cyl) ~ \ar@<0.5ex>[l]^{~~~~r^{\cE}~~~~} ~ \ar@<0.5ex>[u]^{P_\perp} ~
}}~~.
\een

\subsection{Relating CGK pinors and CGK forms on $M$ and ${\hat M}$}
\label{sec:coneCGK}

\paragraph{Lifting the CGK pinor equations from $M$ to ${\hat M}$.}
Consider the CGK pinor equations defined on $M$ by some connection
$D$ on $S$ and by a single endomorphism $Q\in \Gamma(M,\End(S))$ as
well as the CGK pinor equations defined on ${\hat M}$ by the lifts
${\hat D}$ and ${\hat Q}$, as before. Denoting the corresponding $\K$-vector spaces of solutions by $\cK(D,Q)=\cK(D)\cap
\cK(Q)$ and $\cK({{\hat D},{\hat Q}})=\cK({\hat D})\cap \cK({\hat Q})$,
the observations of the previous subsections imply that we have the inclusion: 
\be
\cK({\hat D},{\hat Q})\subset \Gamma^\vt({\hat M},{\hat S})
\ee
and that we have mutually-inverse isomorphisms of $\K$-vector spaces: 
\ben
\label{diagram:cKDQ}
\scalebox{1.2}{
\xymatrix@1{
\cK(D,Q)~~ \ar@<0.7ex>[r]^{~~~\Pi^\ast|_{\cK(D)}~~~} &~~ \cK({\hat D},{\hat Q}) \ar@<0.7ex>[l]^{~~~j^\ast|_{\cK({\hat D},{\hat Q})}~~~}}} 
~~.
\een
This observation allows us to lift the CGK pinor equations from $M$
to ${\hat M}$. 

\paragraph{Lifting CGK forms to the cylinder and cone.}
When using the cylinder or cone metric on ${\hat M}$,
we can of course apply the formalism of \cite{ga1} on
${\hat M}$, thereby working with the dequantized connections
$(\D)^\cyl$ and $(\D)^\cone$ which were discussed above. 
Combining the observations of the previous subsections shows that the $\K$-algebras of 
CGK forms on the cylinder and cone are related to the $\K$-algebra of CGK forms on $(M,g)$ through: 
\be
\check{\cK}^\cyl_{{\hat D},{\hat Q}}=(P_\epsilon^\cyl\circ \Pi^\ast)(\check{\cK}_{D,Q})~~,
~~\check{\cK}^\cone_{{\hat D},{\hat Q}}=(P_\epsilon^\cone\circ r^\cE\circ \Pi^\ast)(\check{\cK}_{D,Q})=r^\cE(\check{\cK}^\cyl_{{\hat D},{\hat Q}})
\ee
while the flat Fierz $\K$-algebras determined on the cylinder and cone by the ${\hat \cB}$-flat subspace $\cK({\hat D},{\hat Q})\subset \Gamma({\hat M},{\hat S})$ are related to 
the flat Fierz $\K$-algebra determined on $M$ by the $\cB$-flat subspace $\cK(D,Q)\subset \Gamma(M,S)$ through:
\be
\check{\cK}^\cyl({\hat D},{\hat Q})=(P_\epsilon^\cyl\circ \Pi^\ast)(\check{\cK}(D,Q))~~,
~~\check{\cK}^\cone({\hat D},{\hat Q})=(P_\epsilon^\cone\circ r^\cE\circ \Pi^\ast)(\check{\cK}(D,Q))=r^\cE(\check{\cK}^\cyl({\hat D},{\hat Q}))~~.
\ee
Since $P_\epsilon^\cyl\circ \Pi^\ast$ and $P_\epsilon^\cone\circ r^\cE\circ \Pi^\ast$ are isomorphism of $\K$-algebras, 
the relations above show that $(\check{\cK}^\cyl_{{\hat D},{\hat Q}},\diamond^\cyl)$, $(\check{\cK}^\cone_{{\hat D},{\hat Q}},\diamond^\cone)$ and $(\check{\cK}^\cyl({\hat D}, {\hat Q}),\diamond^\cyl)$, 
$(\check{\cK}^\cone({\hat D}, {\hat Q}),\diamond^\cone)$ provide isomorphic models of $\check{\cK}_D$ and $\check{\cK}(D)$, respectively. The situation is summarized in the commutative diagrams: 
\!\!\!\!\!\!\!\!\!\!\!\!\!\!\!\!\!\!\!\ben
\label{diagram:cCGK}
\scalebox{1.1}{\xymatrix{
(\check{\cK}_{D,Q},\diamond) \ar@<0.5ex>[d]^{} ~ \ar@<0.5ex>[r]^{\Pi^\ast} ~ & ~
(\check{\cK}_{{\hat D},{\hat Q}},\diamond^\cyl) ~ \ar@<0.5ex>[l]^{j^\ast} ~ \ar@<0.5ex>[d]^{P_\epsilon^\cyl} ~\\
~ (\check{\cK}^\cone_{{\hat D},{\hat Q}},\diamond^\cone) ~ \ar@<0.5ex>[u]^{}  \ar@<0.5ex>[r]^{~~~~r^{-\cE}~~~}  & ~
(\check{\cK}^\cyl_{{\hat D},{\hat Q}},\diamond^\cyl) ~ \ar@<0.5ex>[l]^{~~~~r^{\cE}~~~~} ~ \ar@<0.5ex>[u]^{P_\perp} ~
}}~~~~~~~~~\scalebox{1.1}{\xymatrix{
(\check{\cK}(D,Q),\diamond) \ar@<0.5ex>[d]^{} ~ \ar@<0.5ex>[r]^{\Pi^\ast} ~ & ~
(\check{\cK}({\hat D},{\hat Q}),\diamond^\cyl) ~ \ar@<0.5ex>[l]^{j^\ast} ~ \ar@<0.5ex>[d]^{P_\epsilon^\cyl} ~\\
~ (\check{\cK}^\cone({\hat D},{\hat Q}),\diamond^\cone) ~ \ar@<0.5ex>[u]^{}  \ar@<0.5ex>[r]^{~~~~r^{-\cE}~~~}  & ~
(\check{\cK}^\cyl({\hat D},{\hat Q}),\diamond^\cyl) ~ \ar@<0.5ex>[l]^{~~~~r^{\cE}~~~~} ~ \ar@<0.5ex>[u]^{P_\perp} ~
}}~~~~~~
\een

\subsection{Lifting truncated models}
\label{sec:conetruncated}

Other isomorphic models --- which are particularly convenient for computer
computations --- are obtained upon applying the isomorphisms of algebras given
in Section 3 of \cite{ga1}.  To describe this, we define: 
\beqa
\check{Q}^{<,\cyl}\eqdef P_<(\check{Q}^\cyl)\in \Omega^{<,\cyl}_\K({\hat M})~~&,
&~~\check{Q}^{<,\cone}\eqdef P_<(\check{Q}^\cone)\in \Omega_\K^{<,\cone}({\hat M})~~,\\
~~\check{A}_a^{<,\cyl}\eqdef P_<(\check{A}_a^\cyl)\in \Omega^{<,\cyl}_\K({\hat M})~~&,
&~~\check{A}_a^{<,\cone}\eqdef P_<(\check{A}_a^\cone) \in \Omega^{<,\cone}_\K({\hat M})~~,\\
D_a^{\ad,<,\cyl}\eqdef \nabla_a^\cyl+2[A_a^{<,\cyl},~~]_{-,\bdiamond^\cyl_\epsilon}~~&,
&~~D_a^{\ad,<,\cone}\eqdef \nabla_a^\cone+2[A_a^{<,\cone},~~]_{-,\bdiamond^\cone_\epsilon}~~
\eeqa
as well as: 
\beqa
\check{\cK}_{{\hat D},{\hat Q}}^{<,\cyl}\eqdef P_<(\check{\cK}_{{\hat D},{\hat Q}}^\cyl)~~&,&
~~\check{\cK}_{{\hat D},{\hat Q}}^{<,\cone}\eqdef P_<(\check{\cK}_{{\hat D},{\hat Q}}^\cone)~~,\nn\\
\check{\cK}^{<,\cyl}({\hat D},{\hat Q})\eqdef P_<(\check{\cK}^\cyl({\hat D},{\hat Q}))~~&,&
~~\check{\cK}^{<,\cone}({\hat D},{\hat Q})\eqdef P_<(\check{\cK}^\cone({\hat D},{\hat Q}))~~.
\eeqa
Using the results above, it is not hard to see that the isomorphisms of algebras (see diagrams \eqref{diagram:cylforms} and \eqref{diagram:coneforms}): 
\beqa
(\Xi_\epsilon^\cyl)^{-1}~~&\eqdef&~ 2 P_< \circ P_\epsilon^\cyl\circ \Pi^\ast: (\Omega_\K(M),\diamond)\stackrel{\sim}{\longrightarrow}(\Omega_\K^{<,\cyl}({\hat M}),\bdiamond_\epsilon^\cyl)
~~,\\
(\Xi^\cone_\epsilon)^{-1} &\eqdef& 2 P_<\circ P_\epsilon^\cone \circ r^{\cE}\circ  \Pi^\ast:(\Omega_\K(M),\diamond)\stackrel{\sim}{\longrightarrow}(\Omega_\K^{<,\cyl}({\hat M}),\bdiamond_\epsilon^\cyl)
\eeqa
satisfy:
\beqa
2 L_{\check{Q}^{<,\cyl}}^\cyl\circ (\Xi_\epsilon^\cyl)^{-1}=(\Xi_\epsilon^\cyl)^{-1}\circ L_{\check{Q}}~~&,
&~~2 R_{\tau_{\hat \cB}(\check{Q}^{<,\cyl})}^\cyl\circ (\Xi_\epsilon^\cyl)^{-1}=(\Xi_\epsilon^\cyl)^{-1}\circ 
R_{\tau_\cB (\check{Q})}~~,\\
2 L_{\check{Q}^{<,\cone}}^\cone \circ (\Xi_\epsilon^\cone)^{-1}=(\Xi_\epsilon^\cone)^{-1}\circ L_{\check{Q}}~~&,
&~~2 R_{\tau_{\hat \cB}(\check{Q}^{<,\cone})}^\cone\circ (\Xi_\epsilon^\cone)^{-1}=(\Xi_\epsilon^\cone)^{-1}\circ 
R_{\tau_\cB(\check{Q})}\nn~~
\eeqa
as well as:
\beqa
D_m^{\ad,<,\cyl}\circ (\Xi_\epsilon^\cyl)^{-1}=(\Xi_\epsilon^\cyl)^{-1}\circ D_m^\ad~~&,&~~D_m^{\ad,<,\cone}\circ (\Xi_\epsilon^\cone)^{-1}=(\Xi_\epsilon^\cone)^{-1}\circ D_m^\ad~~.
\eeqa
Together with the discussion of the previous subsections, this implies:  
\beqa
\check{\cK}_{{\hat D},{\hat Q}}^{<,\cyl}= (\Xi_\epsilon^\cyl)^{-1}(\check{\cK}_{D,Q})~~&,&~~\check{\cK}_{{\hat D},{\hat Q}}^{<,\cone}= (\Xi_\epsilon^\cone)^{-1}(\check{\cK}_{D,Q})~~,\\
\check{\cK}^{<,\cyl}({\hat D},{\hat Q})= (\Xi_\epsilon^\cyl)^{-1}(\check{\cK}(D,Q))~~&,&~~\check{\cK}^{<,\cone}({\hat D},{\hat Q})= (\Xi_\epsilon^\cone)^{-1}(\check{\cK}(D,Q))~~.
\eeqa
Therefore, $(\check{\cK}_{{\hat D},{\hat Q}}^{<,\cyl},\bdiamond_\epsilon^\cyl)$, $(\check{\cK}_{{\hat D},{\hat Q}}^{<,\cone},\bdiamond_\epsilon^\cone)$ 
and $(\check{\cK}^{<,\cyl}({\hat D},{\hat Q}),\bdiamond_\epsilon^\cyl)$, $(\check{\cK}^{<,\cone}({\hat D},{\hat Q}),\bdiamond_\epsilon^\cone)$ provide isomorphic models
for the $\K$-algebras $(\check{\cK}_{D,Q},\diamond)$ and $(\check{\cK}(D,Q),\diamond)$, respectively. The collection of isomorphic models of the latter $\K$-algebras which 
were discussed in this Section is summarized in the two commutative diagrams:
\be
\scalebox{1.02}{
\!\!\!\!\!\!\!\!\!\!\! \xymatrix{
& (\check{\cK}^{<,\cyl}_{{\hat D},{\hat Q}},\bdiamond_\epsilon^{\cyl})  \ar@<0.5ex>^{P_\epsilon^{\cyl}}[rr]  \ar@{.}[d]<0.5ex> \ar@<-0.5ex>@{<.}_>>>>>>>{(\Xi^{\cyl}_\epsilon)^{-1}}[d] \ar@<0.5ex>[dl]^{~~r^\cE}
& & (\check{\cK}^\cyl_{{\hat D},{\hat Q}},\diamond^{\cyl})  \ar@<0.5ex>^{2P_\perp}[dd] \ar@<0.5ex>^{r^\cE}[dl] \ar@<0.5ex>^{2P_<}[ll] \\
~~~~~~~(\check{\cK}^{<,\cone}_{{\hat D},{\hat Q}},\bdiamond_\epsilon^{\cone})  \ar@<0.5ex>[ur]^{~r^{-\cE}}\ar@<0.5ex>^{~~~~~~~~~~~~~~~~~~~P_\epsilon^{\cone}}[rr] \ar@<0.5ex>^{\Xi^{\cone}_\epsilon}[dd]
& &  (\check{\cK}^\cone_{{\hat D},{\hat Q}},\diamond^{\cone})   \ar@<0.5ex>^{r^{-\cE}}[ur]{~~}\ar@<0.5ex>^>>>>>>{2P_\perp}[dd] \ar@<0.5ex>^{2P_<~~~~~~~~~~~~~~~}[ll] \\
& (\check{\cK}_{D,Q},\diamond)  \ar@<0.5ex>@{.}[r]\ar@<-0.5ex>@{<.}[r]_{j^\ast}\ar@<0.5ex>@{.}[u]\ar@<-0.5ex>@{<.}_{\Xi^{\cyl}_\epsilon}[u]
& &  (\check{\cK}_{{\hat D},{\hat Q}},\diamond^{\cyl})  \ar@<0.5ex>@{.}[l]\ar@<-0.5ex>@{<.}_{\Pi^\ast}[l] \ar@<0.5ex>^{P_\epsilon^{\cyl}}[uu] \ar@<0.5ex>^{~~r^{\cE}}[dl]\\
(\check{\cK}_{D,Q},\diamond) \ar@<0.5ex>^{r^\cE \circ~\Pi^\ast}[rr] \ar@<0.5ex>^{(\Xi^{\cone}_\epsilon)^{-1}}[uu] \ar@{.>}^{\id}[ur]<0.5ex>\ar@<-0.5ex>@{<.}_{\id}[ur]
& & (r^\cE(\check{\cK}_{{\hat D},{\hat Q}}),\diamond^{\cone})  \ar@<0.5ex>^{r^{-\cE}}[ur] \ar@<0.5ex>^>>>>>>{P_\epsilon^{\cone}}[uu] \ar@<0.5ex>^{j^\ast\circ~r^{-\cE}}[ll]
}} ~~~ \nn
\ee
and: 
\be
\scalebox{0.99}{
\!\!\!\!\!\!\!\!\!\!\!\!\!\!\!\!\! \xymatrix{
& (\check{\cK}^{<,\cyl}({\hat D},{\hat Q}),\bdiamond_\epsilon^{\cyl})  \ar@<0.5ex>^{P_\epsilon^{\cyl}}[rr]  \ar@{.}[d]<0.5ex> \ar@<-0.5ex>@{<.}_>>>>>>>{(\Xi^{\cyl}_\epsilon)^{-1}}[d] \ar@<0.5ex>[dl]^{~~r^\cE}
& & (\check{\cK}^\cyl({\hat D},{\hat Q}),\diamond^{\cyl})  \ar@<0.5ex>^{2P_\perp}[dd] \ar@<0.5ex>^{r^\cE}[dl] \ar@<0.5ex>^{2P_<}[ll] \\
~~~~~~~(\check{\cK}^{<,\cone}({\hat D},{\hat Q}),\bdiamond_\epsilon^{\cone})  \ar@<0.5ex>[ur]^{~r^{-\cE}}\ar@<0.5ex>^{~~~~~~~~~~~~~~~~~~~P_\epsilon^{\cone}}[rr] \ar@<0.5ex>^{\Xi^{\cone}_\epsilon}[dd]
& &  (\check{\cK}^\cone({\hat D},{\hat Q}),\diamond^{\cone})   \ar@<0.5ex>^{r^{-\cE}}[ur]{~~}\ar@<0.5ex>^>>>>>>{2P_\perp}[dd] \ar@<0.5ex>^{2P_<~~~~~~~~~~~~~~~}[ll] \\
& (\check{\cK}(D,Q),\diamond)  \ar@<0.5ex>@{.}[r]\ar@<-0.5ex>@{<.}[r]_{j^\ast}\ar@<0.5ex>@{.}[u]\ar@<-0.5ex>@{<.}_{\Xi^{\cyl}_\epsilon}[u]
& &  (\check{\cK}({\hat D},{\hat Q}),\diamond^{\cyl})  \ar@<0.5ex>@{.}[l]\ar@<-0.5ex>@{<.}_{\Pi^\ast}[l] \ar@<0.5ex>^{P_\epsilon^{\cyl}}[uu] \ar@<0.5ex>^{~~r^{\cE}}[dl]\\
(\check{\cK}(D,Q),\diamond) \ar@<0.5ex>^{r^\cE \circ~\Pi^\ast}[rr] \ar@<0.5ex>^{(\Xi^{\cone}_\epsilon)^{-1}}[uu] \ar@{.>}^{\id}[ur]<0.5ex>\ar@<-0.5ex>@{<.}_{\id}[ur]
& & (r^\cE(\check{\cK}({\hat D},{\hat Q})),\diamond^{\cone})  \ar@<0.5ex>^{r^{-\cE}}[ur] \ar@<0.5ex>^>>>>>>{P_\epsilon^{\cone}}[uu] \ar@<0.5ex>^{j^\ast\circ~r^{-\cE}}[ll]
}} ~~~~~~ \nn
\ee

\section{Application to ${\cal N}=2$ flux compactifications of M-theory on
eight-manifolds}
\label{sec:application}

In this Section, we apply our methods to the study of the most general ${\cal N}=2$
warped compactification of eleven-dimensional supergravity on eight-manifolds
down to an $\AdS_3$ space. After some basic preparations in Subsection
\ref{sec:applprep}, Subsection \ref{sec:applcone} explains how the cone
formalism of Section \ref{sec:cone} can be applied to this example and gives a
brief explanation of the reasons for relying on the cone
construction. Subsection \ref{sec:applCGK} gives our translation of the
generalized Killing pinor equations into a system of algebraic and
differential constraints on differential forms defined on the cone as well as
a brief analysis of the structure of those equations. While these equations
serve only an illustrative purpose in the present paper, they will be analyzed
in  detail in upcoming work.

\subsection{Preparations}
\label{sec:applprep}

As in \cite{MartelliSparks,Tsimpis, ga1}, we start with eleven dimensional
supergravity on an 11-manifold endowed with a spinnable Lorentzian metric 
of `mostly plus' signature. As in loc. cit.,  we consider compactification down to an
$\AdS_3$ space of cosmological constant $\Lambda=-8\kappa^2$, where
$\kappa$ is a positive real parameter --- this includes the Minkowski
case as the limit $\kappa\rightarrow 0$.  Thus ${\tilde M}=N\times M$,
where $N$ is an oriented 3-manifold diffeomorphic with $\R^3$ and carrying the
$\AdS_3$ metric while $M$ is an oriented Riemannian eight-manifold whose metric
we denote by $g$. The metric on ${\tilde M}$ is a warped product 
whose warp factor $\Delta$ is a smooth function defined on $M$. 
For the field strength ${\tilde G}$, we use the ansatz:
\be
{\tilde G} = e^{3\Delta} G~~~{\rm with}~~~ G = {\rm vol}_3\wedge f+F~~,
\ee
where $f=f_m e^m\in \Omega^1(M)$, $F=\frac{1}{4!}F_{mnpq} e^{mnpq}\in \Omega^4(M)$ and
$\vol_3$ is the volume form of $N$. Small Latin indices from the middle of the alphabet run
from $1$ to $8$ and correspond to a choice of frame on $M$. For the
eleven-dimensional supersymmetry generator ${\tilde \eta}$, we
use the ansatz:
\be
{\tilde \eta}=e^{\frac{\Delta}{2}}\eta~~~{\rm with}~~~ \eta=\psi\otimes \xi~~,
\ee
where $\xi$ is a real pinor of spin $1/2$ on the
internal space $M$ and $\psi$ is a real pinor on the $\AdS_3$
space $N$. Mathematically, $\xi$ is a section of the pinor bundle of
$M$, which is a real vector bundle of rank $16$ defined on $M$,
carrying a fiberwise representation of the Clifford algebra
$\Cl(8,0)$. Since $p-q\equiv_8 0$ for $p=8$ and $q=0$, this
corresponds to the simple normal case of \cite{ga1}. In
particular, the corresponding morphism $\gamma:(\wedge T^\ast
M,\diamond)\rightarrow (\End(S),\circ)$ of bundles of algebras is an
isomorphism, i.e. it is bijective on the fibers. We set
$\gamma^m=\gamma(e^m)$ and $\gamma^{(9)}:=\gamma^1\circ\ldots \circ\gamma^8$ for
some local orthonormal frame $e^m$ of $M$. In dimension eight with Euclidean signature,
there exists an admissible \cite{AC0, AC1} (and thus $\Spin(8)$-invariant)
bilinear pairing $\cB$ on the pin bundle $S$, which is a scalar product. 
Assuming that $\psi$ is a Killing pinor on the $\AdS_3$ space, the
supersymmetry condition amounts to the following {\em constrained generalized
Killing (CGK)} pinor equations \cite{ga1} for $\xi$:
\ben
\label{par_eq}
D_m\xi = 0~~,~~Q\xi = 0~~,
\een
where $D_m$ is a linear connection on $S$ and $Q\in \Gamma(M,\End(S))$ is
a globally-defined endomorphism of the vector bundle $S$. As in \cite{MartelliSparks, Tsimpis} (and in contradistinction with
\cite{Becker}) {\em we do not require that $\xi$ has definite
chirality}. As we shall see in a moment, this seemingly trivial
generalization has drastic consequences, leading to a problem which is
technically much harder than that solved in the celebrated work of
\cite{Becker}. The space of solutions of \eqref{par_eq} is a
finite-dimensional $\R$-linear subspace $\cK(D,Q)$ of the space $\Gamma(M,S)$ of
smooth sections of $S$. The problem of interest is to find those metrics and fluxes on $M$ for
which some fixed amount of supersymmetry is preserved in three
dimensions, i.e.  for which the space $\cK(D,Q)$ has some given
non-vanishing dimension, which we denote by $s$. 
The case $s=1$ (which corresponds to ${\cal N}=1$ supersymmetry in three dimensions) was studied in
\cite{MartelliSparks, Tsimpis} and reconsidered in \cite{ga1} by using
geometric algebra techniques.  The case $s=2$ (which leads to ${\cal N}=2$ supersymmetry in three
dimensions) was studied in \cite{Becker}, but
considering only Majorana-{\em Weyl} solutions $\xi$ of \eqref{par_eq}, i.e.
only the case when $\cK_{D,Q}$ is a subspace of the kernel $\cK(\id_S-\gamma_{(9)})$ or of the
kernel $\cK(\id_S+\gamma_{(9)})$. Here, we consider the case when no extraneous
chirality constraint is imposed on the solutions of \eqref{par_eq}.

Direct computation gives the following expressions for $D$ and $Q$ (see \cite{ga1,MartelliSparks}):
\ben
\label{Dappl}
D_m=\nabla^S_m+A_m~~,~~A_m= \frac{1}{4}f_p\gamma_{m}{}^{p}\gamma^{(9)}
+\frac{1}{24}F_{m p q r}\gamma^{ p q r}+\kappa \gamma_m\gamma^{(9)}
\een
and
\ben
\label{Qappl}
Q=\frac{1}{2}\gamma^m\partial_m\Delta -\frac{1}{288}F_{m p q r}\gamma^{m p q r}
-\frac{1}{6}f_p \gamma^p \gamma^{(9)}
-\kappa\gamma^{(9)} ~~.
\een
In the present paper, we are interested in the case $s=2$
(${\cal N}=2$ supersymmetry in three dimensions), so we require that \eqref{par_eq}
admits {\em two} linearly independent solutions $\xi_1$ and $\xi_2$. The formed-valued
pinor bilinears $\bcE^{(k)}_{\xi_i,\xi_j}=\frac{1}{k!}\bcE^{(k)}_{m_1\ldots
  m_k}(\xi_i,\xi_j) e^{m_1\ldots  m_k}
\in \Omega^{k}(M)$ with $i,j=1,2$ are constrained by Fierz identities
which play a crucial role below. As we shall see in a moment,
these identities are much more involved in our case (even after reformulating
them on the cone) than the identities which were encountered in \cite{MartelliSparks} and
\cite{Tsimpis}. The translation of \eqref{par_eq} into equations on the differential forms
$\bcE^{(k)}_{\xi_i,\xi_j}$ could be achieved starting from the
following equivalent reformulation of the algebraic constraints
$Q\xi_1=Q\xi_2=0$:
\be
\cB( \xi_i,\left(Q^t\gamma_{m_1\ldots m_k}\pm \gamma_{m_1\ldots m_k}Q\right) \xi_j) =0~~
\ee
and treating the differential constraints $D_m \xi_1=D_m\xi_2=0$ through the
method outlined in \cite{MartelliSparks}. This direct approach due to
\cite{MartelliSparks} is discussed in detail in the Appendices of
\cite{ga1}, where it was also shown how that method is equivalent with the
formalism developed in loc. cit. As it turns out, the direct approach is
computationally quite impractical in our case and we have to rely on the
methods and techniques of \cite{ga1}.

\subsection{The cone construction}
\label{sec:applcone}

Before attempting to solve \eqref{par_eq}, one can ask whether the
mere assumption of existence of two independent solutions
$\xi_1,\xi_2$ provides some useful constraints on the geometry. The
$D$-flatness conditions $D_m\xi_1=D_m\xi_2=0$ imply that the values of the
sections $\xi_1,\xi_2$ at two different points $x,y$ on the internal
manifold are related through the parallel transport of the connection
$A_m$, which is an invertible linear operator between the fibers of
$S$ at $x$ and $y$. In turn, this shows that two solutions which are
linearly independent over $\R$ as sections of $S$ must be linearly
independent everywhere, i.e. the vectors $\xi_1(x), \xi_2(x)\in S_x$
must --- for any point $x$ of $M$ --- be linearly independent in the
fiber $S_x\approx \R^{16}$ (and hence determine a point
$(\xi_1(x),\xi_2(x))$ in the second Stiefel manifold $V_2(S_x)$ of
$S_x$).  Using this observation, one finds that solutions of
\eqref{par_eq} can be classified according to the orbit passing through
$(\xi_1(x),\xi_2(x))$ of the action (induced by restricting
$\gamma_x$) of the group $\Spin(8)\subset \Cl(8,0)\approx (\wedge
T^\ast_x M,\diamond_x)$ on $V_2(S_x)$ --- an orbit which is
independent of $x$ up to the action induced by the parallel transport
of $D$. However, it turns out that the action of $\Spin(8)$ on the
second Stiefel manifold of $\R^{16}$ fails to be transitive, which
leads to complications when attempting to classify solutions in this
manner. In particular (and unlike what happens in many other cases),
two generic solutions of \eqref{par_eq} fail to determine a {\em
global} reduction of the structure group $\SO(8,\R)$ of $(M,g)$ --- a
phenomenon which (as discussed in \cite{Tsimpis}) also occurs for the
case $s=1$ (the case of ${\cal N}=1$ supersymmetry in three dimensions).  Due
to this fact, it is convenient instead to consider $\xi_1$ and $\xi_2$
up to an action (induced by restricting $\gamma_x$) of the group
$\Spin(9)$ --- which can be viewed in a natural manner as a subgroup of
$\Cl(8,0)$.  As pointed out in \cite{Tsimpis}, this $\Spin(9)$ action
can be geometrized by introducing an extra dimension --- for example,
by passing to the metric cylinder (as in \cite{Tsimpis}) or to the
metric cone (as we shall do below) over $M$. The fact that certain
aspects of the simplest flux compactifications (such as Freund-Rubin
compactifications on squashed spheres) can be better understood by
passage to the metric cone is of course well-known --- as is the fact
that the (ordinary) Killing pinor equations of a Riemannian manifold
can be reformulated as parallel pinor equations by passage to the
metric cone (see \cite{Bar}). In {\em general} flux compactifications,
the simplification which one obtains through this construction is
less drastic, though still quite useful both from a
computational and conceptual standpoint.

Let us therefore consider the metric cone $({\hat M},g_\cone)$
(cf. Section \ref{sec:cone}) and the lift of $D$ to the connections
${\hat D}$ on the pin bundle ${\hat S}$ of ${\hat M}$ (see Subsection
\ref{sec:conedeq}).  Since the metric cone over $(M,g)$ has
signature $(9,0)$ and since $9-0\equiv_8 1$, the Clifford algebra
$\Cl(9,0)$ corresponds to the normal non-simple case discussed in \cite{ga1}.  
We have two inequivalent\footnote{Inequivalent
in the sense of the representation theory of Clifford algebras.}
choices for the fiberwise pin representation of the \KA bundle of
${\hat M}$, which are distinguished by the signature $\epsilon\in
\{-1,1\}$. Both choices correspond to representations which are
surjective but non-injective on each of the fibers $(\wedge T^\ast_x
{\hat M},\diamond_x) \approx \Cl(9,0)$. As in Section \ref{sec:cone},
the pin bundle ${\hat S}$ of ${\hat M}$ can be constructed as the
pullback of $S$ through the natural projection of ${\hat M}$ to $M$,
while the morphism $\gamma_\cone:(\wedge T^\ast {\hat
M},\diamond^\cone) \rightarrow (\End({\hat S}),\circ) $ is completely
determined by the morphism $\gamma:(\wedge T^\ast M,
\diamond)\rightarrow (\End(S),\circ)$ once the signature $\epsilon$
has been chosen. In the following, we shall work with the choice
$\epsilon=+1$. Setting $d=8$, $\epsilon=+1$ and $\rho=2\kappa$ in the
equations of Subsection \ref{sec:coneconnections} and rescaling the
metric on $M$ as $g\rightarrow (2\kappa)^2 g$ ({\em without} changing
the local orthonormal frame $e_a^\cone$ of the cone or the
local orthonormal frame $e_m$ of $(M,g)$) gives ${\hat
D}_a=\nabla^{{\hat S},\cone}_a+A_a^\cone$, where:
\beqan
\label{scondplus1}
\nabla^{{\hat S},\cone}_{\partial_r} = \cL_r^S~~&,&~~\nabla^{{\hat S},\cone}_{e_m} =\nabla^S_{e_m}+\kappa \gamma_{m9}~~,\\ 
A^\cone_9=0~~~&,&~~A^\cone_m=\frac{1}{4}f^p\gamma_{m p 9}+\frac{1}{24}F_{m p q r}\gamma^{p q r}~~,\nn
\eeqan
where $\cL_r^S$ in the right hand side denotes the
Kosmann-Schwarzbach derivative \cite{KS} on sections of ${\hat S}$, taken with
respect to the vector field $\partial_r$. Here and below, indices from
the middle of the Latin alphabet run from $1$ to $8$ and those from
the beginning of the Latin alphabet run from $1$ to $9$. The latter
correspond to frames $e_a^\cyl$ and $e_a^\cone$ chosen
as in \eqref{cylconeframe}. 

\paragraph{Notational convention.} We do not indicate some pullbacks explicitly --
in particular, we use the same notation for $\cB$, $\gamma^m$ and for their
pullbacks ${\hat \cB}$, ${\hat \gamma}^m=\gamma^m_\ast$ from $S$ to ${\hat S}$,
with a similar convention for differential forms. The bilinear pairing has the properties 
 $\sigma_\cB=+1$ and $\epsilon_\cB=+1$.

\

As in Subsection \ref{sec:coneconnections}, the generalized Killing
pinor equations $D_m\xi=0$ ($m=1\ldots 8$) for pinors $\xi$ defined on
$M$ amount to the flatness conditions ${\hat D}_a {\hat \xi}=0$
($a=1\ldots 9$) for pinors ${\hat \xi}$ defined on ${\hat M}$. Indeed,
the last of the flatness equations ${\hat D}_9 {\hat \xi}=0$ is
equivalent with the requirement that the section ${\hat \xi}$ of
${\hat S}$ is the pullback of some section $\xi$ of $S$ through the
natural projection map from ${\hat M}$ to $M$, while the remaining
equations amount to the generalized Killing conditions $D_m\xi=0$ for
the pinor $\xi$ defined on the manifold $M$. Also recall from Section
\ref{sec:cone} that the algebraic constraints are equivalent with the
following equations for ${\hat \xi}$:
\be
{\hat Q} {\hat \xi}=0~~,
\ee
where ${\hat Q}\in \Gamma({\hat M},\End({\hat S}))$ is the pullback of
$Q\in \Gamma(M,\End(S))$ to ${\hat M}$. With our notational
conventions (in which we won't explicitly indicate the pullback),
${\hat Q}$ has the same form \eqref{Qappl} as $Q$ in the appropriate
local frame on the cone.

\subsection{The $\K$-algebra of constrained generalized Killing forms}
\label{sec:applCGK}

For the reasons outlined above, we consider the CGK pinor equations
formulated on the metric cone over $M$. As explained in \cite{ga1} (see also
Subsection \ref{sec:conetruncated}), we realize the algebra $(\Omega^{+,\cone}({\hat M}), \diamond^\cone)$
(the effective domain of definition of $\gamma_\cone$) as the algebra
$(\Omega^<({\hat M}), \bdiamond_+^\cone)$. We have $\Omega^<({\hat
M})=\oplus_{k=0}^{4}\Omega^k({\hat M})$, so we are interested in
pinor bilinears $\bcE^{(k)}_{{\hat \xi}_1,{\hat \xi}_2}$ with
$k=0\ldots 4$ for two independent solutions ${\hat \xi}_1, {\hat \xi}_2$ of the CGK
pinor equations lifted to the cone:
\ben
\label{coneq}
{\hat D}_a {\hat \xi}={\hat Q}{\hat \xi}=0~~,
\een
which are equivalent with the original CGK pinor equations on $M$.

To extract the translation of these equations into constraints on
differential forms, we implemented certain procedures within the
package {\tt Ricci} \cite{Ricci} for tensor computations in
{\tt Mathematica}\textsuperscript{\textregistered}. We also implemented similar
procedures in {\tt Cadabra} \cite{Cadabra}. The dequantizations:
\be
\check{A}_a^\cone=\gamma_\cone^{-1}(A_a^\cone)\in \Omega^<({\hat M})~~,
~~\check{Q}^\cone =\gamma_\cone^{-1}({\hat Q})\in \Omega^<({\hat M})~~,
\ee
of $A^\cone$ and ${\hat Q}$ are given by $\check{A}_9^\cone=0$ and
(recall that $\theta\eqdef \dd r=e^9_\cone$):
\beqa
\check{A}_m^\cone &=&\frac{1}{4} \iota^\cone_{e_m^\cone}  F_\cone +
\frac{1}{4}(e_m^\cone)_{\hash,\cone} \wedge f_\cone \wedge \theta~~\in \Omega^<({\hat M})~~, \nn\\
\check{Q}^\cone&=&\frac{1}{2} r \dd
\Delta-\frac{1}{6}f_\cone \wedge\theta-\frac{1}{12}F_\cone -\kappa \theta ~~\in \Omega^<({\hat
M})~~,\nn
\eeqa
while the $\cB$-transpose of ${\hat Q}$ dequantizes to the reversion of $\check{Q}^\cone$:
\be
{\hat \tau}(\check{Q}^\cone)=\frac{1}{2} r \dd \Delta+\frac{1}{6}f_\cone
\wedge\theta-\frac{1}{12}F_\cone -\kappa \theta ~~. \nn
\ee
Here, ${\hat \tau}$ is the main anti-automorphism \cite{ga1} of
$(\Omega(M),\diamond^\cone)$ (which coincides with the main anti-automorphism
of $(\Omega(M),\diamond^\cyl)$). The forms $f_\cone$ and $F_\cone$ above are
the cone lifts (see definition \eqref{cone_lift}) of $f$ and $F$ respectively,
while (in accordance with our notational conventions) $\Delta$ stands for the
pullback $\Pi^\ast(\Delta)=\Delta\circ \Pi$, even though the notation does not
show this explicitly.

A basis for the space spanned by $\bcE^{(k),\cone}_{{\hat \xi}_1,{\hat
\xi}_2}\eqdef \frac{1}{k!}(\epsilon_\cB)^k\cB({\hat \xi}_1, {\hat \gamma}_{a_1\ldots a_k}{\hat
\xi}_2)e^{a_1}_\cone\wedge \ldots \wedge e^{a_k}_\cone \in
\Omega^<({\hat M})$ (of rank at most 4) which can be constructed
on the cone from ${\hat \xi}_1$ and ${\hat \xi}_2$ is
given by (where we have raised all indices using the cone metric in order to avoid
notational clutter) :
\beqan
\label{forms}
V_1^a=\cB({\hat \xi}_1, {\hat \gamma}^a {\hat \xi}_1) ~~,~~ V_2^a&=&\cB({\hat
\xi}_2, {\hat \gamma}^a {\hat \xi}_2) ~~,~~ V_3^a = \cB({\hat \xi}_1, {\hat \gamma}^a {\hat
\xi}_2)~,\nn\\
K^{ab} = \cB({\hat \xi}_1, {\hat \gamma}^{ab} {\hat \xi}_2)~~&,&~~\Psi^{abc} =
\cB({\hat \xi}_1, {\hat \gamma}^{abc} {\hat \xi}_2)~, \nn\\
\Phi_1^{abce}=\cB({\hat \xi}_1,{\hat \gamma}^{abce} {\hat \xi}_1) ~~,
~~ \Phi_2^{abce}&=&\cB({\hat \xi}_2,{\hat \gamma}^{abce} {\hat \xi}_2)
~~,~~\Phi_3^{abce} = \cB({\hat \xi}_1,{\hat \gamma}^{abce} {\hat \xi}_2)~,
\eeqan
Here and below, we have taken ${\hat \xi}_1$ and ${\hat \xi}_2$ to form an
orthonormal basis of the $\R$-vector space $\cK({\hat D},{\hat Q})$ on the cone:
\be
\cB({\hat \xi}_i,{\hat \xi}_j)=\delta_{ij}~~,~~\forall i,j=1,2~~.
\ee
To arrive at \eqref{forms}, we used the identity $\cB({\hat \xi}_i,
{\hat \gamma}^{a_1\ldots a_k}{\hat \xi}_j)=(-1)^{\frac{k(k-1)}{2}}\cB({\hat \xi}_j,
{\hat \gamma} ^{a_1\ldots a_k}{\hat \xi}_i)$, which follows from
$({\hat \gamma}_a)^t={\hat \gamma}_a$ and implies that certain of the forms
$\bcE^{(k),\cone}_{{\hat \xi}_i,{\hat \xi}_j}$ vanish identically.

\

\paragraph{Notational convention.} From now on --- in order to avoid
notational clutter --- we shall suppress the superscripts and subscripts
``$\cone$''. In particular, we shall denote the
cone lifts $F_\cone=r^4 \Pi^\ast(F)$ and $f_\cone=r\Pi^\ast(f)$ simply by
$F$ and $f$. 
We remind the reader of our notations in \cite{ga1} for the basis elements $\check{E}_{ij}$ of the algebra
$(\Omega^{+}({\hat M}), \diamond)$ on the cone:
\be
\check{E}_{ij}\eqdef\check{E}_{\xi_i,\xi_j}=\frac{1}{2^{[\frac{d+1}{2}]}}\bcE_{\xi_i,\xi_j}=
\frac{1}{2^{[\frac{d+1}{2}]}}\sum_{k=0}^{d}\bcE^{(k)}_{\xi_i,\xi_j} ~\in\Omega^+(M)~~,
\ee
where we know that $\check{E}_{ij}$ are twisted selfdual forms, thus $\check{E}_{ij}=\check{E}^<_{ij}+\tilde\ast\check{E}^<_{ij}$, 
where $\check{E}^<_{ij}\in\Omega^<(\hat{M})$. 
Hence the basis elements $\check{E}^<_{ij}$ of the truncated Fierz algebra 
$(\Omega^<({\hat M}),\bdiamond_+)$ read:
\be
\check{E}^<_{ij}=\frac{1}{2^{[\frac{d+1}{2}]}}\sum_{k=0}^{[\frac{d}{2}]}\bcE^{(k)}_{\xi_i,\xi_j} ~\in\Omega^<(M)~~,
\ee
where $d=9$. With the notations and conventions above, the truncated model of the flat Fierz $\K$-algebra $(\check{\cK}^{<}({\hat D},{\hat Q}),\bdiamond_+)$
on the cone admits the basis:
\beqan
\label{Basis}
\check{E}^<_{12} = \frac{1}{32}(V_3+K+\Psi+\Phi_3)~~&,&~~ \check{E}^<_{21} = \frac{1}{32}(V_3-K-\Psi+\Phi_3)~~, \nn\\
\check{E}^<_{11} = \frac{1}{32}(1+V_1+\Phi_1)~~&,&~~\check{E}^<_{22} = \frac{1}{32}(1+V_2+\Phi_2)~~
\eeqan
and can be generated by two elements (see Subsection 5.10 of \cite{ga1}), which we choose to be:
\be
 \check{E}^<_{12}=\frac{1}{32}(V_3+K+\Psi+\Phi_3)~~~\mathrm{and}~~~
\check{E}^<_{21}= {\hat \tau}(\check{E}_{12})=\frac{1}{32}(V_3-K-\Psi+\Phi_3)~. \nn
\ee
In order to avoid notational clutter, we shall henceforth use $\bdiamond$ instead of $\bdiamond_+$.
As exlained in general in \cite{ga1}, the Fierz relations for the truncated model amount to:
\ben
\label{FierzRel}
\check{E}^<_{ij}\bdiamond\check{E}^<_{kl}=\frac{1}{2} \delta_{jk}\check{E}_{il}~~,~~\forall i,j,k,l=1,2~~,~~
\een
while the ideal of relations corresponding to $\check{E}^<_{12}$ and
$\check{E}^<_{21}$ is generated by:
\beqan
\label{Fierz1}
 && \check{E}^<_{12}\bdiamond \check{E}_{12}=0 \;\; \;\;
\bigg(\Longleftrightarrow {\hat \tau}(\check{E}^<_{12})\bdiamond
{\hat \tau}(\check{E}_{12})=0\bigg)~~, \\
 \label{Fierz2} && \check{E}^<_{12}\bdiamond {\hat \tau}(\check{E}^<_{12})\bdiamond
\check{E}^<_{12}= \frac{1}{4}\check{E}^<_{12} \;\; \;\; \bigg(\Longleftrightarrow
{\hat \tau}(\check{E}^<_{12})\bdiamond \check{E}_{12}\bdiamond
{\hat \tau}(\check{E}^<_{12})= \frac{1}{4}{\hat \tau}(\check{E}^<_{12}) \bigg)~~.
\eeqan
On the other hand, the algebraic constraints in \eqref{coneq} amount to the following two relations for $\check{E}^<_{12}$:
\ben
\label{AConstr}
 \check{Q}\bdiamond \check{E}_{12}\mp \check{E}_{12}\bdiamond {\hat \tau}(\check{Q})=0 ~~,
\een
while the differential constraints of \eqref{coneq} give the equations
$\D_a \check{E}^<_{12}=0\Leftrightarrow \D_a
\check{E}^<_{21}=0$, which in turn imply:
\ben
\label{DConstr}
 \dd \check{E}^<_{12}= e^a\wedge\nabla_a \check{E}^<_{12}=-2e^a\wedge[\check{A}_a,\check{E}^<_{12}]_{-,\bdiamond}~~.
\een
As explained in Subsection 5.10 of \cite{ga1}, it is enough to consider the
constraints \eqref{AConstr} and \eqref{DConstr} for the generators
$\check{E}^<_{12}$ and $\check{E}^<_{21}={\hat \tau} (\check{E}^<_{12})$, since
the corresponding constraints for
$\check{E}^<_{11}=2\check{E}^<_{12}\bdiamond \check{E}^<_{21}$ and
$\check{E}^<_{22}=2\check{E}^<_{21}\bdiamond \check{E}^<_{12}$ follow from
those.

\paragraph{Algebraic constraints.}
Using the procedures which we have implemented and the package {\tt
Ricci} for tensor computations in {\tt
Mathematica}\textsuperscript{\textregistered} (see \cite{Ricci}), we
find that the first equation (with the minus sign) in \eqref{AConstr} amounts to the
following system when separated on ranks:
\beqa
 \iota_{f\wedge\theta} K &=& 0 ~~,\nn\\
 r\iota_{\dd \Delta} K+\frac{1}{3}\iota_{f\wedge\theta}\Psi - \frac{1}{6}\iota_{\Psi} F -2 \kappa~\iota_\theta K  &=& 0 ~~,\nn\\
 \frac{1}{3} \iota_{f\wedge\theta}\Phi_3-\frac{1}{6}F\bigtriangleup_3 \Phi_3 +r(\dd \Delta)\wedge V_3+2\kappa~ V_3\wedge\theta &=& 0 ~~,\nn\\
 r\iota_{\dd \Delta} \Phi_3 -\frac{1}{3}V_3\wedge f\wedge\theta+\frac{1}{6}\iota_{V_3} F
 -\frac{1}{6}{\ast}(F\bigtriangleup_1 \Phi_3) +\frac{1}{3}{\ast}(f\wedge\theta\wedge \Phi_3 )-2\kappa~\iota_\theta\Phi_3 &=& 0 ~~,\nn\\
 r(\dd \Delta)\wedge\Psi-\frac{1}{3}f\wedge\theta\wedge K-\frac{1}{6} K\bigtriangleup_1 F
 -\frac{1}{3}{\ast}(f\wedge\theta\wedge\Psi) +\frac{1}{6}{\ast}(\Psi\bigtriangleup_1 F)+2\kappa~\Psi\wedge\theta &=& 0~~, \nn
\eeqa
while the second equation (with the plus sign) in \eqref{AConstr} amounts to:
\beqa
 -\frac{1}{6} \iota_F \Phi_3 +r\iota_{\dd \Delta} V_3 -2\kappa~\iota_\theta V_3 &=& 0 ~~,\nn\\
 \frac{1}{3} \iota_{V_3}(f\wedge\theta)-\frac{1}{6}{\ast}( F\wedge  \Phi_3 ) &=& 0 ~~,\nn\\
 r\iota_{\dd \Delta}\Psi +\frac{1}{3} (f\wedge\theta)\bigtriangleup_1 K+\frac{1}{6}  \iota_{K} F
 +\frac{1}{6}{\ast} (F\wedge\Psi)-2\kappa~ \iota_\theta \Psi &=& 0 ~~,\nn\\
 \frac{1}{3} (f\wedge\theta)\bigtriangleup_1\Psi +\frac{1}{6}\Psi\bigtriangleup_2 F +\frac{1}{6}{\ast}( K\wedge F)
 +r(\dd \Delta)\wedge K - 2\kappa~ K\wedge\theta &=& 0 ~~,\nn\\
 \frac{1}{3} (f\wedge\theta)\bigtriangleup_1 \Phi_3 + \frac{1}{6} F\bigtriangleup_2  \Phi_3
 -\frac{1}{6}{\ast}(F\wedge V_3) +{\ast}(r(\dd \Delta)\wedge \Phi_3 )-2\kappa\ast(\Phi_3\wedge\theta) &=& 0~~. \nn
\eeqa
\paragraph{Differential constraints.}
Using the same {\tt Mathematica}\textsuperscript{\textregistered}
package, we can extract the differential constraints given by
\eqref{DConstr}, which --- when separated on ranks --- amount to:
\beqa
 \dd V_3 &=&  \Phi_3 \bigtriangleup_3 F + 2\iota_{f\wedge\theta} \Phi_3 ~~,  \nn\\
 \dd K &=& 2(f\wedge\theta)\bigtriangleup_1 \Psi + 2\Psi\bigtriangleup_2 F ~~,  \nn\\
 \dd \Psi &=& 3F\bigtriangleup_1 K - F\bigtriangleup_3{\ast}\Psi
 +4{\ast}(f\wedge\theta\wedge\Psi) -2f\wedge\theta\wedge K ~~, \nn\\
 \dd \Phi_3  &=&-4 F\wedge V_3 + e^m\wedge{\ast}((\iota_{e^m}F)\bigtriangleup_1 \Phi_3 )
  - e^m\wedge{\ast}(((e_m)_\hash \wedge f\wedge\theta)\bigtriangleup_1 \Phi_3 ) ~~.\nn
\eeqa
According to our notational conventions, $e^m$ in the equations above stands for $e^m_\cone$
while $e_m$ stands for $e_m^\cone$. Furthermore, $\ast \eqdef  \ast_\cone$ is the
(ordinary\footnote{As opposed to the {\em twisted} Hodge operator of \cite{ga1}.}) Hodge operator of $(M,g_\cone)$ and $\iota$ stands for $\iota^\cone$. The
generalized products $\bigtriangleup_p\eqdef \bigtriangleup_p^\cone$ are
constructed with the cone metric. 

\paragraph{Fierz relations.}
Let us consider the Fierz identities \eqref{FierzRel} for the basis elements $\check{E}_{ij}$
($i,j=1,2$) of the truncated model of the flat Fierz algebra $(\check{\cK}^{<}({\hat D},{\hat Q},\bdiamond)$:

\be
\begin{array}{ccccccc}
({\rm F1}):&~\check{E}^<_{12}\bdiamond\check{E}^<_{12}=0~~~~~&~,~&({\rm F2}):&~\check{E}^<_{12}\bdiamond\check{E}^<_{21}=\frac{1}{2}\check{E}^<_{11}~~,\\
({\rm F3}):&~\check{E}^<_{12}\bdiamond\check{E}^<_{22}=\frac{1}{2}\check{E}^<_{12} &~,~&({\rm F4}):&~\check{E}^<_{12}\bdiamond\check{E}^<_{11}=0~~~~~~~,\\
({\rm F5}):&~\check{E}^<_{11}\bdiamond\check{E}^<_{11}=\frac{1}{2}\check{E}^<_{11} &~,~&({\rm F6}):&~\check{E}^<_{11}\bdiamond\check{E}^<_{12}=\frac{1}{2}\check{E}^<_{12}~~,\\
({\rm F7}):&~\check{E}^<_{11}\bdiamond\check{E}^<_{21}=0~~~~~&~,~&({\rm F8}):&~\check{E}^<_{11}\bdiamond\check{E}^<_{22}=0~~~~~~~,\\
({\rm F9}):&~\check{E}^<_{21}\bdiamond\check{E}^<_{12}=\frac{1}{2}\check{E}^<_{22} &~,~&({\rm F10}):&~\check{E}^<_{21}\bdiamond\check{E}^<_{11}=\frac{1}{2}\check{E}^<_{21}~~,\\
({\rm F11}):&~\check{E}^<_{21}\bdiamond\check{E}^<_{21}=0~~~~~&~,~&({\rm F12}):&~\check{E}^<_{21}\bdiamond\check{E}^<_{22}=0~~~~~~~,\\
({\rm F13}):&~\check{E}^<_{12}\bdiamond\check{E}^<_{11}=0~~~~~&~,~&({\rm F14}):&~\check{E}^<_{22}\bdiamond\check{E}^<_{12}=0~~~~~~~,\\
({\rm F15}):&~\check{E}^<_{22}\bdiamond\check{E}^<_{21}=\frac{1}{2}\check{E}^<_{21}&~,~&({\rm F16}):&~\check{E}^<_{22}\bdiamond\check{E}^<_{22}=\frac{1}{2}\check{E}^<_{22}~~.
\end{array}
\ee
We note that some of these conditions are equivalent through reversion
(namely (F1)$\Leftrightarrow$(F11), (F3)$\Leftrightarrow$(F15),
(F4)$\Leftrightarrow$(F7), (F6)$\Leftrightarrow$(F10),
(F8)$\Leftrightarrow$(F13) and (F12)$\Leftrightarrow$(F14), while
relations (F2), (F5), (F9), (F16) are selfdual under reversion).
After expanding the geometric product and separating ranks, we find
independent relations only from certain rank components of (F1) --
(F6), (F8), (F9), (F12) and (F16). Namely, equation (F1) (which
coincides with \eqref{Fierz1}) takes the form:
\be
(V_3+K+\Psi+\Phi_3)\bdiamond(V_3+K+\Psi+\Phi_3)=0
\ee
and gives the following relations when separated into rank components:
\beqa
-\vectornorm{K}^2 + || \Phi_3||^2 - || \Psi ||^2 + || V_3 ||^2 &=& 0
~~, \nn\\ -2\iota_K\Psi +{\ast}(\Phi_3\wedge\Phi_3) &=& 0 ~~, \nn\\
\iota_{V_3} \Psi - {\ast}(\Phi_3\wedge\Psi)-\iota_K \Phi_3 &=& 0
~~, \nn\\ K\wedge V_3 -{\ast}(K\wedge\Phi_3) -
\Psi\bigtriangleup_2\Phi_3 &=& 0 ~~, \nn\\ \Psi\bigtriangleup_1\Psi
-\Phi_3\bigtriangleup_2\Phi_3 + 2{\ast}(K\wedge\Psi) +
2{\ast}(V_3\wedge\Phi_3) +K\wedge K &=& 0~~. \nn
\eeqa
Separating (F2) into rank components gives the following nontrivial relations:
\beqa
|| K||^2 + || \Phi_3||^2 + || \Psi ||^2 + || V_3 ||^2 &=& 16 ~~, \nn\\
2\iota_{K}\Psi -2\iota_{V_3} K-2\iota_\Psi\Phi_3+{\ast}(\Phi_3\wedge\Phi_3)-16V_1 &=& 0 ~~,
\nn\\ -\Psi\bigtriangleup_1\Psi -\Phi_3\bigtriangleup_2\Phi_3 -
2{\ast}(K\wedge\Psi) + 2{\ast}(V_3\wedge\Phi_3) -K\wedge K -16\Phi_1
&-&\nn\\ -2K\bigtriangleup_1\Phi_3
-2V_3\wedge\Psi+2{\ast}(\Psi\bigtriangleup_1\Phi_3) &=& 0~~ \nn
\eeqa
and similarly for (F3):
\beqa
\langle\Phi_2,\Phi_3\rangle+\langle V_2,V_3\rangle &=& 0  ~~, \nn\\
 -\iota_{V_2}K-\iota_{\Psi}\Phi_2 +{\ast}(\Phi_2\wedge\Phi_3)-15V_3 &=& 0  ~~, \nn\\
 -15K -V_2\wedge V_3-\Phi_2\bigtriangleup_3\Phi_3-\iota_{K}\Phi_2 +\iota_{V_2}\Psi-{\ast} (\Phi_2\wedge\Psi) &=& 0  ~~, \nn\\
 K\wedge V_2 +\iota_{V_3}\Phi_2-\iota_{V_2}\Phi_3-{\ast}(K\wedge\Phi_2) -15\Psi-{\ast}(\Phi_2\bigtriangleup_1\Phi_3) -
\Psi\bigtriangleup_2\Phi_2 &=& 0  ~~, \nn\\
 -K\bigtriangleup_1\Phi_2  -\Phi_2\bigtriangleup_2\Phi_3 +{\ast}(\Psi\bigtriangleup_1\Phi_2)+
{\ast}(V_2\wedge\Phi_3)+{\ast}(V_3\wedge\Phi_2)-15\Phi_3-V_2\wedge\Psi &=& 0~~, \nn
\eeqa
for (F4):
\beqa
\langle\Phi_1,\Phi_3\rangle+\langle V_1,V_3\rangle &=& 0  ~~, \nn\\
-\iota_{V_1} K-\iota_{\Psi}\Phi_1 +{\ast}(\Phi_1\wedge\Phi_3)+V_3 &=& 0 ~~,  \nn\\
 K -V_1\wedge V_3-\Phi_1\bigtriangleup_3\Phi_3-\iota_{K}\Phi_1 +\iota_{V_1}\Psi- {\ast}(\Phi_1\wedge\Psi) &=& 0 ~~,  \nn\\
 K\wedge V_1 +\iota_{V_3}\Phi_1-\iota_{V_1}\Phi_3-{\ast}(K\wedge\Phi_1) +\Psi-{\ast}(\Phi_1\bigtriangleup_1\Phi_3)-
\Psi\bigtriangleup_2\Phi_1 &=& 0 ~~,  \nn\\
 -K\bigtriangleup_1\Phi_1  -\Phi_1\bigtriangleup_2\Phi_3 +{\ast}(\Psi\bigtriangleup_1\Phi_1)+{\ast}(V_1\wedge\Phi_3)+{\ast}(V_3\wedge\Phi_1)+\Phi_3-V_1\wedge\Psi  &=& 0~~, \nn
\eeqa
(F5):
\beqa
|| \Phi_1||^2 + || V_1 ||^2 &=& 15 ~~, \nn\\
{\ast}(\Phi_1\wedge\Phi_1)-14V_1 &=& 0  ~~, \nn\\
-\Phi_1\bigtriangleup_2\Phi_1 + 2{\ast}(V_1\wedge\Phi_1) -14\Phi_1&=& 0~~, \nn
\eeqa
(F6) (not writing the rank 0 component, equal with the rank 0 component of (F4)):
\beqa
 \iota_{V_1} K+\iota_{\Psi}\Phi_1 +{\ast}(\Phi_1\wedge\Phi_3)-15V_3 &=& 0  ~~, \nn\\
 -15 K+V_1\wedge V_3+\Phi_1\bigtriangleup_3\Phi_3-\iota_{K}\Phi_1 +\iota_{V_1}\Psi- {\ast}(\Phi_1\wedge\Psi) &=& 0  ~~, \nn\\
 K\wedge V_1 -\iota_{V_3}\Phi_1+\iota_{V_1}\Phi_3-{\ast}(K\wedge\Phi_1) -15\Psi+{\ast}(\Phi_1\bigtriangleup_1\Phi_3) -
\Psi\bigtriangleup_2\Phi_1 &=& 0 ~~,  \nn\\
 K\bigtriangleup_1\Phi_1 -\Phi_1\bigtriangleup_2\Phi_3
-{\ast}(\Psi\bigtriangleup_1\Phi_1)+{\ast}(V_1\wedge\Phi_3)+{\ast}(V_3\wedge\Phi_1)-15\Phi_3+V_1\wedge\Psi  &=& 0~~, \nn
\eeqa
(F8):
\beqa
 1+\langle\Phi_1,\Phi_2\rangle+\langle V_1,V_2\rangle &=& 0  ~~, \nn\\
 {\ast}(\Phi_1\wedge\Phi_2)+V_1+V_2 &=& 0  ~~, \nn\\
 V_1\wedge V_2+\Phi_1\bigtriangleup_3\Phi_2 &=& 0  ~~, \nn\\
 \iota_{V_1}\Phi_2-\iota_{V_2}\Phi_1+{\ast}(\Phi_1\bigtriangleup_1\Phi_2) &=& 0  ~~, \nn\\
 -\Phi_1\bigtriangleup_2\Phi_2 +{\ast}(V_1\wedge\Phi_2)+{\ast}(V_2\wedge\Phi_1)+\Phi_1 +\Phi_2&=& 0~~, \nn
\eeqa
(F9) (not writing the rank 0 component, equal with the rank 0 component of (F2)):
\beqa
2\iota_{K}\Psi +2\iota_{V_3} K+2\iota_{\Psi}\Phi_3+{\ast}(\Phi_3\wedge\Phi_3)-15V_2 &=& 0  ~~, \nn\\
 -\Psi\bigtriangleup_1\Psi -\Phi_3\bigtriangleup_2\Phi_3 - 2{\ast}(K\wedge\Psi) -K\wedge K -15\Phi_2 +2{\ast}(V_3\wedge\Phi_3)&+&\nn\\
         +2K\bigtriangleup_1\Phi_3 +2V_3\wedge\Psi-2{\ast}(\Psi\bigtriangleup_1\Phi_3) &=& 0~~, \nn
\eeqa
(F12):
\beqa
\langle\Phi_2,\Phi_3\rangle+\langle V_2,V_3\rangle &=& 0  ~~, \nn\\
 \iota_{V_2}K + \iota_{\Psi}\Phi_2 +{\ast}(\Phi_2\wedge\Phi_3)+V_3 &=& 0  ~~, \nn\\
 -\Phi_2\bigtriangleup_3\Phi_3-\iota_{V_2}\Psi+ {\ast}(\Phi_2\wedge\Psi)+\iota_K\Phi_2-K-V_2\wedge V_3 &=& 0~~, \nn\\
 -K\wedge V_2 +\iota_{V_3}\Phi_2-\iota_{V_2}\Phi_3+{\ast}(K\wedge\Phi_2) -\Psi-{\ast}(\Phi_2\bigtriangleup_1\Phi_3) +
\Psi\bigtriangleup_2\Phi_2 &=& 0  ~~, \nn\\
 K\bigtriangleup_1\Phi_2  -\Phi_2\bigtriangleup_2\Phi_3 -{\ast}(\Psi\bigtriangleup_1\Phi_2)+
{\ast}(V_2\wedge\Phi_3)+{\ast}(V_3\wedge\Phi_2)+\Phi_3+V_2\wedge\Psi &=& 0~~, \nn
\eeqa
and finally for (F16):
\beqa
|| \Phi_2||^2 + || V_2 ||^2 &=& 15 ~~, \nn\\
{\ast}(\Phi_2\wedge\Phi_2)-14V_2 &=& 0 ~~,  \nn\\
-\Phi_2\bigtriangleup_2\Phi_2 + 2{\ast}(V_2\wedge\Phi_2)-14\Phi_2&=& 0~~. \nn
\eeqa
The system of equations given above can be studied by elimination. Its
detailed analysis and implications are taken up in a forthcoming
publication.

\section{Conclusions and further directions}
\label{sec:conclusions}

We studied the \KA algebra of metric cones and cylinders and certain
subalgebras thereof, constructing a
number of isomorphic models which can be used to study the \KA algebra of
their unit sections and to lift generalized Killing equations,
Fierz isomorphisms etc. from a (pseudo-)Riemannian manifold to its metric
cylinder or cone. These results provide a toolkit for the study of
generalized Killing spinor equations on (pseudo-)Riemannian manifolds, 
being especially relevant to problems which arise in flux compactifications.
Our formulation is highly amenable to implementation in
various symbolic computation systems specialized in tensor algebra,
and we touched on two particular implementations which we have carried
out using {\tt Ricci} \cite{Ricci} and {\tt Mathematica}\textsuperscript{\textregistered} as well as {\tt
Cadabra} \cite{Cadabra}. We illustrated our techniques for the case of the most general flux
compactifications of M-theory which preserve ${\cal N}=2$ supersymmetry in
three dimensions, a class of compactifications whose most general
members were not studied before --- obtaining a complete description
of the differential and algebraic constraints on pinor bilinears and
uncovering the underlying algebraic structure.  A detailed analysis of
the resulting equations, geometry and physics is the subject of
ongoing work.

\acknowledgments{This work was supported by the CNCS projects
PN-II-RU-TE, contract number 77/2010, and PN-II-ID-PCE, contract
numbers 50/2011 and 121/2011. The work of C.I.L. was also supported by the 
Research Center Program of IBS (Institute for Basic Science) in Korea
(grant CA1205-01). The Center for Geometry and Physics is supported by the Government of 
Korea through the Research Center Program of IBS (Institute for Basic Science).  C.I.L. thanks
Perimeter Institute for hospitality and for providing an excellent and
stimulating research environment during the preparation of this paper. Research at Perimeter Institute is
supported by the Government of Canada through Industry Canada and by
the Province of Ontario through the Ministry of Economic Development
and Innovation. He also thanks Lilia Anguelova for interest and
for stimulating discussions as well as for critical input.}


\end{document}